\documentclass[12pt]{iopart}
\usepackage{iopams}
\usepackage{amssymb,amsfonts,mathrsfs}
\usepackage[toc,page]{appendix}
\usepackage{graphicx}
\usepackage{etoolbox}
\usepackage{dsfont}
\usepackage{stackrel}
\usepackage{enumitem}
\usepackage[T1]{fontenc}
\usepackage{xcolor}
\usepackage{etoolbox}

\definecolor{customgreen}{HTML}{0A982B}
\definecolor{seagreen}{HTML}{2e8b57}
\definecolor{teal}{HTML}{48d1cc}
\definecolor{olive}{HTML}{808000}
\definecolor{darkorange}{HTML}{ff8c00}
\definecolor{carnationpink}{HTML}{ffa6c9}
\definecolor{crimson}{HTML}{dc143c}
\usepackage[colorlinks = true,
            linkcolor = seagreen,
            urlcolor  = crimson,
            citecolor = darkorange,
            anchorcolor = blue]{hyperref}

\def\ket#1{\mathinner{|{#1}\rangle}}
\def\expect#1{\langle#1\rangle}

\def\ol#1{\bar{#1}}

\newcommand{\ii}{{\rm i}}
\newcommand{\dd}{{\rm d}}
\newcommand{\pp}{\xi}

\makeatletter
\long\def\@makefntext#1{\parindent 1em \noindent
 \makebox[1.5em][l]{\footnotesize\rm$\m@th{\arabic{footnote}}$}%
 \footnotesize\rm #1}
\def\@thefnmark{\textsuperscript{\arabic{footnote}}}
\def\@makefnmark{\textsuperscript{\arabic{footnote}}}
\makeatother

\makeatletter
\def\@mkboth#1#2{}
\newlength\appendixwidth
\preto\appendix{\addtocontents{toc}{\protect\patchl@section}}
\newcommand{\patchl@section}{%
  \settowidth{\appendixwidth}{\textbf{Appendix }}%
  \addtolength{\appendixwidth}{1.5em}%
  \patchcmd{\l@section}{1.5em}{\appendixwidth}{}{\ddt}%
}
\makeatother

\begin{document}

\title[Popcorn Drude weights]{Popcorn Drude weights from quantum symmetry}
\author{E Ilievski$^{1}$}
\address{$^{1}$ Faculty of Mathematics and Physics, University of Ljubljana, Jadranska 19, 1000 Ljubljana, Slovenia}
\ead{enej.ilievski@fmf.uni-lj.si}

\begin{abstract}
Integrable models provide emblematic examples of non-ergodic phenomena. One of
their most distinguished properties are divergent zero-frequency conductivities signalled by finite Drude weights.
Singular conductivities owe to long-lived quasiparticle excitations that propagate ballistically through the system
without any diffraction. The case of the celebrated quantum Heisenberg chain, one of the best-studied many-body paradigms, turns out
to be particularly mysterious. About a decade ago, it was found that the spin Drude weight in the critical phase
of the model assumes an extraordinary, nowhere continuous, dependence on the anisotropy parameter in the shape of a `popcorn function'.
This unprecedented discovery has been afterwards resolved at the level of the underlying deformed quantum symmetry algebra which
helps explaining the erratic nature of the quasiparticle spectrum at commensurate values of interaction anisotropy.
This work is devoted to the captivating phenomenon of discontinuous Drude weights,
with the aim to give a broader perspective on the topic by revisiting and reconciling various perspectives from the previous studies.
Moreover, it is argued that such an anomalous non-ergodic feature is not exclusive to the integrable spin chain
but can be instead expected in a number of other integrable systems that arise from realizations of the quantum
group $\mathcal{U}_{q}(\mathfrak{sl}(2))$, specialized to unimodular values of the quantum deformation parameter $q$.
Our discussion is framed in the context of gapless anisotropic quantum chains of higher spin and the sine-Gordon quantum field theory
in two space-time dimensions.
\end{abstract}

\maketitle

\tableofcontents

\eqnobysec

\section{Introduction}

Nonequilibrium phenomena in many-body quantum systems have been at the forefront of theoretical research for several decades now.
Theoretical interest has recently been shifted quite heavily towards low-dimensional models where effects of quantum correlations play
a crucial role. Further inspiration comes from numerous recent experimental advancements, particularly with cold-atom technologies, with
numerous landmark achievements \cite{Schemmer19,Jepsen20,Scheie21,Malvania21,Bloch_KPZ}.
At the same time, our understanding of non-ergodic dynamical systems and their transport properties
has also advanced quite tremendously \cite{transport_review}, particularly since the inception of the generalized hydrodynamics (GHD)
\cite{GHD_Doyon,GHD_Italy}. In the past couple of years, the formalism of GHD has established itself as a powerful and
versatile computational tool for investigating one-dimensional integrable systems \cite{Doyon_lectures,GHD_review}, including systems with weakly broken integrability \cite{Bastianello_review}.

Integrable models typically display behavior that is markedly different from that of generic interacting many-body systems.
Their unconventional dynamical behavior can be attributed to the lack of ergodicity which is implied by the presence of infinitely many local conservation laws in involution, arguably the most renowned hallmark feature of completely integrable systems.
Absence of ergodicity can likewise be associated with existence of stable interacting quasiparticles that undergo elastic collisions.
Using that such quasiparticles spread ballistically through the system without experiencing diffraction, one could readily conclude
that, at least qualitatively, integrable systems cannot be ordinary (i.e. diffusive) conductors.
Indeed, by computing the linear-response conductivities, one indeed finds a (delta-peak) singularity at zero frequency coupled to a
finite \emph{Drude weight} \cite{CZP95,transport_review}. In the setting of the linear-response theory, finite Drude weights are
conventionally related to incomplete relaxation of current autocorrelation functions at late times \cite{Mazur69,Suzuki71},
see also \cite{Ilievski12,Doyon_projections}. The fact that conservation laws preclude generic current-carrying perturbations from
fully dissipating signifies the lack of ergodicity.

It nonetheless came as a surprise that in certain situations the outlined simple qualitative picture cannot fully capture
the phenomenology of spin or charge transport in an integrable model. This in fact happens already in the
well-known anisotropic Heisenberg spin-$1/2$ chain~\footnote{The antiferromagnetic exchange coupling $J$ has been set to unity, $J=1$, leaving
anisotropy $\Delta$ as the free parameter of the model.}
\begin{equation}
{\rm H} = \sum_{i=1}^{L}\left({\rm S}^{x}_{i}{\rm S}^{x}_{i+1} + {\rm S}^{y}_{i}{\rm S}^{y}_{i+1}
+ \Delta\,\Big({\rm S}^{z}_{i}{\rm S}^{z}_{i+1}-\frac{1}{4}\Big)\right),
\label{eqn:XXZ_Hamiltonian}
\end{equation}
which is widely regarded as one of the most rudimentary integrable quantum systems out there.

The elusive nature of spin transport in the Heisenberg spin chain has been attracting attention for quite some time now.
It required more than two decades worth of effort since the seminal work \cite{CZP95,ZCP97,Zotos99} before (most of) the aspects
have been pieced together, see for example Refs.~
\cite{Prosen11,Marko11,PI13,DeLuca17,IN_Drude,IN17,Bulchandani18,Ilievski18,Gopalakrishnan18,Ljubotina19,GV19,GVW19,Urichuk19,LZP19,Urichuk21}.
As it turned out, the nature of spin transport in the Heisenberg chain is more nuanced than one might have expected.
Depending on the value of anisotropy parameter, one indeed finds \emph{three} distinct types of transport laws in equilibrium ensemble
at finite temperature (with no average magnetization): in the gapped phase, the spin Drude weight at half-filling is identically zero, with magnetization undergoing diffusive spreading \cite{DeNardis18,Gopalakrishnan18,DeNardis_SciPost,GHD_review},
by crossing into the critical phase, the spin Drude weight is rendered finite \cite{Zotos99,Prosen11};
These two markedly different dynamical regimes meet together at the isotropic (quantum critical) point where,
due to the restoration of nonabelian rotational symmetry, there is an onset of superdiffusive spin transport \cite{Marko11,Ilievski18,superuniversality} characterized by the fractional dynamical exponent of the
Kardar--Parisi--Zhang equation \cite{KPZ}, see Refs.~\cite{Ljubotina19,GV19,NGIV20,Vir20,superdiffusion_review}.
Finally, in the gapless phase the spin Drude weight was found to exhibit an unexpected, completely discontinuous, dependence on the anisotropy parameter \cite{PI13,IN_Drude,LZP19}. The latter feature, which is probably the most puzzling of all, will be the central subject of the ongoing discussion.

\medskip

In oder to reconcile the elusive aspects of such anomalous Drude weights, it is necessary to have a look `under the hood'
and systematically explore the exotic features that the underlying quantum symmetry has in store.
This line of research begun by Prosen's seminar work \cite{Prosen11} on the so-called quasilocal charges \cite{QLreview}, by demonstrating that the most salient features of finite-temperature magnetization transport in the Heisenberg XXZ chain
cannot be simply explained on the basis of standard local conservation laws known from textbooks \cite{Faddeev95,Faddeev16}. This pivotal insight led to a number of milestones. Firstly, an entire family of hitherto unknown
quasilocal charges have been algebraically constructed in Ref.~\cite{PI13} from commuting transfer matrices with \emph{non-unitary} auxiliary spin representations, facilitating a direct computation the
Mazur--Suzuki bound \cite{Mazur69,Suzuki71,Ilievski12} on the spin Drude weight \cite{PI13}. Apart from improving on the (suboptimal) bound of Ref.~\cite{Prosen11}, the result of Ref.~\cite{PI13} was found to comply with the earlier analytic result by Zotos \cite{Zotos99} obtained using the approach of Thermodynamic Bethe Ansatz (TBA),
later confirmed in Ref.~\cite{Benz05}. A few years later, a new protocol for computing exact Drude weights has been proposed in Refs.~\cite{IN_Drude,Bulchandani18} by leveraging the formalism of GHD \cite{GHD_Doyon,GHD_Italy}.
The hydrodynamic description of the Drude weights has been subsequently developed in Refs.~\cite{DS17,IN17}.
The still puzzling discontinuous behavior of the spin Drude weight in the gapless Heisenberg spin-$1/2$ chain has eventually been
demystified in Ref.~\cite{IN_Drude} which explained the hitherto missing connection between the magnonic bound states and the aforementioned
quasilocal charges by invoking the `string-charge duality' \cite{StringCharge,DeLuca17}.
Meanwhile, Ref.~\cite{IN_Drude} also elaborated on the discontinuous character of the finite-temperature spin Drude weight from the viewpoint of quasiparticle excitations. The exact high-temperature limit computed in Ref.~\cite{PI13}
has been afterwards recovered by other related analytical techniques \cite{Collura_wall,LZP19,Urichuk19}. Just recently, the computation also been extended to nonlinear spin Drude weights \cite{Urichuk22}. 
For a historical overview and other perspectives, we can recommend the reader a recent review article \cite{transport_review}. It would not be out of place to highlight that importance of the quasilocal conservation laws \cite{Prosen14,Pereira14,Zadnik16,Zadnik17}
reaches beyond the transport phenomena (see Ref.~\cite{QLreview} for a review). For instance, quasilocal charges also proved indispensable for constructing complete generalized Gibbs ensembles \cite{Ilievski_GGE,Landscape} and for computing the steady-state values of local observables following a global quench \cite{StringCharge,PVC16,IQC17} to explain anomalous thermalization in
integrable models \cite{VR16}.

\medskip

In this work, we revisit the enchanting phenomenon of discontinuous Drude weights.
We set out to discuss various interconnected aspects that are presently dispersed across several papers, and to highlight some of the open ends. Our discussion will be focused on the relevant regime of compact deformations with $q\in U(1)$,
where the model features an abrupt change in the structure of magnonic bound states upon (continuously) varying $q$ throughout the entire
critical interval. Such an anomaly is, as we shall explain in more detail, a direct corollary of eigenspectrum \emph{refragmentation},
deeply rooted in the fusion rules for the commuting transfer matrices. Discontinuous Drude weights can be indeed thought of as a physical manifestation of such a refragmentation. We shall also elaborate on a closely related effect of  `quantum truncation' induced at the root-of-unity deformations. In this view, we argue that such the phenomenon of `popcorn Drude weights' cannot be reserved
exclusively to transport coefficients of the Heisenberg spin-$1/2$ chain in the critical regime. With this in mind, we initiate the search for similar behavior in other solvable models and explore various possibilities
of realizing the $q$-deformed Hopf algebra $\mathcal{U}_{q}(\mathfrak{sl}(2))$.

The subsequent analysis is mostly framed in the language of generalized hydrodynamics.
Our aim is to keep our presentation focused on the physical aspects and hence we do not include comprehensive (or self-contained) exposition of the framework and the models in this paper. For additional information on the analytical techniques we can recommend the reader a series of recent review articles on the subject \cite{GHD_review,superdiffusion_review,Alba_review} intended for non-specialists.
We also decided to completely avoid explicit algebraic constructions of transfer matrices and conservation laws since, in our view, this extra layer of technicality would hardy add any extra value to the core message. Likewise, we shall not attempt to reproduce (often quite lengthy) derivations and recycle some of the results from the previous works. In particular, the general formulae for computing the Drude weights (within GHD) have already been obtained in Refs.~\cite{DS17,IN17}.
To employ these formulae, one requires some minimal amount of knowledge in the form of the quasiparticle spectrum and the associated kinematic data in the form of two-body scattering amplitudes. The state-dependent information is then supplied in the form of mode occupation functions of the reference equilibrium state. With this information at hand, one typically proceeds to solve the TBA dressing equations to deduce the dressed dispersion relations and dressed charges. The dressing transformation constitutes a system of coupled Fredholm-type integral equations. Analytic solutions to such equations are in practice accessible only in certain limits, while in general one has to resort to numerical solutions. Apart from the ground-state, the only major simplification can be achieved in the high-temperature limit where the dressing equation is rendered algebraic. For this reason, we will perform the calculations in the high-temperature regime.

\medskip

In this work, our nomenclature occasionally deviates from the previously used one. This specifically pertains to the discontinuous Drude weights which have previously been loosely referred to as `fractal'. Our perspective is that a different term would be more appropriate, given that fractals are conventionally refer to sets whose Hausdorff dimension is strictly lower that the topological dimension,
representing everywhere \emph{continuous} functions that are nowhere differentiable on their domain (e.g. the blancmange curve, to give an example). By contrast, the spin Drude weight in the gapless anisotropic Heisenberg chain is instead
given by a nowhere-continuous function of parameter $q=\exp{(\ii \gamma)}$ on a dense set of rational (also called commensurate) points $\gamma/\pi \in \mathbb{Q}$, which is continuous for irrational (incommensurate) values
$\gamma/\pi \in \mathbb{R}\setminus \mathbb{Q}$ -- functions of this type are known in the literature under the name of \emph{Thomae functions}, or more colloquially are known as `popcorn functions' \cite{Abbott_book} (since they remind of how popcorn bounces on a frying pan).

\paragraph*{\bf Outline.}
The paper is structured as follows. In \Sref{sec:popcorn}, we familiarize the reader with the most essential concepts give a non-technical introduction to the notion of popcorn Drude weights. Moving on, we touch upon certain aspects of practical relevance
and discuss the notion of refragmentation. In \Sref{sec:TBA}, we proceed by shortly introducing the basics of the TBA approach and summarize how to compute the Drude weights in the scope of generalized hydrodynamics. Section \Sref{sec:Heisenberg} is devoted to the Heisenberg spin-$1/2$ chain, where we dive deeper into discussing various concepts revolving around the popcorn spin Drude weight.
In \Sref{sec:XXZ} we compute the exact spin Drude weight in the XXZ Heisenberg spin-$1/2$ chain in the high-temperature limit by taking advantage of the hydrodynamic mode decomposition of the Drude weights and proceed to adapt the method for computing the spin Drude weight in anisotropic integrable chains of higher spin.
In \Sref{sec:sG}, we switch gears and focus our attention to the quantum sine-Gordon model. From the algebraic point of view, the model is closely related to the Heisenberg XXZ chain in the critical regime. We confine our analysis to the special discrete values of the coupling parameter at which the reflection amplitude of the scattering matrix vanishes. By invoking the exact correspondence between auxiliary topologically charge pseudo-particles and magnonic bound states in the spin chain, we obtain explicit formulae for computing the topological Drude weight. We also include some remarks regarding the case of general couplings.
In \ref{app:SUSY}, we briefly outline how the $q$-deformed quantum symmetry can be realized in certain exactly solvable one-dimensional
models of correlated electrons. We conclude in \Sref{sec:conclusion} by summarizing the key points.

\section{Popcorn Drude weights}
\label{sec:popcorn}

The peculiar findings of Ref.~\cite{PI13} have sparked animated discussions regarding the prospects of observing such discontinuous dependence in an experimental setting, with many arguing that the effect is no more than just a `mathematical artifact' of little physical significance. Before delving into technical aspects, we would like to share our perspectives to address this concern.

Discontinuous behavior of physical quantities is most typically associated with the notion of criticality and phase transitions.
In the case of thermodynamic phase transitions, for example, free energy develops a non-analyticity at a critical point in the limit of large system size. It is thus important to note that discontinuous behavior observed in the Drude peak is different in a fundamental way. To begin with, transport coefficients do not represent physical observables. Since conducitvities are defined in terms of time-integrated correlation functions (or their limiting values at late times), they may be perceived as `ergodic observables'. In particular, the Drude weights quantify non-ergodicity through the late-time decay of dynamical susceptibilities. Given that in an experiment (or even in simulations on a classical computer) we can only probe relaxation over a finite window of time, it can be very challenging to estimate the DC conductivities with a degree of precision. There will invariably be some degree smearing, causing the delta peak to `spill' over to small frequencies. One expects, to give a rough estimate, the zero-frequency peak to broaden over a narrow frequency window of size $\omega \sim 1/\tau$, where $\tau$ is the duration of an experiment.

The described general qualitative picture however changes quite radically when there are additional `commensurability effects' at play, referring to a `secret' hierarchy of timescales inherent to the eigenspectrum of the model. In such a scenario, only a tiny change in the model's parameter can suffice to appreciably affect the structure of thermodynamic eigenstates and, in effect, significantly impact the asymptotic behaviour of dynamical correlation functions. Indeed, the the critical regime in the Heisenberg XXZ spin chain is plagued by such commensurability effects,
being inherent to the underlying $\mathcal{U}_{q}(\mathfrak{sl}(2))$ quantum symmetry. To give a more physical perspective, we note that
such commensurability effects manifest themselves as the stability criterion for the formation of the magnonic bound states, depending very delicately on the deformation parameter $q=\exp{(\ii \gamma)}$ that parametrizes the axial anisotropy $\Delta = \cos{(\gamma)}$ of spin interactions, cf. \Eref{eqn:XXZ_Hamiltonian}. Additionally, commensurability effects are likewise responsible for pronounced finite-size effects, making it very difficult to reliably compute the spin Drude weight (or diffusion constant) by exact diagonalization and related techniques \cite{Herbrych11,transport_review,Mierzejewski21}. This is mainly the reason why the majority of numerical computations are restricted only to a set of discrete (isolated) points at $\pi/\gamma = \ell$ (for moderately small $\ell \in \mathbb{Z}_{+}$ ($\ell>2$) \cite{Karrasch12,Karrasch13,Karrasch17,Collura_wall}; at these particular values the largest stable bound state is composed of $\ell-1$ magnons, and dynamical correlations decay relatively rapidly. For certain other values of $\gamma$, the size of bound state can become exceedingly large in size. In practice, such behavior is nicely captured in the numerical simulation done in Ref.~\cite{LZP19}, which
arguably provides the most accurate numerical validation of the popcorn spin Drude weight to this date: by constructing a discrete-time version
of the Heisenberg chain, the authors have demonstrated how finite-time approximations to the spin Drude weight gradually collapse onto the limiting `popcorn function' as the simulation time becomes larger. It would certainly be very interesting to attempt detecting such a `silhouette' of the limiting function in an experiment using state-of-the-art quantum simulators \cite{Jepsen20,Bloch_KPZ}.

\medskip

Concerning experimental detection of commensurability effects, it is also worth pointing out certain conspicuous similarities to the famous Azbel--Hofstadter problem \cite{Azbel64,Hofstadter76}, which reveals a `fractal' eigenspectrum in the shape of a butterfly. As we clarify below, the phenomenon of popcorn Drude weights is indeed very intimately related to the discernible pattern of energy levels -- the Hofstadter butterfly.

The Azbel--Hofstadter model describes Bloch electrons hopping freely on a two-dimensional square lattice threaded by a magnetic flux.
The intricate structure of the eigenspectrum owes to the group of magnetic translations, referring precisely to the aforementioned
$\mathcal{U}_{\mathfrak{q}}(\mathfrak{sl}(2))$ Hopf algebra. In this particular incarnation, the deformation parameter $\mathfrak{q}$ controls the magnetic flux through the elementary plaquette. The model can be conveniently mapped to the Harper equation, a one-dimensional difference equation in a quasiperiodic potential (known as the almost Matthieu equation)
\begin{equation}
\psi_{n-1} + 2\lambda \cos{(k_{y}+2\pi n \phi)}\psi_{n} + \psi_{n+1} = E \psi_{n},
\end{equation}
where here $\phi$ is the flux piercing through the elementary plaquette (in units of the flux quantum). When flux $\phi$ is rational, namely for $\phi = \mathfrak{p}/\mathfrak{q}$, the spectrum comprises of $\mathfrak{q}$ bands.
At incommensurate points, referring to irrational multiples of $\pi$ with $\mathfrak{p},\mathfrak{q}\to \infty$,
the spectrum becomes an infinite Cantor set. At the critical point $\lambda=1$,
the spectrum is purely singular continuous with multifractal wavefunctions.

Even though the Harper equation represents a single-particle problem, it has been discovered that its eigenfunctions can be characterized in terms of solutions to the Bethe equations \cite{WZ94,FK95}. Time time, the deformation parameter $\mathfrak{q}$ that plays the role of system length. More specifically, the model corresponds to the two-site spin chain where the band number $\mathfrak{q}$ sets
the size of spins. For $\mathfrak{q}=e^{\ii\,\phi}$ and $\phi \in \mathbb{Q}$, the Harper equation may be recast as a functional equation in form of the Baxter's TQ-relation by identifying the wavefunction $\psi$ with the $Q$-function \cite{ATW98}.
In the `thermodynamic limit' $\mathfrak{q}\to \infty$, the solutions to the Bethe quantization equations can be fully described in
terms of the so-called Bethe strings and labelled by the Takahashi--Suzuki (TS) quantum numbers \cite{TS72}.\footnote{For finite $\mathfrak{p}$
and $\mathfrak{q}$, the exact analysis is obstructed by the fact the strings are no longer exact.}

In the Azbel--Hofstadter model, Hall conductance is a distinguished quantity of main physical interest. Hall conductance of the $k$th gap can be expressed as $\sigma_{k}=\delta N_{k}/\delta \phi$, where $\delta N_{k}$ is the number of energy levels that cross the energy $E$ inside the the $k$th gap \cite{ATW98}. Remarkably, the values of Hall conductances are in precise agreement
with the TS numbers. Remarkably, the TS numbers are in one-to-one correspondence with the unitary irreducible representations of
$\mathcal{U}_{q}(\mathfrak{su}(2))$ of finite dimension and definite parity when $q$ is specialized to a root of unity (see \ref{app:irreps} for details). As we discuss next, it is precisely the same TS numbers that govern refragmentation of the quasiparticle spectra in many-body integrable models featuring popcorn Drude weights.

\paragraph*{\bf Popcorn Drude weights from refragmentation.}
Figuratively speaking, one could say that the spin Drude weight of the gapless Heisenberg spin chain is `fingerprint'
of the $\mathcal{U}_{q}(\mathfrak{sl}(2))$ representation theory when restricted to the compact regime $q\in U(1)$.
Purely on formal grounds it is therefore quite reasonable to anticipate that similar effects will take place also in other
quantum integrable systems governed by same quantum symmetry. We accordingly propose the following conjecture:

\begin{quote}
\emph{In any quantum integrable system constructed from a quantum $R$-matrix invariant under $\mathcal{U}_{q}(\mathfrak{sl}(2))$
with $q\in U(1)$, the Drude weights, with the exception of those associated to exactly conserved currents, admit a nowhere-continuous dependence upon varying $q=e^{\ii \gamma}$, implying popcorn Drude weights for commensurate values
$\gamma/\pi \in \mathbb{Q}$.}
\end{quote}

\medskip

In order to unpack the above claim, we shall first exhibit here the main characteristics of the eigenspectrum in the gapless Heisenberg spin chain. In this respect, it proves vital to relate the quasiparticle content to the exceptional structure of the underlying symmetry algebra. For establishing a precise identification, it will be necessary to obtain complete classification of the unitary finite-dimensional irreducible representations of $\mathcal{U}_{q}(\mathfrak{sl}(2))$ for the particular case of root-of-unity deformations.
For compactness, we shall relegate the technical exposition to \ref{app:irreps} and here only attempt to give
a succinct summary of the most essential properties which will be of relevance for the ongoing discussion.

For non-compact deformations, namely for $q\in \mathbb{R}$, the finite-dimensional unitary irreducible representations of $\mathcal{U}_{q}(\mathfrak{sl}(2))$ are in a \emph{bijective} correspondence with the unitary irreducible representations
of `classical' Lie algebra $\mathfrak{su}(2)$ and, correspondingly, to evaluation representations of the associated undeformed Hopf algebra,
$Y(\mathfrak{su}(2))$ known as the Yangian. Such a one-to-one correspondence however no longer holds in the compact regime of deformations with $q$ on a unit circle, $q \in U(1)$. In other words, conditions for irreducibility depend crucially on the value of $q$. Another key distinction to semi-simple Lie algebras is that one no longer has a uniform fusion rule for constructing higher-dimensional representations in the compact regime. There are, in particular, three special features of particular significance:
\begin{itemize}
\item irreducible representations that are both of finite dimension and definite parity depend on deformation parameter $q \in U(1)$,
\item when $q=e^{\ii \gamma}$, with $\gamma/\pi \in \mathbb{Q}$, there is only a \emph{finite} number of unitary finite-dimensional
irreducible representations,
\item upon varying $q\in U(1)$, the number of unitary irreducible representations changes in a \emph{discontinuous} manner.
\end{itemize}

The aforementioned commensurability effects are indeed just the admissibility criteria for the unitary representations in disguise. It has also been found that the same condition governs the structure of the fusion relations for formation of thermodynamic quasiparticles in the form of magnonic bound states (the Bethe strings).

\subsection{Thermodynamic Bethe Ansatz}
\label{sec:TBA}

To set the stage for later considerations we proceed with an elementary introduction to the TBA formalism. We shall not attempt here to give a general (abstract) formulation of the TBA equations. Moreover, we shall
assume there is no internal degrees of freedom that induce nested spectra. Rather, we consider the anisotropic Heisneberg spin-$S$ quantum chains \cite{KR87I,KR87II} as our working example. Afterwards, we briefly explain how the Drude weights can be computed in the formalism of GHD and quote the universal formula \cite{DS17,IN17} that requires the TBA data as an input.

Any admissible local equilibrium state in the model can be described by a generalized Gibbs ensemble \cite{Ilievski_GGE,IQC17}. In a finite volume, an unnormalized density matrix $\varrho_{L}$ takes the form
\begin{equation}
\varrho_{L} = \exp{\left[-\sum_{i}\beta_{i}{\rm Q}^{(i)}+h\,{\rm S}^{z}_{\rm tot}\right]},
\end{equation}
where ${\rm Q}^{(i)}$ represent the local (quasilocal included) conserved charges of the model (enumerated by a discrete label $i$) while ${\rm S}^{z}_{\rm tot}$ is the conserved total projection of spin onto the anisotropy axis $z$.
To each charge ${\rm Q}^{(i)}$ we have assigned a chemical potential $\beta_{i}$ (adopting the usual convention in which
$\beta_{1}\equiv \beta=1/T$ and ${\rm Q}^{(1)}\equiv {\rm H}$).

The TBA approach, pioneered in the seminal paper by Yang and Yang \cite{YY69} and subsequently developed largely by Takahashi \cite{Takahashi_Heisenberg,TS72} and Gaudin \cite{Gaudin71}, enables to evaluate the equilibrium free energy exactly.
Denoting the GGE partition function by $\mathcal{Z}_{L} = \Tr(\varrho_{L})$, the free energy density is defined via
$-\beta\,f=\lim_{L\to \infty}\case{1}{L}\log \mathcal{Z}_{L}$.
The computation is performed with aid of a functional-integral representation of $\mathcal{Z}_{L}$ by using the densities of Bethe roots (associated to the quasiparticles of the model) as the variational basis.

\paragraph*{\bf Thermodynamic state functions.}

Thermodynamic ensembles represent local equilibrium states compatible with the maximum entropy principle. Employing the `rapidity representation', they can be described in terms of quasiparticle densities $\rho_{s}(\theta)$. Here and subsequently we make use of the following notation: an integer `type index' $s$ corresponds to the number of (magnon) quanta, while $\theta \in \mathbb{R}$ is the rapidity variable parametrizing the quasiparticles' bare momenta. The complete set of densities $\{\rho_{s}(\theta)\}$  store enough information to (uniquely) determine all the local correlation functions in generalized Gibbs ensembles. Moreover, this information cannot in general
be further compressed \cite{IQC17,Landscape}, and hence $\rho_{s}(\theta)$ provide a convenient parametrization of equilibrium macrostates, i.e. they can be understood as continuous coordinates of a manifold of maximum-entropy equilibrium states the model \cite{IQC17}.\footnote{Such manifolds arise as a limit of two-dimensional torus partition sums, see e.g. Ref.~\cite{Landscape}}

Following the lines of the Yang--Yang's approach \cite{YY69}, we then introduce the densities of unoccupied modes $\ol{\rho}_{s}(\theta)$, called \emph{hole densities}, and accordingly define the total \emph{density of states}
\begin{equation}
\rho^{\rm tot}_{s}(\theta)=\rho_{s}(\theta)+\ol{\rho}_{s}(\theta).
\end{equation}
Due interaction between magnons, the particle and hole densities are not independent from one another. Interactions are taken into account by the quantization condition; in a chain of length $L$, rapidities $\{\theta_{j}\}_{j=1}^{M}$ of $M$ magnons subjected to the periodic boundary conditions satisfy the celebrated Bethe Ansatz equations
\begin{equation}
e^{\ii p(\theta_{j}) L} \prod^{M}_{j\neq k}S(\theta_{j},\theta_{k}) = 1.
\label{eqn:Bethe_equations}
\end{equation}total
By introducing $p^{\prime}_{s}(\theta)\equiv \partial_{\theta}p_{s}(\theta)$ and
the associated $\mathbb{Z}_{2}$ parities $\kappa_{s}\equiv {\rm sign}[p^{\prime}_{s}(\theta)]$, the large-volume limit of \Eref{eqn:Bethe_equations} yields the following Bethe--Yang integral equations
\begin{equation}
\rho^{\rm tot}_{s} = \frac{\kappa_{j}}{2\pi}p^{\prime}_{s} - \mathcal{K}_{s,s'}\star \rho_{s'}.
\label{eqn:Bethe-Yang}
\end{equation}
Here and the rest of the way we employed the economic notation
$\mathcal{K}_{s,s'}\star g_{s'}\equiv \sum_{s'}\int_{\mathbb{R}} \dd \theta\,\mathcal{K}_{s,s'}(\theta-\theta^{\prime})g_{s'}(\theta^{\prime})$
for the convolution-type integrals (with respect to dummy functions $g_{s}(\theta)$, with implicit range of summation over $s^{\prime}$.

By exciting an unoccupied quasiparticle, say of type $s$ with rapidity $\theta$, the energy of a reference states will be increased by an amount $e_{s}(\theta)$. The latter are regarded as bare energies (measured relative to the quasiparticle pseudovacuum state). In the thermodynamic limit, the energy density of a macrostate is therefore given by the following mode resolution $e=\sum_{s}\int \dd \theta\,\rho_{s}(\theta)e_{s}(\theta)$, mirroring the Fourier representation of the energy dispersion in noninteracting theories.
Switching from microscopic quasiparticle configurations to the respective rapidity distributions can be seen as a coarse-graining procedure with the associated combinatorial weight
$\mathfrak{s}=\sum_{s}\int \dd \theta\,\mathfrak{s}_{s}(\theta)$, otherwise known
as the \emph{Yang--Yang entropy} \cite{YY69}. We shall assume that the excitations of the model obey the Fermi--Dirac statistics, with each quasiparticle specie contributing to the total entropy an amount
$\mathfrak{s}_{s}(\theta)=\rho_{s}\log{[1+\ol{\rho}_{s}(\theta)/\rho_{s}(\theta)]}+\ol{\rho}_{s}(\theta)\log{[1+\rho_{s}(\theta)/\ol{\rho}_{s}(\theta)]}$.
The equilibrium partition function can be accordingly cast in the functional-integral form,
\begin{equation}
\mathcal{Z}_{L} = \int {\rm D}[\{\rho_{s}\}]
\exp{\left[-L\sum_{s}\int \dd \theta \Big(\mu_{s}(\theta)\rho_{s}(\theta)+\mathfrak{s}_{j}(\theta)\Big)\right]}.
\end{equation}
The GGE chemical potentials $\beta_{i}$ have been transcribed into mode `fugacities' $\mu_{s}(\theta)$, see e.g. Ref.~\cite{IQC17}. We stress that, apart from fixing the statistical factor, the outlined derivation is entirely general until this point, i.e.
we have not incorporated any extra assumptions about the model. We shall now postpone the discussion of the quasiparticle spectrum to \Sref{sec:Heisenberg} and carry on with general considerations. 

The next step is to evaluate the partition sum in the thermodynamic limit. This amounts to take $L\to \infty$ while keeping the total number of excitations $M\sim \mathcal{O}(L)$ fixed. Since the partition sum $\mathcal{Z}_{L}$ for large $L$
is dominated by a unique saddle point, one seeks the variational minimum of the free-energy functional $f$, given by $f=e-T\,\mathfrak{s}$.
In doing so, one has to account for the fact that hole densities depends on the quasiparticles densities
by virtue of \Eref{eqn:Bethe-Yang}, yielding $\delta \ol{\rho}_{s} = -K_{s,s'}\star \delta \rho_{s'}$.
The ensuing saddle-point equations, $\delta f/\delta \rho_{j}=0$, take the form of an infinite system of coupled non linear
integral equations of the form
\begin{equation}
\log \mathcal{Y}_{s} = \mu_{s} + K_{s,s'}\star \log\big(1+1/\mathcal{Y}_{s'}\big).
\end{equation}
As customary, we will be referring to these as the canonical TBA equations \cite{YY69,Takahashi_Heisenberg}. Above we have
written them in terms of the so-called thermodynamic $\mathcal{Y}$-functions, defined as the ratios
\begin{equation}
\mathcal{Y}_{s}(\theta) = \frac{\ol{\rho}_{s}(\theta)}{\rho_{s}(\theta)}.
\end{equation}
The Fermi mode occupation functions, given by
\begin{equation}
n_{s}(\theta)= \frac{\rho_{s}(\theta)}{\rho^{\rm tot}_{s}(\theta)} = \frac{1}{1+\mathcal{Y}_{s}(\theta)}.
\end{equation}
likewise provide complete information about equilibrium macrostates, namely they are in one-to-one correspondence to the densities $\rho_{s}(\theta)$.
In the Heisenberg spin-$S$ chain, the density of free energy can be obtained by summing over the whole spectrum of physical excitations
\begin{equation}
\mathfrak{f}^{(2S)} \equiv -\beta f^{(2S)} = \sum_{s}\int \dd \theta\,\kappa_{s}K^{(2S)}_{s}(\theta)\log\big(1+1/\mathcal{Y}_{s}(\theta)\big).
\label{eqn:f_canonical}
\end{equation}

\paragraph*{Thermodynamic averages and dressed charges.}
Evaluating the equilibrium expectation values of the (quasi)local charges,
\begin{equation}
q^{(i)}=\lim_{L\to \infty}\frac{1}{L}\expect{{\rm Q}^{(i)}}
= \lim_{L\to \infty}\frac{1}{L}\frac{1}{\mathcal{Z}_{L}}\Tr \left[{\rm Q}^{(i)}\varrho_{L}\right],
\end{equation}
is a fairly straightforward task. The averages $q^{(i)}$ can be most directly computed with help of the mode resolution
\begin{equation}
q^{(i)} = \sum_{s}\int \dd \theta\,\rho_{s}(\theta)q^{(i)}_{s}(\theta)
= \sum_{s}\int \dd \theta\,\rho^{\rm tot}_{s}(\theta)n_{s}(\theta)q^{(i)}_{s}(\theta),
\label{eqn:mode_resolution}
\end{equation}
where $q^{(i)}_{s}(\theta)$ stand for \emph{bare} charges, representing single-particle contributions to total
charge density $q^{(i)}$ of a state. Particularly, the magnetization density
$S^{z}=\lim_{L\to \infty}\case{1}{L}\,\expect{{\rm S}^{z}_{\rm tot}}$, is given by
\begin{equation}
S^{z} = \frac{S}{2} - \sum_{s=1}^{\infty}\int \dd \theta\,m_{s}\, \rho_{s}(\theta),
\label{eqn:magnetization_qp}
\end{equation}
where $m_{s}$ is bare magnetization carried by specie $s$. Particularly, $m_{s}$ is the number of constituent magnons in a bound state.
Importantly however, the internal magnon content of bound states depends on anisotropy $\Delta$. In the gapped regime, $|\Delta|\geq 1$, the $s$th specie is a compound of $s$ bound magnons known as an `$s$-string' \cite{Takahashi_Heisenberg};
in the critical regime, $|\Delta|<1$, $m_{s}$ are instead provided by the Takahashi--Suzuki numbers ${\rm n}_{s}$ \cite{TS72},
that is ${\rm n}_{s}=m_{s}$, to be introduced below in \Sref{sec:Heisenberg}.

\medskip

The ensemble averages of the (quasi)local conserved charges can alternatively be inferred directly from free energy by taking derivative with respect to chemical potentials $\beta_{i}$, that is
\begin{equation}
q^{(i)} = \frac{\partial \mathfrak{f}}{\partial \beta_{i}}.
\label{eqn:q_from_f}
\end{equation}
By computing the change in pseudo-energies $\varepsilon_{s}(\theta)\equiv \log \mathcal{Y}_{s}(\theta)$ upon varying $\beta_{i}$
yields the \emph{dressed} charges ascribed to individual quasiparticles,
\begin{equation}
[q^{(i)}_{s}(\theta)]^{\rm dr} = \frac{\partial \varepsilon_{s}(\theta)}{\partial \beta_{i}}.
\end{equation}
From \Eref{eqn:q_from_f}, we can deduce the following `momentum representation' of the averages
\begin{equation}
q^{(i)} = \sum_{s}\int \dd p_{s}\frac{\kappa_{s}}{2\pi} n_{s}(\theta)[q^{(i)}_{s}(\theta)]^{\rm dr}
= \sum_{s=1}\kappa_{s}\int \frac{\dd \theta}{2\pi}p^{\prime}_{s}(\theta)n_{s}(\theta)[q^{(i)}_{s}(\theta)]^{\rm dr}.
\label{eqn:momentum_resolution}
\end{equation}
By further making use of the symmetry identity,
$\sum_{s}\kappa_{s}\int \dd p_{s}g_{s}n_{s}q^{\rm dr}_{s}=\sum_{s}\kappa_{s}\int \dd p_{s}g^{\rm dr}_{s}n_{s}q_{s}$
for $g_{s}(\theta)\to 1_{s}(\theta)$, we arrive at (cf. Ref.\cite{Doyon_lectures})
\begin{equation}
q^{(i)} = \sum_{s}\kappa_{s}\int \frac{\dd p_{s}}{2\pi}[1_{s}(\theta)]^{\rm dr}n_{s}(\theta)q^{(i)}_{s}(\theta).
\end{equation}
Making a quick comparison to the mode resolution \eref{eqn:mode_resolution}, while recalling the identification
$p^{\prime}_{s}(\theta)=2\pi \kappa_{s}\rho^{\rm tot}_{s}(\theta)$, we conclude that the `dressed identity'\footnote{Here, the dressed identity is to be understood as the formal solution to the (matrix) dressing equation
$({\bf 1}+\boldsymbol{\mathcal{K}}{\bf n})\star {\bf 1}^{\rm dr}={\bf 1}$.} $1^{\rm dr}_{s}$ is the Jacobian of the coordinate transformation $[p^{\prime}_{s}(\theta)]^{\rm dr} \mapsto p^{\prime}_{s}(\theta)$, namely
$[p^{\prime}_{s}(\theta)]^{\rm dr}=p^{\prime}_{s}(\theta)\,[1_{s}(\theta)]^{\rm dr}$.

\medskip

\paragraph*{\bf Refragmentation and local observables.}
Before turning our attention to dynamical quantities, we would like to take a moment to first examine whether refragmentation can have any influence on equilibrium averages of local physical observables.

Using the fact that the total number of quasiparticles in the spectrum, including their internal structure, both change discontinuously with anisotropy $\Delta = \cos{(\gamma)}$ in the regime $\gamma \in (0,\pi/2)$,
it appears entirely plausible (at the level of \Eref{eqn:f_canonical}, at least)
that due to reorganization of the magnon spectrum upon varying the interaction parameter $\gamma$, the free energy function
could experience jumps. In the gapped regime $\Delta = \cosh{(\eta)}$ (corresponding to the `hyperbolic regime' $q = e^{\eta}\in \mathbb{R}$)
and at the isotropic point ($\eta \to 0$), the situation is contrastingly different, we instead the quasiparticle spectrum comprises an infinite tower of magnonic states carrying any integer number of quanta $s \in \mathbb{N}$, thus rendering the expectation values (cf. \Eref{eqn:mode_resolution}) manifestly continuous in interaction parameter $\eta$.

According to the well-known theorem by Araki \cite{Araki69}, quantum lattice Hamiltonians with local and bounded interactions do undergo a phase transition in one spatial dimension. As Araki's result suggests, in spite of an erratic behavior of
$\mathcal{U}_{q}(\mathfrak{sl}(2))$ fusion rules in the critical phase, there apparently exist some underlying mechanism that leads to smooth redistribution of the spectral weight, thereby ensuring continuous dependence of thermodynamic averages.

The expectation values of the (quasi)local charges indeed undergo continuous dependence on anisotropy parameter $\Delta$, including in the regime of compact deformations. To corroborate on this, we exploit a useful `inversion identity' for the TBA integral kernel that reflects the fusion algebra of the model\footnote{The same property is commoly exploited to transform the canonical TBA equations in a more convenient partially decoupled form \cite{Takahashi_Hubbard} which amounts to
find the (pseudo)inverse of the Fredholm scattering kernel \cite{IQC17}.}, enabling explicit resumation over the quasiparticles. The outcome is that in the canonical representation of the free energy, cf. \Eref{eqn:f_canonical}, can be compressed down to
\begin{equation}
\mathfrak{f}^{(2S)} = \int \dd \theta\,s(\theta)\log \big(1+\mathcal{Y}_{2S}(\theta)\big).
\label{eqn:f_compact}
\end{equation}
In this formula, dependence on the interaction parameter $q$ is hidden implicitly in the convolution `$s$-kernel' $s(\theta)$, with the following Fourier representation
$s(\theta)=\case{1}{2\pi}\int_{\mathbb{R}}e^{-\ii\,\omega\,\theta}[2\cosh{(\case{\gamma}{2}\omega)}]^{-1}$. Remarkably, \Eref{eqn:f_compact} involves merely a pseudoenergy of a single quasiparticle mode\footnote{It deserves to mention that the same 
conclusion  can be reached within the `Quantum Transfer Matrix' approach \cite{KP92} which, assuming canonical Gibbs equilibrium, permits to obtain 
a single  nonlinear integral equation for the free-energy density \cite{equivalence}.}; the free-energy density of the system is expressible in 
terms of a  single thermodynamic\footnote{Functions $\mathcal{Y}_{s}(\theta)$ should not be confused with $Y$-functions that encode individual eigenstates of the finite-volume spin chain.} $\mathcal{Y}$-function, that is
$\mathcal{Y}_{2S}(\theta) = \exp{(\varepsilon_{2S}(\theta))}$, where $\varepsilon_{2S}(\theta)$ is the TBA pseudoenergy ascribed to the bound state ($s$-string) comprising of $s=2S$ magnons.

Despite refragmentation, one finds the partial filling fractions
$\int \dd \theta\,\rho_{s}(\theta)$ to smoothly redistribute under continuously changing the anisotropy parameter $\gamma$,
cf. \Eref{eqn:magnetization_qp}.\footnote{Explicit resummation of partial filling fractions restricted to the half-filled sector is given in Appendix D of Ref.~\cite{StringCharge}.} A simple consistency check is to verify that the entropy and magnetization sum rules are satisfied. The latter is most conveniently carried out in the high-temperature limit $\beta \to 0$ where all the $\mathcal{Y}$-functions 
become flat (i.e. independent of rapidity $\theta$). To better corroborate this fact, it proves useful to employ alternative `gauge-covariant representation' in terms of the so-called thermodynamic $\mathcal{T}$-functions $\mathcal{T}_{j}(\theta)$, allowing to recast \Eref{eqn:f_compact} in the following form \cite{Landscape}
\begin{equation}
\mathfrak{f}^{(2S)} = \int \dd \theta\,s(\theta)
\log \left[\frac{\mathcal{T}_{2S}(\theta+\case{\ii}{2})\mathcal{T}_{2S}(\theta-\case{\ii}{2})}{\Phi^{(2S)}(\theta+(s+1)\case{\ii}{2})}\right].
\end{equation}
The $\Phi$-potential $\Phi^{(2S)}$ appearing in the normalization is (besides $\mathcal{T}_{1}(\theta)$) an initial condition, depending only on the physical degrees of freedom (i.e. spin $S$ and, in general, on the inhomogeneities).
The above formula can be most suggestively understood from the viewpoint of an explicit vertex-model realization of the equilibrium torus partition sum, see Ref.~\cite{Landscape}, where functions $\mathcal{T}_{s}(\theta)$ formally represent the analytic continuation of the leading (i.e. dominant) eigenvalue of commuting \emph{column} transfer matrices with a spin-$s/2$ representation in the `physical channel' (bearing a formal similarity the Quantum Transfer Matrix construction of canonical Gibbs states \cite{QTM_review}). The main upshot of here is that so far the $\mathcal{T}$-function $\mathcal{T}_{2S}(\theta)$ remains a part of the fusion hierarchy (also known as the $T$-system), the averages of the (quasi)local charges will assume continuous dependence on the interaction parameter $q$ (resp. anisotropy $\Delta$). In addition, using that $\int \dd \theta\,s(\theta)=1/2$, the $\mathcal{T}$-functions $\mathcal{T}_{j}(\theta)$ reduce
to classical $\mathfrak{su}(2)$ characters of the spin-$S$ representation $\mathcal{V}_{2S}$,
$\chi_{2S}(h)=\Tr[e^{h\,{\rm S}^{z}}]=\sum_{k=-S}^{S}e^{k\,h}$, implying $\mathfrak{f}^{(2S)}(h)=\log \chi_{2S}(h)$ and hence
$S^{z}(h)=\partial_{h} \mathfrak{f}^{(2S)}(h)= \partial_{h}\log \chi_{2S}(h)$.

Strictly speaking, the outlined logic applies only to the conservation laws of the model, and to static susceptibilities associated to higher derivatives of the free energy. Whether also \emph{arbitrary} (quasi)local observables display smooth dependence on $\gamma$ does not seem very obvious. Indeed, explicit expressions for computing equilibrium expectation values of short-range correlators
(valid in any generalized Gibbs ensemble) in the Heisenberg spin-$1/2$ chain presented in Ref.~\cite{Mestyan14} are only applicable in the $|\Delta|\geq 1$ regime. The formulae of Ref.~\cite{Mestyan14} make use of the full set of TBA state functions.
Given that Araki theorem \cite{Araki69} applies to generic local observables (and not just to conserved charges), we find it plausible that such formulae for the short-range correlators may be further compressed.

\subsection{Drude weights from hydrodynamics}
\label{sec:Drude}

We now recapitulate how the Drude weights can be computed exactly in the framework of GHD. For a comprehensive exposition we refer the reader to one of the recent review articles \cite{GHD_review,superdiffusion_review}
or lecture notes \cite{Doyon_lectures}.

In a nutshell, the equations of generalized hydrodynamics govern the dynamics of
infinitely many coupled (quasi)local conservation laws on large spatiotemporal scale.
Note that each of the (quasi)local conserved charges ${\rm Q}^{(i)}=\int \dd x\,{\rm q}^{(i)}(x,t)$ obeys the local continuity equation~\footnote{To keep the presentation simple, we avoid making any explicit distinction between continuous and lattice systems. In lattice models, the same equations hold with spatial derivatives replaced by the first difference.} already at the operatorial level,
\begin{equation}
\partial_{t}\,{\rm q}^{(i)}(x,t)+\partial_{x}\,{\rm j}^{(i)}(x,t)=0,
\end{equation}
where ${\rm j}^{(i)}$ represent the current densities of extensive currents ${\rm J}^{(i)}=\int \dd x\,{\rm j}^{(i)}(x)$.
The hydrodynamic regime only captures the slowest fluctuations of averaged charges in the limit of large wavelengths. Accordingly, the averaged local densities $q^{(i)}$ and $j^{(i)}$ are promoted to dynamical (hydrodynamic) variables that
can be viewed as classical fields, namely $q^{(i)}(x,t)$ and $j^{(i)}(x,t)$, respectively.

In accordance with Kubo formula, the Drude weights $\mathcal{D}_{ij}$ are defined as the long-time limit of the spatially-integrated dynamical two-point current correlation functions
\begin{equation}
\mathcal{D}_{ij} = \lim_{t\to \infty}\int \dd x\,\expect{{\rm j}^{(i)}(x,t){\rm j}^{(j)}(0,0)}^{\rm c},
\label{eqn:Drude_autocorrelator}
\end{equation}
formally representing the coefficients (i.e. matrix elements) of a \emph{positive semi-definite} matrix $\mathcal{D}$. By generalizing the method of hydrodynamic projection developed by Mazur \cite{Mazur69} and Suzuki \cite{Suzuki71} to continuous
spaces \cite{Doyon_projections}, $\mathcal{D}$ can be decomposed as \cite{DS17}
\begin{equation}
\mathcal{D}=\mathcal{B}\,\mathcal{C}^{-1}\,\mathcal{B}^{\rm T},
\label{eqn:hydrodynamic_projection}
\end{equation}
where $\mathcal{B}=\mathcal{B}^{T}$ and $\mathcal{C}$ denote equal-time charge-current and charge-charge susceptibilities
(covariance matrices), respectively,
\begin{equation}
\mathcal{B}_{ij} = \int \dd x\,\expect{{\rm j}^{(i)}(x,0){\rm q}^{(j)}(0,0)}^{c},\quad
\mathcal{C}_{ij} = \int \dd x\,\expect{{\rm q}^{(i)}(x,0){\rm q}^{(j)}(0,0)}^{c}.
\end{equation}
Static susceptibilities assume the following mode decompositions
\begin{equation}
\mathcal{B}_{ij} = -\frac{\partial j^{(i)}}{\partial \beta_{j}}=-\frac{\partial^{2}\mathfrak{g}}{\partial \beta_{i}\partial \beta_{j}},\qquad
\mathcal{C}_{ij} = -\frac{\partial q^{(i)}}{\partial \beta_{j}} = -\frac{\partial^{2}\mathfrak{f}}{\partial \beta_{i}\partial \beta_{j}},
\end{equation}
where we have introduced the `free-energy flux' $\mathfrak{g}$ via
\begin{equation}
j^{(i)} = \frac{\partial \mathfrak{g}}{\partial \beta_{i}}
= \sum_{s=1}^{\infty}\int \dd \theta\,\rho_{s}(\theta)j^{(i)}_{s}(\theta),
\end{equation}
with
\begin{equation}
j^{(i)}_{s}(\theta) \equiv v^{\rm eff}_{s}(\theta)q^{(i)}_{s}(\theta),
\end{equation}
designating the fluxes carried by individual modes\footnote{This relation, originally proposed in Refs.\cite{GHD_Doyon,GHD_Italy},
has been derived from first principles in Refs.~\cite{Pozsgay_PRX,Pozsgay_currents}.} propagating with effective velocities $v^{\rm eff}_{s}(\theta)$. The latter formally correspond to the spectrum of the (linear) propagator,
\begin{equation}
\mathcal{A}_{ij} = \frac{\partial j^{(i)}}{\partial q^{(j)}}.
\end{equation}
The so-called \emph{flux Jacobian} $\mathcal{A}$ can also understood as the charge of variables from the charges to currents. We note that, by the chain rule property,
we have the matrix relation $\mathcal{B}=\mathcal{A}\,\mathcal{C}$. By further taking into account the symmetries $\mathcal{C}=\mathcal{C}^{\rm T}$ and $\mathcal{B}=\mathcal{B}^{\rm T}$ (which is a corollary of Onsager's reciprocity,
see Ref.~\cite{Doyon_lectures}), one can arrive at a useful identity $\mathcal{A}\,\mathcal{C}=\mathcal{C}\,\mathcal{A}^{\rm T}$,
permitting to express the hydrodynamic projection \eref{eqn:hydrodynamic_projection} in a compact form \cite{DS17}
\begin{equation}
\mathcal{D}=\mathcal{A}\,\mathcal{C}\,\mathcal{A}^{\rm T}.
\label{eqn:Drude_Jacobian}
\end{equation}

Eigenfunctions of $\mathcal{A}$ provide the sought-for \emph{normal modes} of GHD.
They can be pictured as small large-scale fluctuations $\delta n_{j}(\theta;x,t)$ of the mode occupation functions functions $n_{j}(\theta;x,t)$\footnote{Fluctuations of the macrostate densities, $\delta \rho_{j}(\theta;x,t)$, are not the normal modes.}.
In terms of the normal modes, the GHD equations take the Riemann normal form
\begin{equation}
\partial_{t}n_{s}(\theta;x,t)+v^{\rm eff}_{s}(\theta)\partial_{x}n_{s}(\theta;x,t)=0.
\end{equation}
Eigenvalues of $\mathcal{A}$ are thus the \emph{effective} velocities of propagation, denoted by $v^{\rm eff}_{j}(\theta)$;
employing matrix notation by flattening the quasiparticle's indices, $(s,\theta)\to i$, we have that $v^{\rm eff} = \mathcal{S}\,\mathcal{A}\,\mathcal{S}^{-1}$ with $\mathcal{S}_{ij}=\partial n_{i}/\partial q_{j}$.

The effective velocities $v^{\rm eff}_{j}(\theta)$ of ballistic modes correspond physically to group velocities of quasiparticles immersed in an equilibrated finite-density background state \cite{GHD_Doyon,GHD_Italy}
\begin{equation}
v^{\rm eff}_{j}(\theta) = \frac{\partial \varepsilon_{j}}{\partial p_{j}}
= \frac{\varepsilon^{\prime}_{j}(\theta)}{p^{\prime}_{j}(\theta)},
\end{equation}
where prime, as usual, stands for rapidity derivative. With aid of the following mode resolution of static susceptibilities,
\begin{eqnarray}
\label{eqn:C_resolution}
\mathcal{C}_{i j} &= \sum_{s}\int \dd \theta\,\chi_{s}(\theta)[q^{(i)}_{s}(\theta)]^{\rm dr}[q^{(j)}_{s}(\theta)]^{\rm dr},\\
\mathcal{B}_{i j} &= \sum_{s}\int \dd \theta\,\chi_{s}(\theta)v^{\rm eff}_{s}(\theta)[q^{(i)}_{s}(\theta)]^{\rm dr}[q^{(j)}_{s}(\theta)]^{\rm dr},
\end{eqnarray}
with `mode susceptibilities'
\begin{equation}
\chi_{s}(\theta)\equiv \rho_{s}(\theta)\big(1-n_{s}(\theta)\big) = \rho^{\rm tot}_{s}(\theta)n_{s}(\theta)\big(1-n_{s}(\theta)\big),
\end{equation}
\Eref{eqn:Drude_Jacobian} can be recast entirely in terms of the `hydrodynamic data' of GHD. The final outcome is the following compact universal formula \cite{DS17,IN17}
\begin{equation}
\mathcal{D}_{ij} = \sum_{s}\int \dd \theta\,\chi_{s}(\theta)
\big(v^{\rm eff}_{s}(\theta)\big)^{2}[q^{(i)}_{s}(\theta)]^{\rm dr}[q^{(j)}_{s}(\theta)]^{\rm dr},
\label{eqn:Drude_weight_resolution}
\end{equation}
where the summation goes over the entire spectrum of thermodynamic excitations and integration is over the respective rapidity domains. The Drude weights have thus been expressed solely in terms of the TBA thermodynamic state functions through the statistical factor $\chi_{s}(\theta)$, the effective velocities, and the dressed charges of the normal modes. Finally, we remark that formula \eref{eqn:Drude_weight_resolution} is completely general, apart from an assumption that quasiparticles obey the Fermi--Dirac statistics (as is typically the case in quantum integrable lattice models and field theories). 

\medskip

Even though the outlined mode resolution of the Drude weights, \Eref{eqn:Drude_weight_resolution}, closely resembles the
mode decompositions of static susceptibilities (cf. \Eref{eqn:C_resolution}) and thermodynamic averages of the (quasi)local charges, \Eref{eqn:mode_resolution}, there is one crucial difference however: the Drude weights are generally not expressible as a variation of an appropriate thermodynamic potential (the exception to this are obviously the conserved currents, $\dd {\rm J}_{i}/\dd t=0$).
Accordingly, the is no reason to expect that \Eref{eqn:Drude_weight_resolution} can be further compressed or possibly even explicitly resummed (as in the case of free energy).

For any generic extensive \emph{non-conserved} current ${\rm J}_{i}$, the autocorrelation function, \Eref{eqn:Drude_autocorrelator}, will undergo a non-trivial relaxation dynamics and, in the long-time limit, saturate to a constant
non-zero value $\mathcal{D}^{(i)}$ (given by the hydrodynamic projection).
From decomposition \eref{eqn:Drude_weight_resolution}, it is manifest that in the gapped regime of the Heisenberg chain, the value of asymptotic plateau has to undergo a continuous change upon continuously varying the anisotropy parameter.
In the gapless regime at rational values of the interaction parameter $\gamma/\pi$,
such asymptotic values instead experience finite jumps as a direct corollary of refragmentation.

\section{Heisenberg spin-S anisotropic chains}
\label{sec:Heisenberg}

We now take a closer look at the integrable spin-$S$ XXZ models and examine the structure of their eigenstates in more detail. The higher-spin variants of Heisenberg XXZ spin chain, see \Eref{eqn:XXZ_Hamiltonian}, can be constructed in a systematic
algebraic fashion from an infinite family of commuting transfer matrices using the methodology of the Algebraic Bethe Ansatz. For every finite-dimensional $\mathfrak{su}(2)$ representation $\mathcal{V}_{2S}$ of dimension $2S+1$, one can construct the associated higher-dimensional quantum $R$-matrix (invariant under the deformed symmetry algebra $\mathcal{U}_{q}(\mathfrak{sl}(2))$)
by employing an iterative procedure called \emph{fusion}. This program has been initiated by
Fateev and Zamolodchikov \cite{FZ82} who first realized it for the $S=1$ case, and afterwards developed in full capacity by Kirillov and Reshetikhin \cite{KRS90}. From the series expansion of fused transfer matrices~\footnote{Commuting transfer matrices that provide Hamiltonians with local interactions are correspond to auxiliary traces of monodromy operators with the same representation in the auxiliary spaces. Other
auxiliary representations yield conserved quantities with quasilocal densities \cite{QLreview}.} acting on the tensor product of physical spin representations one then obtains an infinite sequence of Hamiltonians in involution.
Particularly, the first logarithmic derivative of the fundamental transfer matrix (evaluated at the shift point) yields the (homogeneous) Heisenberg spin-$1/2$ chain. Similarly, transfer matrices with higher (auxiliary) spins yield the spin-$S$ counterparts to \Eref{eqn:XXZ_Hamiltonian}.

Before describing the eigenspectrum of the spin-$S$ Heisenberg chains, we wish to make two remarks that are of particular importance:
\begin{itemize}
\item Even though the $R$-matrices are invariance under the quantum group transformations, the resulting spin-chain Hamiltonians with periodic boundary conditions do not obey global $\mathcal{U}_{q}(\mathfrak{sl}(2))$ invariance.\footnote{Relatedly, Yangian invariance is likewise broken in finite-volume isotropic Heisenberg chains.)} This means particularly that eigenstates of integrable Fateev--Zamolodchikov spin-$S$ chains do not organize into irreducible finite-dimensional multiplets of the quantum algebra.
\item Hermiticity of integrable spin-$S$ Hamiltonians is ensured only for $S\leq 1$ but \emph{not} automatically higher spins $S>1$. To the best of our knowledge, this peculiarity has been first recognized in Refs.~\cite{KR87I,KR87II},
and afterwards examined in greater detail in Ref.~\cite{Frahm90}.
\end{itemize}

\subsection{Quasiparticle content and scattering data}

Below we succinctly describe the structure of magnonic excitations in the XXZ Heisenberg spin-$S$ chains while avoiding algebraic constructions of commuting fused transfer matrices. Most of the relevant information, including the diagonalization procedure, can be found in Refs.~\cite{KR87I,KR87II,Frahm90,FY90}.

The entire algebraic diagonalization can be in fact carried out uniformly for all physical spins $S \in \case{1}{2}\mathbb{Z}_{+}$, taking into the account that bare momenta of elementary single-magnon excitations are of the form
\begin{equation}
p^{(2S)}_{1}(\theta) = -\ii \log \left[\frac{\sinh{(\theta+\ii \gamma S)}}{\sinh{(\theta - \ii \gamma S)}}\right].
\end{equation}
On the other hand, the elementary scattering amplitude $S(\theta)$ (associated to an elastic collision of two magnons) is independent of spin size $S$, reading
\begin{equation}
S(\theta) = \frac{\sinh{(\theta-\ii)}}{\sinh{(\theta + \ii)}}.
\end{equation}
Imposing periodic boundary conditions, the magnon momenta get quantized to discrete values. The Bethe  equations take the form of \Eref{eqn:Bethe_equations}, with bare momenta $p^{(2S)}_{1}(\theta)$. The total momentum and energy of each regular eigenstate with a unique pattern of Bethe roots $\{\theta_{j}\}$ are computed by summing over the magnon rapidities, namely $P=\sum_{j}p^{(2S)}_{1}(\theta_{j})$ and $E=\sum_{j}e^{(2S)}_{1}(\theta_{j})$.
The one-particle bare energies take the form \cite{KR87I,Frahm90}
\begin{equation}
e^{(2S)}_{1}(\theta) = -\frac{\sin{(2S\,\gamma)}}{\sinh{(\theta-\ii \gamma\,2S)}\sinh{(\theta+\ii \gamma\,2S)}}.
\end{equation} 

\paragraph*{\bf Bethe strings.}
Rapidities of individual magnon excitations are complex-valued in general.
In the thermodynamic limit, the overwhelming majority of rapidities arrange themselves into certain regular string compounds. The constituent magnon rapidities of the so-called `$s$-string' form patterns of the form \cite{TS72}
\begin{equation}
\theta^{s}_{\alpha,k} = \theta^{s}_{\alpha} + \ii\left(s+1-2k + \frac{\pi}{2\gamma}(1-\sigma_{2S}\sigma_{s})\right),
\label{eqn:Bethe_string}
\end{equation}
disregarding vanishing finite-size corrections. In \Eref{eqn:Bethe_string}, $\alpha$ is used to designate different strings, while index $k$ runs over internal magnons of an $s$-string, with $s\in \mathbb{N}$. Moreover, each $s$-string is additionally associated a $\mathbb{Z}_{2}$-parity $\sigma_{n} = \pm 1$, while $\sigma_{2S} \in \{\pm 1\}$ is a fixed `spin parity'. As customary, we introduce parameter \cite{TS72,KR87I,KR87II,Frahm90}
\begin{equation}
{\rm p}_{0} \equiv \frac{\pi}{\gamma},
\end{equation}
which is particularly convenient for characterizing the magnon spectrum of the model. The spin-parity is given by
\begin{equation}
\sigma_{2S} = e^{\ii \pi \lfloor 2S\,{\rm p}_{0} \rfloor}.
\end{equation}
The complexified rapidity space of the model thus takes the shape of an infinitely long cylinder of circumference $2\ii {\rm p}_{0}$.

\subsubsection{Roots of unity and Takahashi--Suzuki numbers}

For irrational values of anisotropy parameter ${\rm p}_{0}$, the quasiparticle spectrum of the model involves infinitely many species \cite{TS72}.
For rational values, ${\rm p}_{0} \in \mathbb{Q}$, one however finds only a finite number of species courtesy of the `quantum truncation' \cite{TS72}. Indeed, it is the rational ${\rm p}_{0}$ that are of physical relevance\footnote{Irrational ${\rm p}_{0}$ can be approximated by a rational ${\rm p}_{0}$ to arbitrary prescribed accuracy.} and therefore we subsequently restrict ourselves exclusively
to the root-of-unity values of $q$ and set
\begin{equation}
\gamma = \pi \frac{m}{\ell},\qquad {\rm p}_{0} = \frac{\ell}{m} \geq 2,
\end{equation}
where $m$ and $\ell>m$ two arbitrary co-prime integers.

Complete classification of admissible Bethe strings has been originally obtained by Takahashi and Suzuki in their seminal paper \cite{TS72}. For our purposes, a brief summary will suffice. The Bethe strings, cf. Eq.~\Eref{eqn:Bethe_string}, are uniquely determined by their length ${\rm n}$ and parity $\sigma$ which can be inferred
with aid of the so-called Takahashi--Suzuki numbers. The latter have been deduced from the requirement that in the spin-singlet Fermi sea all the quasiparticle bands are fully occupied. By considering the fundamental chain with $S=1/2$, Takahashi and Suzuki recursively defined three sequences of auxiliary numbers\footnote{Notice that the ${\rm m}$-numbers are unrelated to $m_{s}$, representing bare spin (i.e. the number of quanta) of an $s$-string bound state. In our present notation, the $m_{s}$ are instead given by the ${\rm n}$-numbers, namely $m_{s}={\rm n}_{s}$.}
\begin{eqnarray}
{\rm p}_{i+1} &= {\rm p}_{i-1} - \nu_{i-1}{\rm p}_{i},\quad {\rm p}_{1} = 1,\\
{\rm y}_{i+1} &= {\rm y}_{i-1} + \nu_{i}{\rm y}_{i},\quad {\rm y}_{0} = 1,\quad {\rm y}_{-1} = 0,\\
{\rm m}_{i+1} &= {\rm m}_{i} + \nu_{i},\quad {\rm m}_{1} = \nu_{0},\quad {\rm m}_{0} = 0.
\end{eqnarray}
where $\nu$-numbers specify the continued-fraction expansion of parameter ${\rm p}_{0}$, that is
\begin{equation}
{\rm p}_{0} = [\nu_{0},\nu_{1},\nu_{2},\ldots] = \nu_{0} + \frac{1}{\nu_{1} + \frac{1}{\nu_{2}+\ldots}}.
\end{equation}
In terms of the TS numbers, the admissible string lengths can be expressed recursively as
\begin{equation}
{\rm n}_{s} = {\rm y}_{i-1} + (s-{\rm m}_{i}){\rm y}_{i},\qquad {\rm m}_{i}\leq s < {\rm m}_{i+1},
\end{equation}
while the associated parities are given by (beware the floor function in the exponent)
\begin{equation}
\sigma_{s} = e^{\ii \pi \lfloor ({\rm n}_{s}-1)/p_{0} \rfloor},\qquad s \neq {\rm m}_{1},\qquad \sigma_{{\rm m}_{1}} = -1. 
\end{equation}
When ${\rm p}_{0}\in \mathbb{Q}$, the continued fraction expansion terminates at level $\nu_{b}$, implying there is ${\rm m}_{b}$ distinct species in the spectrum which arrage themselves into $b$ `bands'. It is worth pointing out that there a pair of single (unbound) magnons with ${\rm n}=1$ of opposite parties $\sigma = \pm 1$.

We have already emphasized earlier that for chosen physical spin of size $S \in \case{1}{2}\mathbb{Z}_{+}$,
not every interaction parameter $\gamma$ is admissible. According to the analysis of Refs.~\cite{KR87I,KR87II,Frahm90},
the Hamiltonian ceases to be hermitian on a union of disjoint segments within the compact interval $[0,\pi/2]$.
What this implies, in particular, is that in the regime ${\rm p}_{0}\leq 2S$ there exist no spin-$S$ hermitian XXZ Hamiltonian.
Likewise, for ${\rm p}_{0}\leq 2S$, there is no unitary spin-$S$ representation of $\mathcal{U}_{q}(\mathfrak{sl}(2))$
at values $q=\cos{(\pi/{\rm p}_{0})}$. This fact can be alternatively deduced from the structure of the fused $R$-matrices, yielding
the following requirements
\begin{equation}
\cos{[\gamma(2S+1)]} \lessgtr \cos{(\gamma\,{\rm n})} \quad \forall \, {\rm n} \in \{2S-1,2S-3,\ldots,-2S+1\},
\label{eqn:reality_condition}
\end{equation}
where $\lessgtr$ signifies that either of the two inequalities is sufficient.
By fixing spin $S$, the above condition can likewise only be fulfilled on a unit of disjoint subintervals of $[0,\pi/2]$. The allowed regions of $\gamma$ can be deduced from a remarkably simple criterion \cite{KR87I}, stating that
$2S+1$ has to coincide with one of the TS numbers ${\rm n}_{j}$. What is even more remarkable is that the exact same criterion plays the role of the `stability condition' for the formation of magnonic bound states, which may alternatively be seen
as the condition for normalizability of the wavefunctions\footnote{This condition has been originally found by Korepin \cite{Korepin79} in his study of the massive Thirring model.}
\begin{equation}
\sigma_{\rm n} = \sin{(\gamma s)}\sin{(\gamma ({\rm n}-s))} \qquad \forall \, s \in \{1,2,\ldots, {\rm n}-1\},
\label{eqn:Korepin_condition}
\end{equation}
in agreement with the reality condition \eref{eqn:reality_condition}, upon making
identifications ${\rm n}=2S+1$ and $\sigma_{\rm n}=\sigma_{2S}$.

In summary, the outlined classification scheme can be applied to the entire family of the Heisenberg spin-$S$ XXZ chain provided that $n_{\wp} \equiv 2S+1$, is among the TS numbers (with the corresponding parity $\sigma_{\wp}$). Violation of this condition implies non-hermitian Hamiltonian for that particular value of interaction parameter $\gamma$. Additionally, when $p_{0}$ is a \emph{rational} number, namely for $p_{0} = \ell/m$, there is only finitely many bound states in the spectrum.
In this case $\ell$ is the upper bound for the size of an $n$-string. In fact, the largest, so-called complete `$\ell$-string', represents a \emph{zero mode} in the spectrum, i.e. an excitation that leaves the total energy of a state unchanged.
We further expand on this distinguished property in \Sref{sec:XXZ}.

\subsubsection{Bethe--Yang and TBA equations}

By inserting the Bethe strings into the Bethe equations and introducing the macrostate density $\rho_{s}(\theta)$ in accordance with the standard prescription, one arrives at the Bethe--Yang integral equations of the form
\begin{equation}
\rho^{\rm tot}_{s} = \kappa_{s}\,K_{s} - \kappa_{s}\,\mathscr{K}_{s,s'}\star \rho_{s'},
\label{eqn:Bethe-Yang_XXZ}
\end{equation}
where $s \in \{1,\ldots,{\rm m}_{b}\}$ and $\kappa_{s} = {\rm sign}(K_{s}(\theta)) \in \pm 1$. Convolution kernels $\mathscr{K}_{s,s'}(\theta)$ in \Eref{eqn:Bethe-Yang_XXZ} are defined as logarithmic derivatives of the scattering amplitudes ascribed to the scattering of an $s$-string with an $s^{\prime}$-string with rapidity difference $\theta$. Additional dependence on the physical spin $S$, which has been suppressed for clarity, enters through the elementary kernels
\begin{equation}
K_{s}(\theta) = \frac{1}{2\pi}\partial_{\theta} p^{(2S)}_{s}(\theta).
\end{equation}
We remind that scattering amplitudes do not depend on $S$, but only the scattering data ($S$-matrix) associated with the quasiparticle spectrum. The total state densities $\rho^{\rm tot}_{s}(\theta)$ are identified with the \emph{dressed} quasiparticle momenta via $2\pi \kappa_{s}\rho^{\rm tot}_{s}(\theta) = [\partial_{\theta}p^{(2S)}_{s}(\theta)]^{\rm dr}$.
Special attention should be payed to the presence parities
\begin{equation}
\kappa_{s} = (-1)^{r(s)},\qquad m_{r(s)}\leq s \leq {\rm m}_{r(s)+1}.
\end{equation}
When the reality condition is fulfilled, namely when $2S+1$ is a TS number, parities $\kappa_{s}$ actually do not depend on $S$. For a string of length ${\rm n}$, they are given by
\begin{equation}
\kappa = (-1)^{\lfloor {\rm n}/{\rm p}_{0}\rfloor + \lfloor ({\rm n}-1)/{\rm p}_{0}\rfloor}.
\end{equation}
It is worth reminding here that in the case of $S=1/2$, the parity of the ${\rm n}_{s}$-string equals $\sigma_{s}=(-1)^{\lfloor ({\rm n}_{s}-1)/{\rm p}_{0} \rfloor}$ (with the exception of $s={\rm m}_{1}$), showing that parities $\kappa_{s}$ are generally differ from the string parities $\sigma_{s}$.
Even though there no such distinction is necessary in the case of principal roots of unity with ${\rm m}=1$ (where $\kappa_{s}=\sigma_{s}$), one has to carefully distinguish these parities in the general case. Moreover, the `range function' $r(s)$ specifies to the quasiparticle band of the $s$th quasiparticle.

Another useful sequence of numbers introduced by the Takahashi and Suzuki are ${\rm q}$-numbers,
\begin{equation}
{\rm q}_{s} = (-1)^{i}({\rm p}_{i} - (s-{\rm m}_{i}){\rm p}_{i+1}),\quad
{\rm q}_{0} = {\rm p}_{0},\quad
{\rm q}_{{\rm m}_{b}}=(-1)^{b}{\rm p}_{b},
\end{equation}
with ${\rm m}_{i}<s\leq {\rm m}_{i+1}-1$ for $i=0,1,\ldots,b-1$.
For instance, the elementary kernels associated to the ${\rm n}_{s}$-strings in the fundamental ($S=1/2$) chain read\footnote{It is worth pointing out that our convention deviates slightly from the earlier works in that rapidity variable $\lambda$ used e.g. in Refs.~\cite{TS72,KR87I,KR87II,Frahm90} is linked to our variable $\theta$ via the simple rescaling $\theta=(2/\gamma)\lambda$. Accordingly, the Fourier space representations differ by a multiplicative factor of $\gamma/2$.}
\begin{eqnarray}
K^{(1)}_{s}(\theta) = \frac{1}{2{\rm p}_{0}}\frac{\sin{(\case{\pi {\rm q}_{s}}{{\rm p}_{0}})}}{\cosh{(2\theta)}+\cos{(\case{\pi {\rm q}_{s}}{{\rm p}_{0}})}},\qquad
\hat{K}^{(1)}_{s}(\omega) = \frac{\sinh{(\case{\gamma}{2}{\rm q}_{s}\omega)}}{\sinh{(\case{\pi}{2}\omega)}}.
\end{eqnarray}
The higher spin analogues have been obtained by Kirillov and Reshetikhin \cite{KR87I,KR87II}; here we only quote the final result, which is most economically expressed in the Fourier space $\hat{K}^{(2S)}_{s}(\omega) = \hat{\mathscr{A}}_{s,\wp-1}\hat{s}_{r(s)+1}(\omega) + 2\cosh{(\case{\gamma}{2}q_{\wp}\omega)}$, where $\mathscr{A}$ is the Fredholm kernel,
$\mathscr{A}_{s,s'}(\theta) = \kappa_{s}\delta_{s,s'}\delta(\theta) + \mathscr{K}_{s,s'}(\theta)$, and the sequence of `$s$-kernels' is given explicitly by $\hat{s}_{i}(\omega) = [2\cosh{(\case{\gamma}{2}p_{i}\omega)}]^{-1}$.
In terms of auxiliary functions
\begin{eqnarray}
\hat{\eta}_{s}(\omega) &\equiv \cos{\left[\Big\{\frac{{\rm n}_{j}}{{\rm p}_{0}}\Big\}-\frac{1-\kappa_{s}}{2}\right]} \nonumber \\
&+\sum_{l=1}^{{\rm n}_{s}-1}\cosh{\left[\left(\Big\{\frac{{\rm n}_{s}-1}{{\rm p}_{0}}\Big\}-\Big\{\frac{1}{{\rm p}_{0}}\Big\}\right)\frac{\pi\omega}{2}\right]},
\end{eqnarray}
with $\{\circ\}$ denoting the fractional part of a number, kernel $\mathscr{A}$ admits the following factorizable form (assuming $s' \geq s$)
$\hat{\mathscr{A}}_{s,s'}(\omega) = \hat{\mathscr{A}}_{s',s'}(\omega) =
2\hat{K}_{s'}(\omega)\hat{\eta}_{s}(\omega) + \kappa_{s'}\delta_{s',{\rm m}_{b}}\delta_{s,{\rm m}_{b}-1}$.

\medskip
The energy density in a macrostate of a general spin-$S$ integrable Heisenberg chain admits the following mode resolution
\begin{equation}
e = -4\,p_{0}\sum_{s}\int \dd \theta K^{(2S)}_{s}(\theta)\rho_{s}(\theta).
\end{equation}
The TBA equations for the thermodynamic $\mathcal{Y}$-functions accordingly assume the following canonical form
\begin{equation}
\log \mathcal{Y}_{s} = h\,{\rm n}_{s}-4\beta\,{\rm p}_{0}\,K^{(2S)}_{s} + \kappa_{s}\mathscr{K}_{s,s'}\star\log(1+\mathcal{Y}_{s'}),
\label{eqn:canonical_TBA_S}
\end{equation}
where $h$ is the chemical potential coupling to the total magnetization $S^{z}_{\rm tot}=\sum_{j=1}^{L}S^{z}_{j}$.
The rescaled density of free energy, $\mathfrak{f}=-\beta\,f$, is given the following spectral resolution
$\mathfrak{f} = \sum_{s} \kappa_{s} \int \dd \theta K^{(2S)}_{s}(\theta)\,\log(1+1/\mathcal{Y}_{s}(\theta))$.

\subsubsection{Resolvent formulation.}

Algebraic structure of the scattering kernels becomes more transparent in the resolvent formulation as briefly described below. To facilitate the computation, it is handy to make use of a compact matrix notation.
Firstly, we associate to the scattering kernel $\boldsymbol{\mathcal{K}}$
the \emph{bare} resolvent $\boldsymbol{\mathcal{R}}$ via
$\mathbf{1}-\boldsymbol{\mathcal{R}}=(\mathbf{1}+\boldsymbol{\kappa}\boldsymbol{\mathcal{K}})^{-1}$,
and introduce the Fredholm operator
\begin{equation}
\boldsymbol{\mathcal{F}} \equiv {\bf 1}+\boldsymbol{\kappa}\boldsymbol{\mathcal{K}}{\bf n},
\end{equation}
while suppressing dependence on spin $S$. The Bethe--Yang integral equations can be written in terms of its inverse
\begin{equation}
\boldsymbol{\rho}^{\rm tot} = \boldsymbol{\mathcal{F}}^{-1}\boldsymbol{\kappa}{\bf K}.
\end{equation}
Here $\boldsymbol{\kappa}{\bf K}$ is the vector of elementary kernels $\kappa_{s}K^{(2S)}_{s}(\theta)$.
Operator $\boldsymbol{\mathcal{F}}^{-1}=\mathbf{1}-\boldsymbol{\mathcal{R}}^{\rm dr}$ represents the `dressing kernel',
with \emph{dressed} resolvent
\begin{equation}
\boldsymbol{\mathcal{R}}^{\rm dr} = \mathbf{1} - (\mathbf{1} - \boldsymbol{\mathcal{R}})(\mathbf{1} - \boldsymbol{\kappa}\boldsymbol{\mathcal{K}}{\bf n}) = \boldsymbol{\mathcal{R}}(\mathbf{1}-{\bf n}).
\end{equation}
It is important to note that $\boldsymbol{\mathcal{F}}^{-1}$ depend on both the scattering data via kernel $\boldsymbol{\mathcal{K}}$ and
on the state functions through the Fermi occupation functions ${\bf n}$ (which include also dependence on chemical potential $h$),
but not on physical spin $S$. Now, operating by $\mathbf{1}-\boldsymbol{\mathcal{R}}=({\bf 1}+\boldsymbol{\kappa}\boldsymbol{\mathcal{K}})^{-1}$, we arrive at $(\mathbf{1}-\boldsymbol{\mathcal{K}}^{\rm dr})\boldsymbol{\rho}^{\rm tot} =
(\mathbf{1}-\boldsymbol{\mathcal{R}})\boldsymbol{\kappa}{\bf K}$.
Notice that applying $\mathbf{1}-\boldsymbol{\mathcal{R}}$ to the vector ${\bf K}$ (containing rapidity derivatives of bare momenta)
will produce the unit function (in Fourier space) located at position (node) $s=2S$.

\subsubsection{High-temperature limit.}
In the high-temperature limit (obtained by sending $\beta \to 0$), the thermodynamic $\mathcal{Y}$-functions consequently become flat
(i.e. lose dependence on rapidity variable), and source terms consequently drops out entirely from the TBA equations \eref{eqn:canonical_TBA_S}.
We thus end up with a system of coupled \emph{algebraic} recurrence relations. Moreover, the latter admit a simple explicit solution
in terms of the classical $\mathfrak{su}(2)$ characters $\chi_{s}(h)$ of the unitary $\mathfrak{su}(2)$ representations of spin $S$,
namely\footnote{Notice that functions $\chi_{s}(h)$ are also given by the asymptotic limit of the thermodynamic $\mathcal{T}$-functions
introduced earlier in \Sref{sec:TBA}.}
\begin{equation}
\chi_{s}(h) = \frac{e^{(s+1)h/2}-e^{-(s+1)h/2}}{e^{h/2}-e^{-h/2}}.
\end{equation}
The high-temperature $\mathcal{Y}$-functions $\mathcal{Y}_{s}$ can be expressed as gauge-invariant
ratios of characters, reading explicitly \cite{TS72}
\begin{equation}
1+\mathcal{Y}_{s}(h) = \frac{\chi^{2}_{{\rm n}_{s}+{\rm y}_{i}}(h)}{\chi^{2}_{{\rm y}_{i}}(h)},
\qquad n_{s}(h) = \frac{\chi^{2}_{{\rm y}_{i}}(h)}{\chi^{2}_{{\rm n}_{s}+{\rm y}_{i}}(h)},
\quad 1\leq s \leq {\rm m}_{b}-2,
\end{equation}
Dependence on the band index via TS numbers ${\rm y}_{i}$ indicates that these crucially differ from the $\mathcal{Y}$-functions
of the isotropic and XXZ chains with $q\in \mathbb{R}$.
It is also worth emphasizing that $\mathcal{Y}_{s}$ depend only on interaction parameter ${\rm p}_{0}$
but not on the size of spin $S$. One important corollary of this fact is that at \emph{half filling}, that is for $h\to 0$,
the dressed spin carries by the quasiparticles identically vanishes for all quasiparticle species, with the
exception of the final two species where one instead has
\begin{equation}
\mathcal{Y}_{+}(h) = e^{h\,\ell}\chi_{\ell-2}(h) ,\qquad \mathcal{Y}_{-}(h) = [\mathcal{Y}_{+}(h)]^{-1}.
\end{equation}

\paragraph*{\bf Magnetization sum rule.}

Every admissible solution to the TBA equations has to satisfy the sum rule for magnetization,
\begin{equation}
\frac{1}{2} - \expect{S^{z}} = \sum_{s=1}^{m_{b}}{\rm n}_{s}(1\star \rho_{s}(\theta))
= \sum_{s=1}^{m_{b}}{\rm n}_{s}\rho^{\rm tot}n_{s}(h),
\end{equation}
where the integrated densities $1\star \rho_{s}(\theta)\equiv \int \dd \theta\,\rho_{s}(\theta)$ represent
partial filling fractions associated to individual spaces. In practice, verifying the sum rule serves as a useful consistency check.

\subsection{Gapless Heisenberg spin-1/2 chain}
\label{sec:XXZ}

We have now finally prepared the stage for computing the exact Drude weights.
We shall focus exclusively on the spin Drude weight, quantifying ballistic transport of magnetization in a finite-temperature equilibrium ensemble. We will mainly be concerned with the critical (gapless) regime
(corresponding to $|\Delta|<1$) where the aforementioned exceptional features show up. 
In the following, we shall make use of the standard parametrization $\Delta = \case{1}{2}(q+q^{-1}) = \cos{(\gamma)}$, with $q=e^{\ii \gamma}$.
Moreover, with no loss of generality we can make a restriction to the open interval $\gamma = (0,\pi/2)$.

It is worthwhile to consider first the fundamental spin chain $S=1/2$ \eref{eqn:XXZ_Hamiltonian}. Our main task will be to compute the spin Drude weight in the high-temperature limit\footnote{We make this simplification for purely technical reasons as nonperturbative results at finite temperature are only accessible numerically via iteration which is not in the scope of this study.} explicitly using the outlined hydrodynamic approach of GHD. By employing the universal formula \eref{eqn:Drude_weight_resolution},
we shall retrieve the exact closed-from expression in the form of a popcorn function
obtained originally in Ref.~\cite{PI13} using an alternative (algebraic) computational approach.

In the outlined computational approach, the main steps are in fact identical to those spelled out previously in Ref.~\cite{Collura_wall} where,
using the prescription of Ref.~\cite{IN17} based on the bipartition protocol, the authors retrieve the spin Drude weight from the exact solution
to the quasi-stationary state that emerges from the initial magnetic domain wall at late times (by taking the limit of small magnetization bias at the end of computation). Our approach below in \Sref{sec:spin_Drude} is nonetheless ever more direct as it entirely avoids dealing with nonequilibrium states. Before we present this computation, we would like to spend some time
discussing various aspects related to the exceptional structure of the finite-volume Bethe eigenstates.

\paragraph*{\bf Finite-volume spectrum and zero modes.}

Complete classification of the highely degenerate finite-volume spectrum in the gapless Heisenberg XXZ chain at roots of unity deformations is a long-standing and notoriously difficult problem. In spite of many efforts (see e.g. Ref.~\cite{Miao21} for a major recent progress), the `completeness problem' still remains unresolved with full (mathematical) rigor. What causes trouble are exceptionally large spectral degeneracies stemming from a highly fragile behavior of
$\mathcal{U}_{q}(\mathfrak{sl}(2))$ representation then $q$ is a root of unity \cite{Korff01,Korff03,Deguchi06}.
Besides the ordinary regular Bethe strings, the Bethe equations admit certain special solutions -- in integrability literature mostly known as the Fabricius--McCoy (FM) strings \cite{FM1,FM2}; such `perfect strings', wrapping uniformly around the `rapidity cylinder', are not deformed by finite-volume effects and neither do interact with other types of Bethe strings, i.e. do not pick up
any phase shift \cite{BA99}. Fabricius and McCoy also first observed that spectrum organizes into multiplets of the $\mathfrak{sl}(2)$
loop algebra \cite{Deguchi07}, with descendant eigenstates involving finitely many `zero modes'.

In the most recent attempt \cite{Miao21}, the authors utilized the Wronskian reformulation of the Bethe equations that involves two types of Baxter's $Q$-functions. Each eigenstate of the model is ascribed a unique set of Bethe roots, identified with zeros of the Baxter's $Q$-function. The latter further factorizes into three pieces, comprising the regular Bethe roots, the FM-strings (i.e. zero modes),
and also `roots at infinity' $\pm \infty$ for eigenstates `beyond the equator'.
The non-descendant states are called primitive.\footnote{Even two primitive states can be degenerate  when they are related by spin-reversal symmetry.} Degenerate eigenstates. which all share a common set of regular Bethe roots, form descendant towers which can be presented graphically as Hasse diagrams \cite{Miao21}. Adding roots at infinity amounts to lowers magnetization by one unit. On the other hand, the addition of a FM-string amounts to lower magnetization by $\ell$ units.

The FM-strings also admit the creation and annihilation operators \cite{Miao21} that can be produced transfer matrices with auxiliary semi-cyclic representations \cite{Zadnik16}. Such semi-cyclic transfer matrices either commute of anticommute with the fused unitary commuting transfer matrices but, importantly, do not commute with the total magnetization.\footnote{Semi-cyclic spin representations likewise enable to construct additional quasilocal conservation laws of the model \cite{Zadnik16}. Such charges, commonly referred to as the `$Y$-charges', are nonetheless not in involution with the unitary `$X$-charges' nor the non-unitary quasilocal `$Z$-charges' \cite{QLreview}. They instead enclose, together with the $Z$-charges, the Onsager algebra \cite{Miao21,Miao_Onsager}.}

\paragraph*{\bf Non-unitary quasilocal Z-charges.}

When $q$ is a roots of unity, the descendant eigenstates yield (modulo an overall sign) the same eigenvalues on the unitary (fused) transfer matrices. Since spectral degeneracies cannot be lifted by quasilocal charges associated with the unitary transfer matrices, this leave extra room for quasilocal charges of different type. Although it may not be very obvious, there indeed exist a rather elementary justification for this: in order to guarantee complete factorizability of the many-body $S$-matrix in the large-volume limit, there has to exist enough (quasi)local symmetries to unique determine the macrostate, while keeping in mind that non-local charges contain information that is lost upon coarse-graining. The bijective correspondence formalizing this intuition is known as the string-charge duality \cite{StringCharge}.

To better elaborate on this subtle point, let us have a closer look at the trivial zero-energy eigenstate (representing the completely polarized state). In the framework of the Algebraic Bethe Ansatz, such ferromagnetic eigenstate provides the pseudovacuum for magnon excitations. It is important to note that such a ferromagnetic eigenstate is not unique; for instance, spin-reversal transformation yields another (inequivalent) pseudo-vacuum eigenstate of oppositely polarization.
In the gapped regime, that is for $|\Delta|>1$, such pair of $\mathbb{Z}_{2}$-degenerate pseudovacua produces two independent sets of Bethe eigenstates, in turn implying a global $\mathbb{Z}_{2}$ degeneracy of any (quasi)local charge. This is no longer so at the isotropic point $\Delta = 1$ where, due to restoration of the global $SU(2)$ invariance, the pseudovacuum state becomes an $S^{2}$-valued order parameter, while eigenstates consequently organize into spin multiplets (i.e. irreducible representations of $\mathfrak{su}(2)$). Both ferromagnetic pseudovacua belong to the same (largest) multiplet.
By Nambu--Goldstone theorem, the low-energy spectrum of magnon modes becomes gapless.
Lowering of the total spin amounts to sending individual magnon rapidities to infinity.

In the gapless phase $|\Delta<1|$, the non-abelian continuous symmetry is again reduced down to the residual continuous $U(1)$ symmetry.
This time however, this is accompanied by a dramatic reorganization of the eigenspectrum. Assuming ferromagnetic exchange interaction, the ground state becomes non-trivially filled with quasiparticles that enjoys $U(1)$ degeneracy. In analogy with the isotropic chain, both ferromagnetic pseudovacua belong to the same `vacuum tower' and all the unitary charges act trivially on the pseudovacuum descendants.
The key difference is that in the critical phase the conserved third component of total magnetization is not sufficient to fully resolve the spectral degeneracies. This time, this can only be achieved with additional commuting \emph{non-unitary} quasilocal conservation laws (in involution with all the unitary quasilocal charges and magnetization) identified and constructed\footnote{An algebraic construction of Ref.~\cite{PI13} utilizes transfer matrices built from auxiliary non-unitary highest-weight Verma modules of $\mathcal{U}_{q}(\mathfrak{sl}(2))$. For $q=^{\ii \gamma}$ with $\gamma \in \mathbb{Q}$, such Verma modules
decompose into two submodules. Since one of which has finite dimension, it can be traced over to produce a new transfer matrix.} in Ref.~\cite{PI13}. Despite both pseudovacua remain conjugate to one another under the spin reversal, both yielding zero total energy, only one of the two can be a proper pseudovacuum void of quasiparticles while its counterpart is an eigenstate
filled with the zero modes (alongside the Bethe roots at infinity). This peculiarity can be neatly illustrated in a nonequilibrium setting by initialiing the system in two semi-infinite thermal partitions of opposite magnetization joining at the origin.
In the gapped regime, such an entropic magnetic domain undergoes diffusive spreading.
In the gapless regime however, one observes ballistic melting \cite{IN_Drude,Collura_wall}, telling
that one of the partitions cannot be empty of quasiparticle modes but instead must play a role of an infinite `buffer' quasiparticles.
Moreover, it is not difficult to verify that neither of the two partitions is charged under the unitary (quasi)local conserved charges. This leads to the conclusion that that one of the partitions indeed has to be a sea of zero modes.

The need for additional, non-unitary, quasilocal conservation laws can also be recognized by inspecting the structure of TBA macrostates. Once again invoking the string-charge duality, there should exist sufficiently many (quasi)local conserved quantities to unambiguously determine the complete set of quasiparticle densities that encode equilibrium macrostates. It has been demonstrated in Ref.~\cite{StringCharge} that without incorporating the $Z$-charges, the unitary (fused) charges give access only to a submanifold of macrostates (with vanishing average magnetization), subjected to the additional constraints on the final densities (in the TS enumeration scheme), $\rho_{-}=\ol{\rho}_{+}$ and $\ol{\rho}_{-}=\rho_{+}$. Solely information coming from the unitary conserved charges therefore does not allow to unambiguously determine the total state densities of a macrostate. This was afterwards confirmed explicitly in Ref.~\cite{DeLuca17} which completes the duality transformation with the missing information coming from the non-unitary $Z$-charges.

\medskip

Exponentially large spectral degeneracy linked with the presence of zero modes might seem a bit concerning, particularly because eigenstates that involve such zero modes have not been explicitly accounted for in the construction of equilibrium macrostates despite their number growing exponentially with system size. The complete Bethe strings have been discarded on the basis that they not exert any force on the regular magnons or bound states thereof, reflecting the property of the scattering amplitudes $S_{+}S_{-}=1$. On the other hand, a finite filling fraction of such perfect strings would in principle amount to an extra non-zero contribution
to the (Yang--Yang) entropy density $\mathfrak{s}$. The standard the TBA approach based on the Takahashi--Suzuki's classification, invoking only the regular bound states, nonetheless yields the correct value of free energy.
To resolve this conundrum, one can picture an FM-string as a composite excitation made from the last two magnonic bound states binding together into an exact string. While in any finite volume one can always tell apart a proper zero mode from
two separate species of strings of types $(\ell-1,+)$ and $(1,-)$, this distinction disappears in the thermodynamic limit where quasiparticles represent faithful (i.e. sharply defined) asymptotic excitations. In particular, this means that there is not need of including the FM-strings among other regular (i.e. energy-carrying)
excitations since (at the coarse-grained level of macrostates) they will be automatically accounted for by the quasiparticle densities
of the last two species. Let us also add that the free-fermion point $\Delta=0$ ($\pi/\gamma=2$) is special, since there is no other species besides the pair of distinguished quasiparticles, which are simply unbound magnons of different parities
(i.e. strings of types $(1,+)$ and $(1,-)$) interpreted as two branches of a single fermionic excitation. As we are going to see in a minute,
fermionic character of such distinguished quasiparticles persists at other roots of unity $q$.

\subsubsection{Spin Drude weight}
\label{sec:spin_Drude}

We now turn our attention to the dressing equations. As announced earlier, we shall confine ourselves exclusively to the high-temperature limit where dressing simplifies to a finite system of coupled recurrence relations. Despite that, we suspect that a closed-form expression for the spin Drude weight can only found in the special case of half filling $h=0$, the reason being that the dressed magnetization $m^{\rm dr}_{s}$ of all the quasiparticles with the exception of the last two exactly vanishes in this limit. By contrast, for general $h$ there remains a sum of the magnon spectrum, see \Eref{eqn:Drude_weight_resolution}. 
To proceed, we employ the following strategy: first, we restrict our considerations to the sequence of primitive roots of unity $\gamma/\pi = 1/\ell$; at those isolated points, the spectrum of magnon excitations comprises exactly $\ell$ species, whereas
the final two species are the $(\ell-1)$-string of positive parity at position $s=\ell-1$ and an unbound magnon of negative parity at the end node with $s=\ell$. Dependence on anisotropy $\gamma$ only enters through the dispersion relation
and can be taken care of at the very end.

The computations are most conveniently performed in the resolvent formalism, enabling to express the dressing equations in a locally-coupled form. The bare and dressed Fredholm kernels can be inverted explicitly in Fourier space.
For instance, the bare resolvent reads explicitly
\begin{eqnarray}
\hat{\boldsymbol{\mathcal{R}}}(\omega) &=
\sum_{j=1}^{\ell-2}(\delta_{i-1,j}+\delta_{i+1,j})\hat{s}_{1}(\omega) + \delta_{\ell-2,\ell-2}\hat{s}^{2}_{1}(\omega)
\nonumber \\
&+(\delta_{\ell-1,\ell-2}+\delta_{\ell,\ell-2})\hat{s}_{1}(\omega) + (\delta_{\ell-2,\ell-1}-\delta_{\ell-2,\ell})\hat{s}_{1}(\omega).
\end{eqnarray}
Notice that the self-coupling term that appears at node $\ell-2$ can be removed using
$[\hat{s}_{1}(\omega)]^{2}\hat{\ol{\rho}}_{\ell-2}=\hat{s}_{1}(\omega)(\hat{\rho}_{\ell}(\omega)+\hat{\ol{\rho}}_{\ell}(\omega))$,
yielding the following decoupled dressing equations of the total state densities\footnote{To prevent confusion with the $s$-kernels,
we now label quasiparticle types with $j$ instead of $s$.}
\begin{eqnarray}
\rho^{\rm tot}_{j\leq \ell-3} - s_{1}\star (\ol{n}_{j-1}\rho^{\rm tot}_{j-1}+\ol{n}_{j+1}\rho^{\rm tot}_{j+1}) &= \delta_{j,1}s_{1},\\
\rho^{\rm tot}_{\ell-2} - s_{1}\star (\ol{n}_{\ell-3}\rho^{\rm tot}_{\ell-3}+\ol{n}_{\ell-1}\rho^{\rm tot}_{\ell-1}+n_{\ell}\rho^{\rm tot}_{\ell}) &= 0,\\
\rho^{\rm tot}_{\ell-1} - s_{1}\star \rho^{\rm tot}_{\ell-2} &=0,
\end{eqnarray}
using here $\ol{n}_{j}\equiv 1-n_{j}$ as the abbreviation for the hole occupation functions. It now becomes evident that morphology of the resulting equations is that of the Dynkin graph of type $D_{\ell}$.
The solution can be found explicitly with aid of the Fourier transform,
\begin{equation}
\hat{\rho}^{\rm tot}_{j\leq \ell-2}(\omega) = \frac{\chi_{j}(h)}{\chi_{1}(h)}
\left[\frac{1}{\chi_{j-1}(h)}\frac{\sinh{(\case{\pi}{2}\case{q_{j}}{p_{0}}\omega)}}{\sinh{(\case{\pi}{2}\omega)}} -
\frac{1}{\chi_{j+1}(h)}\frac{\sinh{(\case{\pi}{2}\case{q_{j}-2}{p_{0}}\omega)}}{\sinh{(\case{\pi}{2}\omega)}}\right],
\end{equation}
while (noting that $q_{\ell-1}=1$) the last two densities are of the form
\begin{equation}
\hat{\rho}^{\rm tot}_{\ell-1}(\omega) = \hat{\rho}^{\rm tot}_{\ell}(\omega)
= \Lambda^{(1)}_{\ell}(h)\frac{\sinh{(\case{\pi}{2}\case{\omega}{p_{0}})}}{\sinh{(\case{\pi}{2}\omega)}}
= \Lambda^{(1)}_{\ell}(h)\hat{K}_{(\ell-1,+)}(\omega).
\end{equation}
We have introduced a rescaling factor
\begin{equation}
\Lambda^{(1)}_{\ell}(h) \equiv \frac{\tanh{(h)}}{1-n_{\ell-1}(h)-n_{\ell}(h)},
\end{equation}
where $n_{\ell-1}(h)$ and $n_{\ell}(h)$ are the Fermi occupation function assigned to the last two species.
Recall that $\mathcal{Y}_{\ell-1}(h)\mathcal{Y}_{\ell}(h) = e^{2h\ell}$, implying
\begin{equation}
n_{\ell-1}(h) = \frac{1}{1+\mathcal{Y}_{\ell-1}(h)} = \frac{1}{1+e^{h\ell}\chi_{\ell-2}(h)}.
\end{equation}
Notice that $\Lambda^{(1)}_{\ell}(h)$ remains finite even in the $h\to 0$ limit,
\begin{equation}
\lim_{h\to 0}\Lambda^{(1)}_{\ell}(h) = \frac{\ell}{2(\ell-1)}.
\end{equation}
In the proximity of the bare vacuum we find $\lim_{h\to \infty}\Lambda_{j}(h)=1$ (for all $j$), thereby correctly recovering the bare momenta.

\paragraph*{\bf Pseudo-fermions.}
Using that $p^{\prime}_{j}(\theta) = 2\pi \sigma_{j} K_{j}(\theta)$, we find that the dressed momenta of the last two strings, $p_{+}\equiv p_{\ell-1}$ and $p_{-}\equiv p_{\ell}$, differ from their bare values up to an overall rescaling factor, namely
\begin{equation}
p^{\prime {\rm dr}}_{\pm}(\theta) = \Lambda(h)p^{\prime}_{\pm}(\theta),\qquad
p^{\prime {\rm dr}}_{+}(\theta) = -p^{\prime {\rm dr}}_{-}(\theta).
\end{equation} 
Curiously, momenta of the distinguished quasiparticles get affected by other magnons in a state only through an $h$-dependent rescaling factor, a curious property that one could hardly anticipate. A completely analogous procedure can now be repeated also for the rapidity derivatives of bare energies which are given by
\begin{equation}
e^{\prime}_{j}(\theta) = -\pi \sin{(\gamma)}\,K^{\prime}_{j}(\theta).
\end{equation}
Particularly, the last two species carry bare energies\footnote{This expression holds for all $\gamma \in (0,\pi/2)$,
not only at $\gamma = \pi/\ell$.}
\begin{equation}
e^{\prime}_{\ell-1}(\theta) = -e^{\prime}_{\ell}(\theta) = -\pi\sin{(\gamma)}\,\partial_{\theta}
\left[\frac{1}{2\pi}\frac{2\sin{(\gamma)}}{\cos{(\gamma)}+\cosh{(2\theta)}}\right].
\end{equation}
Once again, the dressed energies coincide with the bare values up to the same rescaling factor $\Lambda^{(1)}_{j}(h)$. The main outcome of this analysis is that the effective velocities of propagation of the
last two quasiparticles in the spectrum (being the only species that retain finite
dressed magnetization in the $h\to 0$ limit) are precisely the bare velocities,
\begin{equation}
v^{\rm eff}_{\pm}(\theta) = \frac{\varepsilon^{\prime}_{\pm}(\theta)}{p^{\prime {\rm dr}}_{\pm}(\theta)}
= \frac{e^{\prime}_{\pm}(\theta)}{p^{\prime}_{\pm}(\theta)} = v_{\pm}(\theta).
\end{equation}
Lastly, let us remind that bare velocities of the strings correspond to the group velocity, reading
\begin{equation}
v_{j}(\theta) = -\frac{\sin{(\gamma)}}{2}\frac{K^{\prime}_{j}(\theta)}{K_{j}(\theta)}.
\end{equation}

We have learned that even for $\ell>2$, the distinguished pair of magnonic species at the end of the truncated quasiparticle spectrum, if seen as a single excitation, bears close formal similarities with a fermionic mode at the noninteracing point ($\ell=2$)
despite non-trivial interaction. The effective velocity of propagation of such a mode is completely unaffected by the finite-density background of other magnon excitations despite non-trivial scattering phase shifts. This observation has already been made earlier in Ref.~\cite{Collura_wall}, where the authors pointed out the fact that
the exceptional quasiparticles disguise themselves as an electron `contracted Brillouin zone'.

\paragraph*{\bf Spin Drude weight at half-filling.}

Upon turning the $U(1)$ chemical potential off, namely sending $h\to 0$, all the regular quasiparticles species exhibit a paramagnetic behavior, i.e. they are stripped off their magnetization while propagating through a non-magnetized thermal background.
The sole contribution to the spin Drude weight, cf. \Eref{eqn:Drude_weight_resolution}, comes therefore from the irregular
doublet of excitations at the end of the spectrum which, in contrast with other regular strings,
retain finite dressed magnetization even as $h\to 0$:
\begin{equation}
n_{+}(0) = 1-n_{-}(0) = \frac{1}{\ell},\qquad m^{\rm dr}_{\pm }(0) = \frac{\ell}{2}.
\end{equation}
The spin Drude weight at half filling, $\mathcal{D}_{0}=\lim_{h\to 0}\mathcal{D}(h)$,
is accordingly given a compact expression nearly identical to that of free fermions,
\begin{equation}
\mathcal{D}_{0} = \frac{\ell-1}{2}\int \dd \theta \,\rho^{\rm tot}_{\ell}(\theta)\big(v^{\rm eff}_{\ell}(\theta)\big)^{2}.
\label{eqn:D0_rapidity}
\end{equation}
The rapidity integral in \Eref{eqn:D0_rapidity} can be performed explicitly as follows:
\begin{eqnarray}
\mathcal{D}_{0} &= \frac{\ell-1}{2}\int \frac{\dd \theta}{2\pi}\sigma_{\ell}[p^{\prime}_{\ell}(\theta)]^{\rm dr}\big(v^{\rm eff}_{\ell}(\theta)\big)^{2} \\
&= \frac{\ell-1}{4\pi \sigma_{\ell}}\Lambda^{(1)}_{\ell}(0) \int^{p_{\ell}(\infty)}_{p_{\ell}(-\infty)} \dd p_{\ell}
\big(v^{\rm eff}(p_{\ell})\big)^{2} \\
&= \frac{\ell-1}{4\pi \sigma_{\ell}}\Lambda^{(1)}_{\ell}(0)\left[\frac{\sin{(\gamma)}}{\sin{(\pi/\ell)}}\right]^{2}
\int^{p_{\ell}(\infty)}_{p_{\ell}(-\infty)}\dd p_{\ell}\sin^{2}(p_{\ell}),
\end{eqnarray}
where in the last line we have used that\footnote{We note that the velocity is a monotonously increasing function of rapidity $\theta$ which at large $\theta$ asymptotes to $\lim_{\theta \to \pm \infty}v_{\pm}(\theta) = \pm \sin{(\pi/\ell)}$.}
\begin{equation}
v^{\rm eff}_{\ell} = \frac{\sin{(\gamma)}}{\sin{(\pi/\ell)}}\sin{(\sigma_{\ell}p_{\ell})},\qquad
p_{\ell}(\pm \infty) = \mp \frac{\pi}{\ell}.
\end{equation}
Finally, by taking into account that $\sigma_{\ell}=-1$, $\Lambda^{(1)}_{\ell}(0)=\ell/[2(\ell-1)]$, we obtain the compact formula
\begin{equation}
\mathcal{D}_{0} = \frac{1}{8}\frac{\sin^{2}{(\gamma)}}{\sin^{2}{(\pi/\ell)}}\left[1-\frac{\sin{(2\pi/\ell)}}{2\pi/\ell}\right],
\end{equation} 
in agreement with the previous results \cite{PI13,IN_Drude,Collura_wall,Urichuk19,LZP19}.

\subsection{Anisotropic chains of higher spin}

Having understood that the only non-trivial contribution to the spin Drude weight at half filling comes from a pseudo-fermionic doublet excitation, it is natural to seek the same feature in the general spin-$S$ XXZ chains. The computation for higher spins $S$ goes essentially analogous, apart from the fact that the `source node' in the (algebraic) dressing equations now jumps higher to position $j=2S$.~\footnote{We are going to make an extra minor assumption that the source node located at $j=2S$ is not any of the last three nodes, i.e. that $j$ is a node `in the bulk' of the Dynkin graph.}

For definiteness, we only consider the case of half filling.\footnote{While general formulae for finite $h$ can be easily evaluated with help of {\tt Mathematica}, they turn out to be rather unwieldy. At any rate, displaying such expressions would not add any substaintial value.} Once again, all the regular Bethe strings behave diamagnetically, $\lim_{h\to 0}m^{\rm dr}_{j<\ell-1}(h)=0$.
The last two magnonic species are different and yield positive contributions to the spin Drude weight for any $S$. It is worth noticing that neither the occupation functions, $n_{\ell-1}\ol{n}_{\ell-1} = n_{\ell}\ol{n}_{\ell} = (\ell-1)/\ell^{2}$,
nor the dressed magnetization, $\lim_{h\to 0}m^{\rm dr}_{\ell-1}(h) = \lim_{h\to 0}m^{\rm dr}_{\ell}(h) = \ell/2$, depend on the size of spin $S$, which only affects the total state densities $\rho_{\ell-1}=\rho_{\ell}$.
Writing $\mathcal{D}^{(2S)}_{0}\equiv \lim_{h\to 0}\mathcal{D}^{(2S)}_{0}(h)$ and using the property $[p^{\prime(2S)}_{\ell}(\theta)]^{\rm dr}=-[p^{\prime(2S)}_{\ell-1}(\theta)]^{\rm dr}$ for all $S\geq 1/2$, we
arrive at the same form as earlier in the case of $S=1/2$,
\begin{equation}
\mathcal{D}^{(2S)}_{0} = \frac{\ell-1}{2}\int \dd \theta \rho^{\rm tot}_{\ell-1}(\theta)\big[v^{\rm eff}_{\ell-1}(\theta)\big]^{2}.
\label{eqn:Drude_2S}
\end{equation}

\paragraph*{\bf Fateev--Zamolodchikov spin-$1$ chain.}
We next explicitly examine the integrable spin-$1$ anisotropic quantum chain introduced by Fateev and Zamolodchikov \cite{FZ82}. A non-trivial (albeit suboptimal) lower bound on the spin Drude weight in the high-temperature limit at half filling
has been computed in Ref.~\cite{PV_XXZ}. To our knowledge, the exact exact spin Drude weight has not been computed thus far.

By inspecting the explicit solution to the algebraic dressing equations, we find that the dressed momenta (and therefore also the total state densities) of the exceptional doublet once again coincide with the rescaled rapidity derivatives
of their bare momenta,
\begin{equation}
\rho^{\rm tot}_{\ell-1}(\theta) = \Lambda^{(2)}_{\ell-1}(h)\,K^{(2)}_{\ell-1}(\theta).
\label{eqn:rescaled_momentum_2}
\end{equation} 
This time, the proportionality factor, denoted by $\Lambda^{(2)}(h)$, acquires an extra multiplicative term compare to
$\Lambda^{(1)}_{\ell-1}(h)$, namely
\begin{equation}
\Lambda^{(2)}_{\ell-1}(h) = \left(1+\frac{1}{e^{2h}+1+e^{-2h}}\right)\left(\frac{\tanh{(h)}}{1-n_{\ell-1}(h)-n_{\ell}(h)}\right).
\end{equation} 
Accordingly, it remains to compute the momentum integral
\begin{equation}
\mathcal{D}^{(2)}_{0} = \frac{\ell-1}{4}\Lambda^{(2)}_{\ell-1}(0)\int^{p_{\ell-1}(\infty)}_{p_{\ell-1}(-\infty)} \dd p_{\ell-1}
\left(v^{\rm eff}_{\ell-1}(p_{\ell-1})\right)^{2},
\end{equation}
where $\Lambda^{(2)}_{\ell-1}(0)=\case{4}{3}\Lambda^{(1)}_{\ell-1}(0)$.
To perform this last step, we adopt a convenient normalization of the Hamiltonian such that the group velocities become of the form\footnote{We note that this normalization differs from the one used in Refs.~\cite{KR87I,KR87II,Frahm90}.}
\begin{equation}
v_{j}(\theta) = -\frac{\sin{(\gamma)}}{2}\frac{\partial_{\theta}K^{(2)}_{j}(\theta)}{K^{(2)}_{j}(\theta)}.
\end{equation}
Using the relation
\begin{equation}
v^{\rm eff}_{\ell-1} = v_{\ell-1} = \frac{\sin{(\sigma_{\ell-1}p_{\ell-1})}}{2\cos{(\gamma)}},
\end{equation}
we finally obtain
\begin{equation}
\mathcal{D}^{(2)}_{0} = \frac{1}{12\cos^{2}(\gamma)}\left[1-\frac{\sin{(4\pi/\ell)}}{(4\pi/\ell)}\right].
\end{equation}

\paragraph*{\bf Higher spins.}
Having confirmed the presence of the pseudo-femionic quasiparticle in the $S=1$ chain, it would seem reasonable to conjecture that the same feature takes place for an arbitrary size of physical spins. However, and to our surprise, this expectation does not materialize. By specializing to the primitive roots of unity $\ell=\pi/\gamma$, we have computed and verified\footnote{We have used \texttt{Mathematica}
to explicitly verify the kernel identities, and also verified that the obtained solutions satisfy the magnetization rule for general $h$.} the solutions for $S=3/2$ and for moderately small $\ell\geq 6$. We have not been able to find any relation analogous to \Eref{eqn:rescaled_momentum_2}. What we have found instead is that the ratio of $K^{(3)}_{\ell}(\theta)$ and $\rho^{\rm tot}_{\ell-1}(\theta)$ involves explicit dependence on rapidity $\theta$. We lack any deeper insight regarding this mysterious observation at this moment, which hopefully is not due to an oversight.\footnote{We can speculate whether this could be indirectly related to the fact that, for $S\geq 3/2$, the XXZ Hamiltonians of higher spin are non-hermitian in the entire range of $\gamma \in (0,\pi/2)$. Equivalently, $n_{\wp}=2S+1$ for $S\geq 3/2$ is \emph{not} a TS number for the entire range of $\gamma \in (0,\pi/2)$.}
It is worth emphasizing though that such absence of such pseudo-fermions does not present any obstruction for implementing the outlined computational scheme, and neither does have any impact on our initial conjecture regarding the popcorn Drude weights. Moreover, in the limit of half-filling $h\to 0$, the contribution is still only due to the exceptional quasiparticle species, in accordance with \Eref{eqn:Drude_2S}. The key difference is in the dressed dispersion relation, which in the case $S>1$ makes the analytic evaluation of the momentum integral \eref{eqn:Drude_2S} somewhat more involved.

\paragraph*{\bf Semi-classical limit.}

Before wrapping this chapter up, we would like to make some insightful remarks regarding the large-$S$ limit. To begin with, we note the spin Drude weight remains strictly positive for all $S$, $\mathcal{D}^{(2S)}>0$, and
as emphasized previously, the only contribution to $\mathcal{D}^{(2S)}$ at half filling is due to the last pair of quasiparticles. Of course, one should not forget about the non-trivial hermiticity constraint which rules out certain domains of interaction parameter $\gamma$. This implies, in particular, that the large-$S$ limit does not exist for arbitrary $\gamma$. Nevertheless, despite this extra technical
condition the spin-$S$ Heisenberg XXZ chain admits a well-defined semi-classical limit.

Integrability of the spin-$S$ XXZ Heisenberg chain remains preserved upon taking the semi-classical limit, yielding the lattice version of the anisotropic Landau--Lifhistz model \cite{Sklyanin79}. Indeed, the one-dimensional Landau--Lifshitz magnet is arguably one of the most renowned completely integrable nonlinear differential equation \cite{Faddeev_book}. In order to access the semi-classical eigenstates of the Heisenberg spin chain, one has to perform a suitable \emph{scaling} limit in which both the effective `Planck constant' $\hbar$ and coupling parameter $\gamma$ are simultaneously sent to zero.\footnote{There exist a similar semi-classical \emph{continuous} limit, yielding an integrable classical field theory of an anisotropic Landau--Lifshitz ferromagnet, where $\gamma$ is sent to zero alongside the lattice spacing parameter, while $S$ is kept fixed (see Ref.~\cite{MIG21} for a derivation).} Presently, the effective Planck constant is just the inverse magnitude of the quantum spin, namely $\hbar \equiv 1/S$. Hence, using the following parametrization of the anisotropy parameter $\gamma = \varrho\,\hbar = \varrho/S$
and $\gamma=\pi/{\rm p}_{0}$, we can readily deduce the following useful relation
\begin{equation}
\frac{{\rm p}_{0}}{S} = \frac{\pi}{\varrho}.
\end{equation}
As it turns out, the rescaled anisotropy parameter $\varrho$ is none other than the axial anisotropy of the classical lattice Landau--Lifshitz model (in the so-called easy-plane regime). One can arrive at the same conclusion on purely algebraic grounds by taking the semi-classical limit of
$\mathcal{U}_{q}(\mathfrak{sl}(2))$ for $q=e^{\ii \gamma}$. This way, one deduces the so-called Sklynin's quadratic Poisson algebra \cite{Sklyanin83,Sklyanin88,Sklyanin89XXZ} (more details can be found e.g. in Appendix A of Ref.~\cite{LLMM} and references therein).
On a related note, the Sklaynin's bracket is only non-degenerate on a symplectic submanifold fixed by a Casimir invariant parametrized by $\varrho$. Such submanifolds are topologically equivalent to a $2$-sphere only when $\varrho$ is restricted
to a compact interval $\varrho \in [0,\pi/2]$. This restriction implies $\pi/\varrho \in (2,\infty)$, which is evidently guaranteed in the large-$S$ limit whenever the hermiticity condition ${\rm p}_{0}>2S$ is respected.
There are no further restrictions apart from ${\rm p}_{0}\sim 1/\hbar$.
The semi-classical regime can therefore be reached simply by putting ${\rm p}_{0}=\ell$ and subsequently considering the sequence of primitive roots of unity. The requirement that $\ell > 2S$ will ensure that the Bethe string of length ${\rm n}=2S+1$ is always present in the quasiparticle spectrum. Hence, the largest Bethe string of size ${\rm n}_{\ell-1}=\ell-1$ and positive parity survives taking the semi-classical limit. While the total number of distinct quasiparticles diverges upon taking the $S\to \infty$ limit, only large bound states
with $S\sim \mathcal{L}$ quanta yield classical solutions that carry finite energy. Indeed, such giant magnonic coherent states composed of ${\rm n}\sim S$ quanta can be seen as an emergent classical mode, representing a precessional soliton of the Landau--Lifshitz equation. Such an identification can be directly established with aid  of the Asymptotic Bethe Ansatz (see Ref.~\cite{Bargheer08,MIG21} for continuous isotropic and easy-axis Landau--Lifshitz field theory).

The above picture suggests that the exceptional pair of excitations that truncates the quantum spectrum of magnonic bound states is a genuinely quantum effect which eventually disappears at the level of semi-classical eigenstates. A na\"{i}ve conclusion would therefore be that the spin Drude weight in the classical lattice Landau--Lifshitz model will be identically vanish at finite temperature and
vanishing magnetization density. However, according to the numerical simulations performed in Refs.~\cite{Bojan,LLMM}, this is clearly not so.
In Ref.~\cite{LLMM} it is found that spin transport in the easy-plane regime is everywhere finite, while displaying smooth dependence on the interaction anisotropy $\varrho$. This apparent mismatch does not necessarily imply a flaw in our logic,
but rather it may be point to a new types of emergent non-linear classical modes.
Just for an illustration, let us mention semi-classically quantized topological kink of the easy-axis Landau--Lifshitz model. As demonstrated in Ref.~\cite{MIG21}, the kink mode, when semi-classically quanitzed, admits an exceptional arrangement of constituent magnons in the form of a `double condensate'. There is evidently no quantum analogue of such a kink among thermodynamic excitations of the gapped Heisenberg spin-$1/2$ chain. For this reason, we find it plausible that next the ordinary solitons the easy-plane regime supports emergent classical current-carrying modes with a non-vanishing contribution to the thermodynamic free energy. In this regard, one should also keep in mind that thermodynamic state functions (i.e. densities and Fermi occupation function) have to be rescaled with an effective Planck parameter \cite{NGIV20} (see e.g. \cite{Koch21} for the attractive Lieb--Liniger Bose gas).

\section{Quantum symmetry in the sine-Gordon model}
\label{sec:sG}

In this chapter, we study the sine-Gordon quantum field theory, another important paradigm of integrable models. While there exist a fairly large body of work devoted to the model and its integrability aspects, the dynamical properties of the model in thermal equilibrium remain virtually unexplored. It nonetheless deserves to highlight the effective semi-classical approach developed by Damle and Sachdev \cite{DamleSachdev05}, and more recent developments in the domain of quantum quenches \cite{KormosZaradi16,Kukuljan18,Horvath19}. Moreover, a recent study \cite{Bertini_sG} focuses particularly to the low-temperature regime by using the approach of GHD.

The quantum sine-Gordon model \cite{Coleman75,Pohlmeyer77,Z77,Pohlmeyer78,ZZ79,Klassen93} is a relativistic quantum field theory for a scalar field $\varphi$, governed by the Lagrangian density
\begin{equation}
\mathcal{L}_{sG} = \frac{1}{2}\varphi^{2}_{\mu} + \frac{m^{2}_{0}}{\beta^{2}_{c}}\cos{(\beta_{c}\varphi)},
\end{equation}
where $\varphi_{\mu}\equiv \partial_{\mu}\varphi$, $\beta_{c}$ denotes the coupling constant and $m_{0}$ is the (dimensionfull) mass.
The Hamiltonian of the model takes the form
\begin{equation}
\mathcal{H}_{sG} = \frac{1}{2}\int \dd x\,\Big(\pi^{2} + \varphi^{2}_{x} - 2\frac{m^{2}_{0}}{\beta^{2}_{c}}
\big(\cos{(\beta_{c}\varphi)}-1\big)\Big),
\end{equation}
where $\pi = \varphi_{t}$ is the canonical momentum. The coupling parameter $\beta_{c}$ is traditionally parametrized in terms of a positive parameter
\begin{equation}
\pp = \frac{(\beta^{2}_{c}/8\pi)}{1 - (\beta^{2}_{c}/8\pi)} \qquad \Leftrightarrow \qquad \frac{\beta^{2}_{c}}{8\pi} = \frac{\pp}{\pp+1}.
\end{equation}
At the critical point, $\pp=1$, the theory becomes equivalent to a free massive Dirac fermion, separating the repulsive regime with $\pp>1$ ($0\leq \beta_{c} \leq \sqrt{8\pi}$) from the attractive regime with $0\leq \beta^{2}_{c}/8\pi < 1/2$.

In the repulsive regime, the $S$-matrix is free of any poles in the `physical strip', signifying absence of bound states. In contrast, in the attractive regime, $0<\pp<1$, $\lfloor 1/\pp \rfloor$ massive breathers appear in the spectrum of elementary
massive excitations, representing bound states of kinks and antikinks with masses
\begin{equation}
M_{j}=2M\sin{\Big(\frac{\pi}{2}j\pp\Big)},\qquad j=1,2,\ldots \lfloor 1/\pp \rfloor,
\end{equation}
where $M$ is the renormalized soliton mass (distinct from the bare mass $m_{0}$).
In the limit $\pp \to \infty$ one approaches the asymptotically free regime.
The semiclassical regime corresponds to taking $\pp \to 0$.\footnote{By virtue of the weak-strong duality, the $\beta^{2}_{c}/8\pi \to \infty$ also admits the semiclassical description.}

\medskip

The excitation spectrum of the sine-Gordon theory has the simplest structure at the so-called \emph{reflectionless points} $\pp = 1/n$ or, equivalently, at values of $\beta^{2}_{c}/8\pi = 1/(n+1)$. These lie precisely at the threshold values for the appearance of new breathers: at $\pp = 1/n$ the spectrum comprises solely of $n+1$ massive excitations, that is $n-1$ bound states alongside the
unbound kink and antikink. At the dual point $\pp=n$ in repulsive regime, one finds instead a kink mode and $n$ types of massless auxiliary magnons. 
Scattering among (anti)kinks with breathers is completely diagonal, referring to absence of back-scattering. This no longer applies to scattering between kinks and antikinks which is instead \emph{non-diagonal}.
Any factorizable scattering process that mixes internal quantum numbers can however be resolved at expense of introducing \emph{auxiliary} massless magnonic degrees of freedom \cite{ZZ79}. The algebraic diagonalization procedure is thus analogous to
that of the Nested Bethe Ansatz.

\paragraph*{\bf Solitons and topological charge.}

Periodicity of the sine-Gordon potential implies $\mathbb{Z}$-fold degeneracy of vacuum states. Different vacua are related to one another via shifts $\varphi \to \varphi + (2\pi/\beta_{c})\mathbb{Z}$.
Solitons are thus regarded as topological kinks, representing field configurations that interpolate between two neighboring vacua.

Topological charge represents a distinguished conserved quantity of sine-Gordon model.
The total topological charge of a given field configuration can be computed by integrating the spatial gradient of $\varphi(x)$, that is
\begin{equation}
{\rm Q}^{\rm top} = \frac{\beta_{c}}{2\pi}\int \dd x\, \partial_{x}\varphi(x).
\end{equation}
Adopting Einstein summation convention, the associated abelian Noether
two-current reads ${\rm j}^{a}=\case{\beta_{c}}{2\pi}\epsilon^{ab}\partial_{b}\varphi$, where $t\equiv x_{0}$, $x\equiv x_{1}$.
Importantly, the topological charge density is a total derivative of a local observable and therefore
the value of $Q^{\rm top}$ depends solely on the difference of the two asymptotic values of $\varphi$,
$Q^{\rm top}=\int_{-\infty}^{\infty}\dd x\,j^{0}=\case{\beta_{c}}{2\pi}\int^{\infty}_{-\infty}\dd x\varphi_{x}=
\case{\beta_{c}}{2\pi}[\varphi(+\infty)-\varphi(-\infty)]$. Kinks carry a finite amount of energy above the bare vacuum and hence can be regarded as quasiparticles. Solitons of the opposite charge, called antikinks, are obtained via conjugation $\varphi \to -\varphi$. They carry a negative unit of topological charge.

\paragraph{\bf Nondiagonal factorizable scattering.}

The sine-Gordon solitons therefore form an $O(2)$ doublet of kinks to which we assign topological quantum numbers $Q^{\rm top}\in \{\pm 1\}$. Kinks experience \emph{non-diagonal} elastic scattering, meaning that their internal quantum numbers can get exchanged during their collision. To employ the powerful machinery of TBA, one has to first identify a complete basis of eigenstates which diagonalize the charges.
In completely factorizable theories with non-diagonal scattering, this is accomplished with help of additional auxiliary pseudoparticles. The entire procedure is rather lengthy and there is little use to fully reproduce it here.
We instead refer our readers to Refs.~\cite{Z77,ZZ79,Feher11,Bertini_sG} and references found therein. Since we mainly wish to highlight certain intimate connections to the gapless Heisenberg chain, it will suffice to briefly sketch the main steps.

The formal similarity between the sine-Gordon model and the Heisenberg XXZ chain can be readily recognized at the level of the two-body $S$-matrix which takes the form
\begin{equation}
\mathbf{S}(\theta) = S_{0}(\theta){\bf R}(\theta)
= \frac{S_{0}(\theta)}{a(\theta)}
\left(
\begin{array}{cccc}
a(\theta) & &  & \\
& b(\theta) & c(\theta) & \\
& c(\theta) & b(\theta) & \\
& & & a(\theta)
\end{array}
\right).
\end{equation}
The multiplicative scalar prefactor $S_{0}(\theta)$ of the $R$-matrix ${\bf R}(\theta)$ is the soliton scattering phase, cf. Refs.~\cite{ZZ79,Feher11}, which admits the following integral representation
\begin{equation}
S_{0}(\theta) = e^{-\ii \Theta(\theta)},\qquad\Theta(\theta) \equiv \int^{\infty}_{0}\frac{\dd \omega}{\omega}
\frac{\sin{(\omega\,\theta)}\sinh[\case{\pi}{2}(1-\pp)\omega]}{\sinh{(\case{\pi}{2}\pp \omega)}\cosh{(\case{\pi}{2}\omega)}}.
\end{equation}
The remaining matrix elements, reading
\begin{equation}
a(\theta) = \sinh{[\case{1}{\pp}(\ii \pi-\theta)]},\qquad
b(\theta) = \sinh{(\theta/\pp)},\qquad
c(\theta) = \ii\,\sin{(\pi/\pp)},
\end{equation}
represent the amplitudes associated to the non-diagonal exchange of charge.
By interpreting them as `Boltzmann weights' of the $6$-vertex $R$-matrix ${\bf R}(\theta)$, we immediately recognize the fundamental trigonometric quantum ${\rm R}$-matrix of the XXZ Heisenberg spin-$1/2$ chain, customarily parametrized in terms of 
\begin{equation}
a_{\rm 6V}(u) = [\lambda+1]_{\gamma},\qquad b_{\rm 6V}(u) = [\lambda]_{\gamma},\qquad c_{\rm 6V} = 1,
\end{equation}
where $[x]_{\gamma}\equiv \sin{(\gamma x)}/\sin(\gamma)$ are the $\gamma$-deformed numbers. To make this identification exact, we multiply $R(u)$ by $\sin{(\gamma)}$ and rescale the rapidity variable as $u = \ii\, \theta/\pi$,
from where we deduce the relation between the coupling parameters, $\gamma=\pi/\pp$.
This connection was discussed explicitly in Ref.~\cite{Haldane80}, although it is implicitly contained already in Ref.~\cite{ZZ79}.
As a consequence, in the in the limit $\beta^{2}_{c}/8\pi \to 1^{-}$ ($\pp \to \infty$), the spectrum of magnon excitations becomes
that of the \emph{isotropic} Heisenberg spin chain.

\medskip

Imagine now a scattering state involving $N_{s}$ solitons with an arbitrary number of kinks and antikinks and
let $\boldsymbol{\theta}\equiv \{\theta_{i}\}_{i=1}^{N_{s}}$ denote the set of rapidities that parametrize momenta of $N_{s}$ solitons.
We then consider an auxiliary particle with rapidity $\lambda$ that loops once around the system of circumference $L$.
The sequence of elastic collisions with all of the physical solitons results in the monodromy matrix
\begin{equation}
{\bf M}_{a}(\lambda|\boldsymbol{\theta}) = \mathbf{S}_{a,N_{s}}(\lambda-\theta_{N_{s}})\cdots
\mathbf{S}_{a,2}(\lambda-\theta_{2})\mathbf{S}_{a,1}(\lambda-\theta_{1}).
\end{equation}
The associated transfer matrix, $T(\lambda|\boldsymbol{\theta})=\Tr_{\mathcal{V}_{a}}{\bf M}_{a}(\lambda|\boldsymbol{\theta})$,
can be diagonalized with the Algebraic Bethe Ansatz procedure \cite{ZZ79}. Notice that $T(\lambda|\boldsymbol{\theta})$ is indeed no other
than the fundamental transfer matrix of an \emph{inhomogeneous} Heisenberg XXZ chain of spin-$1/2$ if one interprets solitons' rapidities as local inhomogeneities for spin waves. In the large-volume limit, such auxiliary magnons bind and form Bethe strings. While the mechanism is analogous to that of the Heisenberg XXZ chain with anisotropy $\Delta=\cos{(\gamma)}$ (upon identifying $\gamma/\pi = 1/\pp$),
there is no strict bijection (at least for general values of coupling $\beta_{\rm c}$) with the magnonic strings in the Takahashi--Suzuki classification due to breathers.

\subsection{Transport of topological charge}

We subsequently discuss transport properties related to the topological charge.
In particular, our aim is to deduce transport properties of topological charge in a equilibrium state with a finite density of solitons and
topological charge $Q^{\rm top}$. For this purpose, we couple the topological two-current to a uniform external abelian gauge potential $A_{a}$ by including
the term $\int \dd^{2}x\,{\rm j}^{a}A_{a}$ into the action. In the Hamiltonian formulation, this amounts to subtracting the term $A\,{\rm Q}^{\rm top}$ where ${\rm Q}^{\rm top}=\int \dd x \, {\rm j}^{0}$.
The inclusion of the gauge potential enables to control the size of the ground state.
For instance, the following equation for the ground state pseudoenergy has been obtained by Zamolodchikov \cite{Z95}
\begin{equation}
A - \int^{B}_{-B}\dd \theta^{\prime}\mathscr{K}(\theta-\theta^{\prime})\varepsilon(\theta^{\prime}) = M\cosh{(\theta)},
\end{equation}
where $\varepsilon(\theta)$ supported compactly on the Fermi sea $[-B,B]$, with scattering kernel
\begin{equation}
\mathscr{K}(\omega) = \int \dd \theta e^{\ii \omega \theta} \mathscr{K}(\theta)
= \frac{\sinh{(\case{\pi}{2}(\pp + 1)\omega)}}{2\cosh{(\case{\pi}{2}\omega)}\sinh{(\case{\pi}{2}\pp \omega)}}.
\end{equation}

We are interested in equilibrium states at finite temperature $T=1/\beta$. In this context, we can regard the gauge potential $A$ as chemical potential ascribed to ${\rm Q}^{\rm top}$. As we now elaborate, there are in fact other formal similarities with magnetization in the Heisenberg spin chain. Firstly, let us recall that the Heisenberg spin chain assumes a global $\mathbb{Z}_{2}$ symmetry corresponding to the spin reversal under which both magnetization and the associated current pick up a minus sign.
In the sine-Gordon model, the combination of `$C$-parity' conjugation, $\varphi \to -\varphi$, and time-reversal ($\varphi \to \varphi$, $\pi \to -\pi$) plays an analogous role.
While the sine-Gordon Hamiltonian together with the entire tower of higher local conserved charges
are invariant under such $\mathbb{Z}_{2}$ `$CT$-symmetry', i.e. they are of even parity, the topological current densities pick up
a minus sign: ${\rm j}^{0}\to -{\rm j}^{0}$ and ${\rm j}^{1}\to +{\rm j}^{1}$.
The hydrodynamic projection of topological current
${\rm J}^{\rm top}=\int \dd x\, {\rm j}^{1}(x)\equiv \case{\beta}{2\pi}\int \dd x\,\pi(x)$ onto the conserved subspace formed by the local charges is therefore going to be identically zero in the `non-topological sector' of the model, $Q^{\rm top}=0$, unless there exist extra `hidden' local charges of \emph{odd} $CT$-parity.
An explicit computation shows, see \Sref{sec:reflectionless}, the Drude weight of the topological charge \emph{always} takes a non-zero value at finite density, including in the topologically neutral sector.
This implicitly validates of our initial expectation: the sine-Gordon model hides
non-unitary (quasi)local conservation laws that have not yet been disclosed.

While identifying and constructing the complete set of (quasi)local conservation laws of the sine-Gordon model is interesting it itself, we save this task for future work. We expect that the full set of (quasi)local charges can be obtained in a systematic fashion from commuting fused transfer matrices with auxiliary representations of higher dimension~\footnote{Existence of additional conservation laws beyond the standard local charges is needed to ensure that all the degrees of freedom (i.e. quantum numbers) of a multisoliton state, including rapidities and internal structure of (anti)kink configuration charge, get preserved throughout
the scattering process.}, by essentially following the lines of Ref.~\cite{IMP15}. The lightcone lattice regularization by Destri and De Vega \cite{DDV87,DeVega90} is perhaps the most  natural setting to carry out this program.
In fact, This strategy has been already initiated in Ref.\cite{Vernier17}, where the authors consider the unitary auxiliary spin representations, for the particular case of principal roots of unity (where the construction mirrors that of the spin chain). In the following, we are going to circumvent explicit algebraic constructions and address the matter from the perspective of quasiparticles.

\medskip

Given that both models, the gapless XXZ Heisenberg chain and the sine-Gordon QTF, share a common quantum symmetry, namely $\mathcal{U}_{q}(\mathfrak{sl}(2))$ with $q \in U(1)$, one can anticipate the that generic Drude weights in the sine-Gordon model
will likewise reveal commensurability effects related to the phenomenon of refragmentation. To corroborate this view, we shall now closely examine the physically distinguished `topological' Drude weight.

Before delving into specifics, there are several remarks in order.
Even though integrable QFTs (Lorentz invariant or non-relativistic ones) can be largely treated in parallel with integrable quantum chains,
there are nonetheless certain technical complications. Primarily, quantum field theories are not UV complete. In distinction to quantum chains, the TBA equations describing equilibrium ensembles do not possess a well-defined high-temperature limit.
As alluded to earlier, there is little hope to obtain closed-form solutions of the dressing equations; apart from well-studied low-temperature limit, where one might set up a perturbative expansion, dealing with finite temperature generally requires to employ numerical iteration schemes. There is yet another, somewhat less severe, technical issue that pertains specifically to the sine-Gordon QFT;
the fully general form of the sine-Gordon TBA equations valid for any coupling $\beta_{c}$ is current not available in the literature.
Unfortunately, this not only obstructs analytical considerations, but also precludes a direct numerical computation of the Drude weights. On the bright side, a purely formal analysis of the dressing equations is already enough to deduce some (but by no means all) of the key properties. This will allow us to further strengthen our conjecture regarding the onset of popcorn Drude weights.

\paragraph{\bf Particle content from continued fractions.}

Reminding ourselves that the kink scattering matrix algebraically coincides with that of the gapless Heisenberg chain, the internal structure of auxiliary magnons can be again inferred with help of the Takahashi--Suzuki numbers, defined
in terms of the continued-fraction expansion of the interaction parameter $\pp$,
\begin{equation}
\pp = [\nu_{1},\nu_{2},\nu_{3},\ldots] = \frac{1}{\nu_{1} + \frac{1}{\nu_{2} + \ldots}}.
\end{equation}
Notice that when $\pp \in \mathbb{Z}$ or $\pp=1/\mathbb{Z}$, the spectrum contains $\pp+1$ quasiparticles. In the repulsive regime, these correspond to a single physical soliton next to $\pp$ auxiliary magnons; in the attractive regime, we instead find $\pp$ breathers and the soliton. When $\pp \in \mathbb{Q}$, the quasiparticle spectrum comprises a finite number of (auxiliary) species.

In what follows, we exploit the bijective correspondence between the quasiparticle spectrum and the $Y$-systems relations.
The sine-Gordon $Y$-system displays an exceptionally rich structure. The general form was conjectured by Tateo \cite{Tateo95},
who also exhibited a duality between the repulsive and attractive regimes under $\pp \to 1/\pp$. Tateo's conjecture has later been proven
in Ref.~\cite{NakanishiStella}, while exhibiting a splendid interplay between the sine-Gordon $Y$-systems, continued fractions, recursive triangulations of polygons and cluster algebras.

\subsubsection{Reflectionless points.}
\label{sec:reflectionless}

The simplest structure arises in the case of reflectionless points,corresponding to a discrete set of couplings where the reflection amplitude vanishes. As we now describe, the quasiparticle spectrum then becomes exceptionally simple.
For $\xi \in \mathbb{N}$ with $\xi\geq 2$, the spectrum of excitations involves precisely $\pp+1$ species enumerated by $j \in \{0,1,\ldots,\pp\}$. The initial species ($j=0$) in this enumeration corresponds to the soliton.
Its bare momentum and energy take the relativistic form
\begin{equation}
p_{0}(\theta) = M \sinh{(\theta)},\qquad e_{0}(\theta) = M \cosh{(\theta)}.
\end{equation}
All the remaining species are massless auxiliary magnons arising from diagonalization of the non-diagonal scattering matrix of the model. In analogy with the spin chains, the number of constituent magnons can be deduced with help of the continued-fraction expansion of $\pp$, yielding the following TS ${\rm q}$-numbers
\begin{equation}
{\rm q}_{i} = i \quad {\rm for} \quad i=1,2,\ldots,\pp-1,\qquad
{\rm q}_{\pp} = 1.
\end{equation}
In addition, the soliton mode is assigned $q_{0}=1$. Notice also that the last quasiparticle is distinguished by the negative `parity',
signifying that its rapidity acquires a finite imaginary part of $\case{\pi}{2}\pp$.

Adopting the conventions from Ref.~\cite{Bertini_sG}, the Bethe--Yang integral equations for the quasiparticle densities take the canonical form
\begin{equation}
\rho^{\rm tot}_{j} = \delta_{j,0}\frac{p^{\prime}_{0}}{2\pi}
- \upsilon_{j}\,\mathscr{K}_{j,k}\star \rho_{k},\qquad j \in \{0,1,\ldots,\pp\},
\label{eqn:BetheYang_SG_canonical}
\end{equation}
where quasiparticles have been assigned parities $\upsilon_{1\leq j\leq \pp-1}=1$ and $\upsilon_{0}=\upsilon_{\pp}=-1$.
The exact form of the scattering kernels $\mathscr{K}_{j,k}$ is not of central importance and we thus suppress it (see e.g. Refs.~\cite{ZZ79,Bertini_sG}). \Eref{eqn:BetheYang_SG_canonical} can be further simplified by employing the resolvent of the Fredholm kernel, similarly as in the case of the spin chain. Assuming $\pp>2$, one then arrives at the following locally-coupled form
\begin{eqnarray}
\rho^{\rm tot}_{0} &= \frac{p^{\prime}_{0}(\theta)}{2\pi} + s \star \rho_{1},\\
\rho^{\rm tot}_{1} &= s\star (\rho_{0} + \ol{\rho}_{2} + \delta_{\pp,3}\rho_{3}),\\
\rho^{\rm tot}_{j} &= s\star (\ol{\rho}_{j-1}+\ol{\rho}_{j+1}),\qquad (1<j<\pp-2)),\\
\rho^{\rm tot}_{\pp-2} &= s\star (\ol{\rho}_{\pp-3}+\ol{\rho}_{\pp-1}+\rho_{\pp}), \qquad (\pp>3),\\
\rho^{\rm tot}_{\pp-1} &= \rho^{\rm tot}_{\pp} = s\star \ol{\rho}_{\pp-2},
\end{eqnarray}
where $s(\theta) = 1/\cosh{(\theta)}$ (see e.g. Ref.~\cite{Bertini_sG} for a derivation). This form neatly reveals the underlying morphology of `TBA incidence matrix': the kink and auxiliary magnons arrange themselves according to the Dynkin diagram of type $D_{\pp}$, mirroring the magnonic TBA equations of the Heisenberg XXZ chain at commensurate points $\Delta=\cos{(\pi/\ell)}$ (with $\ell=\pp+1$), cf. \Sref{sec:XXZ}.

The $\pp=2$ case is exceptional, and the Bethe--Yang equations further simplify
\begin{equation}
\rho^{\rm tot}_{0} = \frac{p^{\prime}_{0}}{2\pi} + s\star (\ol{\rho}_{1} + \rho_{2}),\qquad
\rho^{\rm tot}_{1} = \rho^{\rm tot}_{2} = s\star \rho_{0}. 
\end{equation}
This set of equations is to be compared with the gapless Heisenberg XXZ chain with anisotropy $\Delta=1/2$. There are nevertheless certain minor differences, namely the physical momentum-carrying excitations are here the solitons, next to which we have two massless auxiliary magnons of types $(1,+)$ and $(1,-)$. While solitons carry positive topological charge, both magnon species carry negative $2$ units of topological charge. Moreover, the dispersion entering the source term now takes the relativistic form.

\paragraph{\bf Thermodynamic ensembles.}

Thermodynamic ensembles are identified with macrostates that are uniquely determined by the complete set rapidity densities of quasiparticles, comprising of the total soliton density $n_{s}= \int \dd \theta\, \rho_{0}(\theta)$, momentum and energy densities $p\int \dd \theta\, p_{0}(\theta)\rho_{0}(\theta)$ and $e=\int \dd \theta \,e_{0}(\theta)\,\rho_{0}(\theta)$, respectively, alongside an infinite tower of the higher local and quasilocal charges needed to resolve
the internal spin structure \cite{Vernier17}. On the other hand, the difference of kinks and antikinks per unit volume gives the density of topological charge,
\begin{equation}
q^{\rm top} = \int \dd \theta\, q^{\rm top}_{0} \rho_{0}(\theta) + \sum_{j=1}^{\pp}q^{\rm top}_{j}\int \dd \theta\,\rho_{j}(\theta).
\end{equation}
More specifically, the topological charges carried by individual quasiparticles (above the Fock vacuum) are
\begin{equation}
q^{\rm top}_{0} = 1,\qquad q^{\rm top}_{j>0} = -2q_{j}.
\end{equation}
We have used that creating a magnon amount to convert a kink into an antikink, thus lowering the topological charge by two units.\footnote{Any faithful macrostate has to additionally satisfy the inequality
$\int \dd \theta \rho_{0} \geq \sum_{j=1}^{\pp}q_{j}\int \dd \theta\,\rho_{j}(\theta)$, since
the total number of magnons can never exceed the total number of solitons. The equality holds only in the non-topological sector $Q^{\rm top}=0$).}

\medskip

Introducing chemical potentials $\mu$ and $h$ coupling to the total charge $N$ and topological charge $Q^{\rm top}$, respectively, we consider next the grand-canonical Gibbs ensembles at finite temperature $T=1/\beta$ and with finite density of solitons and topological charge. Variational optimization of the grand-canonical free-energy density $f$, given by
\begin{equation}
\beta f = \beta e-\,s - \mu\,n -A\,q^{\rm top},
\end{equation}
yields the following canonical TBA equations
\begin{eqnarray}
\log \mathcal{Y}_{0} &= \beta e_{0}(\theta) - \mu - A + \upsilon_{j}\mathscr{K}_{j,k}\star \log(1+1/\mathcal{Y}_{k}),\\
\log \mathcal{Y}_{j} &= 2h\,q_{j} + \upsilon_{j}\mathscr{K}_{j,k}\star \log(1+1/\mathcal{Y}_{k}),\quad j \in \{1,2,\ldots,\pp\},
\end{eqnarray}
whereas for $\pp = 2$ one finds
\begin{eqnarray}
\log \mathcal{Y}_{0} &= \beta\,e - \mu - A + \sum_{k=0}^{2}\upsilon_{k}\mathscr{K}_{k,0}\log(1+1/\mathcal{Y}_{k}),\\
\log \mathcal{Y}_{j} &= 2A\,q_{j} + \sum_{k=0}^{2}\upsilon_{k}\mathscr{K}_{k,1}\star \log(1+1/\mathcal{Y}_{k}),\quad j \in \{1,2\},
\end{eqnarray}
with $q_{1}=q_{2}=1$, $\upsilon_{1}=-\upsilon_{0}=-\upsilon_{2}=1$, or in an equivalent decoupled form\footnote{This step is performed by making use of the following kernel identities $\mathscr{K}_{0,0} = s\star K_{1}$, $\mathscr{K}_{0,2} = -\mathscr{K}_{0,1} = K_{1} = s$, $\mathscr{K}_{1,1} = \mathscr{K}_{1,2} = 0$}
\begin{eqnarray}
\log \mathcal{Y}_{0} &= \mu_{0} - s_{1}\star \log(1+1/\mathcal{Y}_{0}) - s\star \log[(1+1/\mathcal{Y}_{1})(1+1/\mathcal{Y}_{2})],\\
\log \mathcal{Y}_{1} &= 2A + s\star \log(1+1/\mathcal{Y}_{0}),\\
\log \mathcal{Y}_{2} &= 2A - s\star \log(1+1/\mathcal{Y}_{0}),
\end{eqnarray}
where $\mu_{0}\equiv \beta\,e - \mu - A$ and $s_{1}\equiv s\star K_{1}$.
By finally removing the self-coupling term on the soliton node using the equation for $\log \mathcal{Y}_{1}$ (at the expense of transforming $1+1/\mathcal{Y}_{1}\rightarrow 1+\mathcal{Y}_{1}$), we arrive at the following standard `$Y$-system form'
\begin{eqnarray}
\log \mathcal{Y}_{0} &= \beta\,e - \mu - s\star \log[(1+\mathcal{Y}_{1})(1+1/\mathcal{Y}_{2})],\\
\log \mathcal{Y}_{1} &= 2A + s\star \log(1+1/\mathcal{Y}_{0}),\\
\log \mathcal{Y}_{2} &= 2A - s\star \log(1+1/\mathcal{Y}_{0}).
\end{eqnarray}
We wish to stress that while chemical potential $\mu$ remains explicitly present, the gauge potential $A$ has disappeared from the first equation as a consequence of $s \, \star \, 2h = h$ (coming from the offset term in the equation for $\log \mathcal{Y}_{1}$). By contrast, the gauge potential remains explicitly coupled to the magnonic nodes, in analogy with the gapless spin chain at roots of unity.

It is worth adding that, by exploiting the kernel identities, one can deduce the standard representation of the free energy density,
$-\beta f = \int \frac{\dd \theta}{2\pi}p^{\prime}_{0}(\theta)\log(1+1/\mathcal{Y}_{0})$,
involving only the Fermi occupation functions of physical solitons. This form also makes it manifest that in $\beta \to 0$ limit (where $\mathcal{Y}_{0}(\theta)$ no longer depends on rapidity $\theta$) the integral becomes UV divergent.

\paragraph*{\bf Ultraviolet and infrared limits.}
The quantum sine-Gordon QFT is not a UV complete theory -- in the high-temperature limit $\beta \to 0$, the energy density blows up due to a divergent integral at large values of quasimomenta. Any ad-hoc regularization in which a UV cutoff is imposed by hand (e.g. by truncating the allowed range of quasimomenta to a compact region) however comes at a high price as it invariably spoils integrability. A proper UV lattice regularization can be achieved by employing the aforementioned lightcone discretization put forward by Destri and De Vega \cite{DDV87}. The main effect of such a regularization is a modified dispersion relation of soliton modes.
Despite such lattice sine-Gordon model assumes a well-defined high-temperature limit, we would still have to navigate around the notoriously difficult problem of classifying all the Bethe-string solutions at general coupling $\beta_{c}$.

In the opposite low-temperature regime, one also encounters certain technical simplifications. As demonstrated in Ref.~\cite{Bertini_sG},
at low temperatures the auxiliary spin-wave excitations completely lose dependence on rapidity $\theta$ and in effect all propagate uniformly with the `fluid velocity' of solitons. While transport properties of the sine-Gordon model in the
low-temperature regime have been for the most part already explored in Ref.~\cite{Bertini_sG}, one should be aware that the low-temperature phenomenology cannot capture generic physics seen at finite temperatures.
Particularly, the discontinuities in the Drude weight are expected to fully smear out in the limit $T\to 0$, as demonstrably happens in the gapless XXZ Heisenberg chain \cite{IN_Drude} (note through that the coefficient of the leading-order term is still described by a popcorn function \cite{Urichuk21}).

\paragraph*{\bf Dressed topological charge.}
We have seen that both the soliton number $N_{s}$ and topological charge $Q^{\rm top}$ take a distinguished role among the conservation laws of the sine-Gordon model. This can be also inferred from the fact that their values are independent from rapidities of quasiparticles. To compute the associated Drude weights, we need to deduce their dressed values in a macrostate. We remind that these correspond, by definition, to derivatives of the quasiparticles' pseudoenergies $\varepsilon_{j}(\theta)=\log \mathcal{Y}_{j}(\theta)$ with respect to chemical potentials $h$ and $A$.

We focus subsequently on the topological charge and restrict our considerations to the high-temperature limit $\beta \to 0$ where the TBA equations become a system of coupled recurrence relations for the constant $\mathcal{Y}$-functions
(denoted hereafter by $\mathscr{Y}_{j}$). In contrast with the Bethe--Yang equations, the TBA equations for the pseudoenergies are not plagued by UV divergencies.

To proceed, we first recast the TBA equations for the $\mathscr{Y}$-functions in an equivalent $Y$-system form.
We then see a general solution to the $Y$-system relations which, in the high-temperature limit, depends on both chemical potentials
$\mu$ and $A$. It is instructive to give a brief look first at the simplest case at $\pp=2$, where the solution has a particularly simple
form
\begin{equation}
\mathscr{Y}^{2}_{0}e^{2\mu} = \frac{1}{(1+\mathscr{Y}_{1})(1+1/\mathscr{Y}_{2})},\quad
\mathscr{Y}^{2}_{1} = e^{4A}(1+1/\mathscr{Y}_{0}),\quad \!\! \mathscr{Y}_{1}\mathscr{Y}_{2} = e^{4A}.
\label{eqn:xi2_solution}
\end{equation}
By furthermore putting $\mu=0$, we immediately recognize the $Y$-system relations of the XXZ chain for the particular value of $\Delta = \cos{(\pi/3)}$, albeit parametrized slightly differently. To clarify the correspondence and examine the role of chemical potentials, we have to remind ourselves of the $Y$-system solution
for the gapless Heisenberg chain for the particular case of $\ell=3$ principal root of unity (involving three $\mathscr{Y}$-functions which we denoted below by $\mathrm{Y}_{j}$ to distinguish them from the sine-Gordon $\mathscr{Y}$-functions).
The $\mathrm{Y}$-functions, associated with strings of the type $(1,+),(2,+),(1,-)$ strings, respectively, satisfy\footnote{We are using the convention in which magnetization $S^{z}$ couples with chemical potential $2h$.}
\begin{equation}
\mathrm{Y}^{2}_{1} = (1+\mathrm{Y}_{2})(1+1/\mathrm{Y}_{3}),\quad
\mathrm{Y}^{2}_{2} = e^{2h\ell}(1+\mathrm{Y}_{1}),\quad
\mathrm{Y}_{2}\mathrm{Y}_{3} = e^{2h\ell}.
\end{equation}
The general solution with $h>0$ can be parametrized in terms of the fundamental character $\chi_{1}(h)=e^{h}+e^{-h}=2\cosh{(h)}$, reading
\begin{equation}
\mathrm{Y}_{1}(h) = \chi^{2}_{1}(h) - 1,\quad
\mathrm{Y}_{2}(h) = e^{h\ell}\chi_{1}(h),\quad
\mathrm{Y}_{3}(h) = e^{h\ell}/\chi_{1}(h).
\end{equation}
As we can see, at half filling $h\to 0$, we have $\mathrm{Y}_{1}=3$ and $\mathrm{Y}_{2}=1/\mathrm{Y}_{3}=2$.
By performing the particle-hole transformation on $\mathscr{Y}_{0}$, $\mathscr{Y}_{0}\to 1/\mathscr{Y}_{0}$,
a direct comparison to the sine-Gordon model (at $\pp=\ell-1=2$) reveals the the two solutions agree apart from the mismatch which is due to the extra chemical potential $\mu$ (and rescaling of the twist $e^{2h\ell}\leftrightarrow e^{2A\pp}$).
Notice that chemical potential $\mu$ cannot be `gauged away'; if we absorbed into a redefinition of function $\mathscr{Y}_{0}$,
it would reappear in the second equation.

Another thing worth noticing is that at zero density, obtained by sending $\mu \to +\infty$, $\mathscr{Y}_{0}$ diverges to $+\infty$
while the occupation function for solitions consequently decays to zero. For any finite $\mu$, there is thus a finite density of solitons
in the state. Indeed, by setting $\mu=0$, the solution to the sine-Gordon TBA equations exactly matches that of the Heisenberg chain,
provided the $U(1)$ chemical potential is set to $h = \case{\pp}{\ell}A$. We remind that the $\ell$-string in
the XXZ spin chain is mapped to an $\pp=\ell-1$ string in the sine-Gordon model,
\begin{equation}
\mathscr{Y}_{0}(A) = \frac{1}{2\cosh{[2\xi A/(\xi+1)]}+1},
\end{equation}
reducing to $\lim_{A\to 0}\mathscr{Y}_{0}(A)=1/3$ for vanishing topological charge density.

The main upshot of the preceding analysis is that the \emph{dressed} topological charges carried by the sine-Gordon exicitations all vanish upon turning the gauge potential off ($A\to 0$), apart from the final two auxiliary magnonic species.
Specifically, for rational values of the coupling constant $\xi$ we find that
$q^{\rm dr}_{\pp-1} = \partial_{A}\log \mathscr{Y}_{\pp-1}(A) = \case{2\pp}{\pp+1} + \mathcal{O}(A)$
and $q^{\rm dr}_{\pp} = \partial_{A}\log \mathscr{Y}_{\pp}(A) =  -\case{2\pp}{\pp+1} + \mathcal{O}(A)$, whereas for all the remaining auxiliary magnons, alongside the physical excitations (solitons), the dressed charges
diminish as $\sim A$ at small $A$ (i.e. get perfectly screened by the thermal background). We therefore deal with a nearly a perfect analogy with the situation in the gapless Heisenberg chain which we thoroughly
examined in the previous section. This leads us to conclude that, in the grand-canonical Gibbs ensemble with equal densities of kinks and antikinks, the topological charges is transported ballistically through the system.

\paragraph*{\bf Truncated sine-Gordon model.}
We mention in passing that there exist a particular restriction of the sine-Gordon model, often referred to as the `quantum group truncation', with a reduced space of states; with the final three nodes of the $Y$-system dropping out,
one ends up with a Dynkin graph of type ${\rm A}$. In compare to the unrestricted case, the chief difference is that in the `restricted sine-Gordon model' \cite{LeClair89,BL90,Babelon96,Balog04} the topological Drude weight will necessary
vanish in the non-topological sector. Nevertheless, discontinuous dependence of the Drude weights upon continuous variation of coupling $\beta_{c}$ will still remain as a consequence of refragmentation.

\subsubsection{General coupling.}

Having established a direct correspondence between magnonic bound states of the gapless XXZ chain and auxiliary magnons of the sine-Gordon model at reflectionless points, one might immediately ask whether there is a similar identification
for other values of coupling $\beta_{c}$. To properly answer to that question, we would require the knowledge of complete quasiparticle spectrum for arbitrary value of interaction parameter $\xi$. Surveying the literature, it does not appear that such a classification has been worked out, at least not in an explicit form.
The overwhelming majority of works devoted to the sine-Gordon model and other massive integrable QFTs are interested primarily in understanding the RG flows and related concepts, with focus specifically on the IR and UV fixed points and the corresponding effective central charges. While the spectrum of physical (i.e. momentum-carrying) excitations comprising of solitons and massive breathers
is known for any value of $\xi$, the key piece of information that seems to be missing is the internal structure of magnonic nodes
(outside, of course, of the reflectionless points where there is a bijection with the spin-chain bound states). To the authors' knowledge, explicit solutions to the sine-Gordon $Y$-system relations, include full dependence on both $\mu$ and the
gauge potential $A$, are unavailable in the literature. This means that numerical evaluation of the topological Drude weight along the lines of \Sref{sec:XXZ} is currently out of reach.

A possible route to proceed without dealing with ab-initio computations based on the $S$-matrix is to start from the sine-Gordon $Y$-system and apply the standard procedure (exemplified in, for example, Refs.~\cite{IQC17,Landscape} on the isotropic Heisenberg chain) to convert it to the corresponding set of integral equations. The caveat with this strategy is that the functional relations of the $Y$-system admit more solutions that the TBA equations, namely there is infinitely many sets
of $\mathcal{Y}$-functions with different analytic properties obeying the same $Y$-system; the state-dependent information enters via analytic structure of the $\mathcal{Y}$-function entering the TBA equations
in the form of source terms. Besides specifying the dispersion relations for all the massive nodes, this also entails introducing the gauge potential that couples to the number of magnonic quanta ascribed to massless $\mathcal{Y}$-functions $\mathcal{Y}_{j}$. To better elucidate the technical issue, we consider an explicit example.

\paragraph*{\bf Second generation.}
For demonstrative purposes, it will be sufficient to examine the `second generation' of couplings,
\begin{equation}
\pp = [\nu_{1},\nu_{2}] = \frac{1}{\nu_{1} + \frac{1}{\nu_{2}-1}} = \frac{\nu_{2}}{\nu_{1}\nu_{2}+1},
\end{equation}
which reside within the attractive zone $\pp < 1$. The quasiparticle spectrum for this particular sequence comprises of $\nu_{1}+\nu_{2}$ species, consisting of $\nu_{1}$ breathers and the soliton mode, while the remaining (higher) nodes
all pertain to the auxiliary magnons.

Indeed, obtaining the TBA equations for the second generation would already be sufficient for a direct numerical validation of popcorn (topological) Drude weight: by implementing the computational procedure of Ref.~\cite{IN_Drude}:
tracing a discrete sequence of points $[\nu_{1},\nu_{2}]$ by progressively increasing the number of magnons via $\nu_{2}$ would allow to approach, in a pointwise manner, the reflectionless points $\pp = 1/\nu_{1}$ in the limit $\nu_{2}\to \infty$.

\medskip

Considering the \emph{constant} $\mathcal{Y}$-system (by dropping imaginary rapidity shifts, cf. Ref.~\cite{Tateo95}), the second generation pertains to two-parameter sequences $\pp = [\nu_{1},\nu_{2}]$ with general $\nu_{1}\geq 1$ and $\nu_{2}\geq 4$,
\begin{eqnarray}
\label{eqn:Y_2nd_generation}
\mathscr{Y}^{2}_{i} &= \prod_{j;j\sim i}(1+\mathscr{Y}_{j}),\qquad i=1,\ldots,\nu_{1}-1,\\
\mathscr{Y}^{2}_{\nu_{1}} &= (1+\mathscr{Y}_{\nu_{1}-1})\prod_{j=\nu_{1}+1}^{\nu_{1}+\nu_{2}-2}(1+\mathscr{Y}_{j})^{2}
(1+\mathscr{Y}_{\nu_{1}+\nu_{2}-1})(1+\mathscr{Y}_{\nu_{1}+\nu_{2}}),\\
\mathscr{Y}^{2}_{\nu_{1}+1} &= \frac{1+\mathscr{Y}_{\nu_{1}}}{1+1/\mathscr{Y}_{\nu_{1}+2}},\\
\mathscr{Y}^{2}_{i} &= \prod_{j;j\sim i}(1+1/\mathscr{Y}_{j})^{-1},\qquad i=\nu_{1}+2,\ldots,\nu_{1}+\nu_{2}.
\end{eqnarray}
The $\nu_{1}+1$ massive nodes are arranged on the initial nodes in the range $j=1,\ldots,\nu_{1}+1$, with the soliton node at position $\nu_{1}+1$ coupling to the first breather node at position $\nu_{1}$.
With the exception of two nodes at positions $\nu_{1}$ and $\nu_{1}+1$, all the remaining nodes in the $\mathscr{Y}$-system indeed couple among each other the same way the ordinary bound states of magnons in the spin chain. The equivalent $T$-system representation, written below in terms of constant functions $\mathscr{T}_{j}$,
represents a system of coupled recurrence relations of the form
\begin{eqnarray}
\label{eqn:T_2nd_generation}
\mathscr{T}^{2}_{i} &= 1+\prod_{j;j\sim i}\mathscr{T}_{j},\qquad i=1,2,\ldots,\nu_{1},\\
\mathscr{T}^{2}_{\nu_{1}+1} &= \mathscr{T}^{2}_{\nu_{1}} + \mathscr{T}_{\nu_{1}+2},\\
\mathscr{T}^{2}_{i} &= \mathscr{T}^{2}_{m} + \prod_{j;j\sim i}\mathscr{T}_{i},\qquad i=\nu_{1}+2,\ldots,\nu_{1}+\nu_{2}-2,\\
\mathscr{T}^{2}_{\nu_{1}+\nu_{2}-1} &= \mathscr{T}^{2}_{\nu_{1}+\nu_{2}} = \mathscr{T}_{\nu_{1}} + \mathscr{T}_{\nu_{1}+\nu_{2}-2}.  
\end{eqnarray}
The general solution for $\mathscr{T}$-functions $\mathscr{T}_{i}$ with $i\leq \nu_{1}+1$, \Eref{eqn:T_2nd_generation}, associated with massive excitations, subjected to the `initial condition' $\mathscr{T}_{0}=1$, takes the form
\begin{equation}
\mathscr{T}_{i\leq \nu_{1}} = \chi_{i}(a),
\end{equation}
where parameter $a\in \mathbb{R}$ is still free.
All the remaining higher $\mathscr{T}$-functions can then be determined in a recursive fashion from the above algebraic relations. Evidently however, the solution at hand cannot describe the most general (constant) solutions to the $Y$-system as it involves only a single free parameter $a$. Instead, we are seeking a more general two-parameter family of solutions depending on chemical potentials $\mu$ and $A$ (similarly as in \Eref{eqn:xi2_solution} for a reflectionless point $\xi=2$).
Dependence on $\mu$ will plausibly appear explicitly at the soliton node in the form of multiplicative fugacity factor $e^{\mu}$ while
the constant offsets by $2\mu$ in the canonical TBA source at the breather nodes should one the other hand disappear upon transforming the equations in the $Y$-system form (similarly to what happens with chemical potential $h$ in the Heisenberg chain).
At the same time, we also anticipate that dependence of the gauge potential $A$ always remains present in the source terms at the last two nodes even for general couplings (i.e. away from the reflectionless points), once again mirroring the situation encountered in the gapless Heisenberg chain. As an outcome, one would find a non-zero dressed topological charge even in the absence of the gauge field, yielding a finite topological Drude weight in the topologically neutral sector of the sine-Gordon theory.

We currently do not see how to fill in the missing information without explicitly referencing the structure of the Fredholm kernel and the associated resolvent. Resolving this technical step however extends slightly out of 
the main scope of this paper and for this reason we prefer to save this problem for future. Our view is that the problem of reconstructing the general sine-Gordon TBA equations from the $Y$-system relations that includes the presence of all chemical potentials $h$ and $A$ is important on its own merit.

\section{Conclusion}
\label{sec:conclusion}

We devoted this work to the so-called `popcorn Drude weights', one of the most perplexing features of the critical regime of integrable anisotropic Heisenberg spin chains. After briefly surveying the recent progress in understanding this captivating phenomenon, we walked the reader through various interrelated algebraic aspects. By expounding on the earlier findings of Ref.\cite{IN17}, we elaborated
on the discontinuous dependence of the Drude weights and linked it with refragmentation of the quasiparticle spectrum. The internal structure of quasiparticle excitations is inherently related to the quantum symmetry algebra
$\mathcal{U}_{q}(\mathfrak{sl}(2))$ at the roots of unity deformations, with a bijective correspondence to its finite-dimensional irreducible representations.
By invoking formal symmetry arguments, we then argued that the phenomenon of popcorn Drude weights is a general feature of the particular quantum symmetry in the regime of compact deformations $q\in U(1)$ and should therefore be expected
to take place in many other integrable systems with the same underlying symmetry.

In our analysis, we were primarily concerned with the family of quantum Heisenberg spin-$S$ chains with axial anisotropy in the critical regime. Choosing the deformation parameter $q$ to be a roots of unity, the quasiparticle content truncates to a finite number of magnonic species whose internal structure can be inferred from the Takahashi--Suzuki enumeration based on the continued fraction of the quantum deformation parameter $q$. There is a doublet of quasiparticle species generated by the quantum truncation with distinguished properties. These special quasiparticles carry non-unitary (quasi)local conserved charges and, unlike other regular magnonic bound states, do not behave paramagnetically in a non-magnetized background, implying non-zero spin Drude weight at half filling. We have also discussed the presence and meaning of complete exact Fabricius--McCoy strings representing spin-carrying zero energy modes.

Equipped with explicit knowledge of the quasiparticle spectrum and the associated scattering data, we recapitulated how to evaluate exact Drude weights using the framework of generalized hydrodynamics, while illustrating the procedure on the quantum Heisenberg XXZ spin-$S$ chains. More specifically, by specializing to the high-temperature limit and half filling, we retrieved the analytic formula derived originally in Ref.~\cite{PI13} and later corroborated in other works \cite{IN_Drude,Collura_wall,LZP19,Urichuk19}. In the presence of finite chemical potential, the have not managed to compress the final result beyond providing an explicit mode resolution. We highlighted a peculiar `pseudo-fermionic' nature of the distinguished quasiparticles species which, curiously, is no longer present in chains of higher spin with $S\geq 3/2$.

As our second explicit example, we picked the sine-Gordon model -- another popular integrable relativistic quantum field theory with an underlying $\mathcal{U}_{q}(\mathfrak{sl}(2))$ symmetry that displays many discernible similarities to integrable Heisenberg gapless chains. Our interest was primarily in transport properties of the conserved topological charge which, as we elaborated,
is intimately related to the total projection of magnetization in the spin chains.
The sine-Gordon model is also subject to regramentation effects stemming from
the underlying $\mathcal{U}_{q}(\mathfrak{sl}(2))$ symmetry in the regime of compact deformations. Accordingly, we argued that the finite temperature topological Drude weight should also yield a popcorn function, including in
ensembles with a vanishing topological charge. We substantiated our claims by examining the structure of the dressing equations for the kinks and auxiliary massless quasiparticles for couplings at which soliton scattering becomes reflectionless. We have included some additional remarks regarding the general case and outlined some of the technical challenges.

\paragraph*{Acknowledgements.}

The author acknowledges support by the Slovenian Research Agency (ARRS) under the Programs P1-0402 and J1-2480.

\appendix

\section{Unitary irreducible representations at roots of unity}
\label{app:irreps}

The $q$-deformed Hopf algebra $\mathcal{U}_{q}(\mathfrak{sl}(2))$ is generated by the $\{S^{+},S^{-},q^{\pm 2S^{z}}\}$, satisfying the following $q$-deformed defining commutation relations
\begin{equation}
[S^{z},S^{\pm}] = \pm S^{\pm},\qquad
[S^{+},S^{-}] = \frac{q^{2S^{z}}-q^{-2S^{z}}}{q-q^{-1}} \equiv [2S^{z}]_{q}.
\end{equation}
with the Casimir invariant reads $C=S^{+}S^{-}+([S^{z}-\case{1}{2}]_{q})^{2}$.
In general, $q\in \mathbb{C}$. In the $q\to 1$ limit, the above relations reduce to those of the Lie algebra $\mathfrak{sl}(2)$.
In the following we confine our analysis to the regime of unimodular deformations $q\in U(1)$ and accordingly set $q=e^{\ii \gamma}$.

Our first task is to classify all the \emph{unitary} irreducible finite-dimensional representations. By further assuming $\gamma/\pi \in \mathbb{Q}$, there turns out to be only finitely many such representations. In what follows, we revisit the analysis of Ref.\cite{Mezincescu90}. Unitary irreducible representations are labelled by their dimension $n$ and an extra discrete `parity' label $\sigma_{n} \in \pm$, defined via
\begin{equation}
(S^{z})^{\dagger} = S^{z},\qquad S^{+} = \sigma_{n}S^{-},\qquad (S^{-})^{\dagger} = \sigma_{n}S^{+}.
\end{equation}
Representations of negative parity are unitary representations of $\mathcal{U}_{q}(\mathfrak{su}(1,1))$.
Any finite-dimensional irreducible representation of $\mathcal{U}_{q}(\mathfrak{su}(2))$ of the highest-weight type is equivalent (up to a similarity transformation) to either a representation of parity $\sigma$ or has no definite parity.

Following Ref.~\cite{Mezincescu90}, we subsequently denote by $\ket{s,m}$ the simultaneous eigenstates of $S^{z}$ and Casimir $C$,
$S^{z}\ket{s,m}=m\ket{s,m}$ (where $m$ is an integer or half-integer).
The highest-weight vectors $\ket{s,s}$ satisfy $S^{+}\ket{s,s}=0$, whereas the descendant states are reached by iterative application of the lowering generator $S^{-}$. Writing,
\begin{equation}
S^{+}\ket{s,m} = a_{s,m}\ket{s,m+1},\qquad S^{-}\ket{s,m} = b_{s,m}\ket{s,m-1},
\end{equation}
one can deduce
\begin{equation}
a_{s,m-1}b_{s,m} = [s+m]_{q}[s-m+1]_{q}.
\label{eqn:truncation_hws}
\end{equation}
When $q$ is \emph{not} a root of unity, \Eref{eqn:truncation_hws} vanishes only after we descending down to state with $m=-s$. If we additionally require that the representation is of finite dimension and irreducible,
it has to hold that $a_{s,-(s+1)}=b_{s,-s}=0$. The corresponding dimension thus equals $n=2s+1$. Conversely, for $n\in \mathbb{Z}_{\geq 0}$, there exist a unitary irreducible representation of dimension $n$. With a suitably chosen (diagonal) similarity transformation, one can achieve that $a_{s,m-1}=b_{s,m}$ with
$m \in \{-s+1,-s+2,\ldots,s\}$, and
\begin{equation}
a^{2}_{s,m-1}b^{2}_{s,m}=\frac{\sin{[\gamma(s+m)]}\sin{[\gamma(s-m+1)]}}{\sin^{2}{(\gamma)}}.
\end{equation}
We can recognize than when both amplitudes $a$s and $b$s on the left-hand side are real, the right-hand side will be positive for the entire range of $m$. Likewise, when both $a$s and $b$s are purely imaginary, the right-hand side
always stays negative. When neither of these scenarios apply, there is no definite parity.

On the other hand, if there exist an integer $m$ such that $q^{d}= \pm 1$ then $[d]_{q}=0$ and, to avoid truncation in \Eref{eqn:truncation_hws} before descending to the state of the lowest weight, one has the condition that $n\leq d$.

\medskip

Let us now consider ${\rm n}=2s+1$ and $\sigma_{\rm n}$. The corresponding unitary irreducible representation has to obey the condition
\begin{equation}
\sigma_{\rm n}\sin{(\gamma\,k)}\sin{(\gamma({\rm n}-k))} > 0,\qquad k=1,\ldots,{\rm n}-1.
\label{eqn:Takahashi_condition}
\end{equation}
On this occasion, it is instructive to mention several other reincarnations of the very same condition which are all intimately related.
Inequality \eref{eqn:Takahashi_condition} provides 
\begin{itemize}
\item a `reality condition', ensuring hermiticity of the higher-spin Hamiltonians arising from fusion of the fundamental transfer matrix \cite{KR87I,KR87II,Frahm90},
\item an admissibilty condition of the Bethe-string solutions in the gapless spin-$1/2$ Heisenberg chain according to classification by Takahashi and Suzuki \cite{TS72},
\item a condition for normalizability of the wavefunctions \cite{Korepin79}
\item an irreducibility condition for the unitary representations of $\mathcal{U}_{q}(\mathfrak{su}(2))$ of
definite parity \cite{Mezincescu90}.
\end{itemize}

By invariance of \Eref{eqn:Takahashi_condition} under integral shifts by $2\pi$ and reflection property $\gamma \to -\gamma$, it is sufficient to analyze the range $\gamma \in [0,\pi]$. By further exploiting that $\gamma \to \pi-\gamma$ corresponds to the sign flip in \Eref{eqn:Takahashi_condition}, we subsequently specialize to the compact interval $\gamma \in [0,\pi/2]$ (corresponding to
$\Delta \in [0,1]$ in the spin-$1/2$ anisotropic Heisenberg chain). Excluding the free-fermionic point, we employ the conventional parametrization ${\rm p}_{0}=\pi/\gamma \in (2,\infty)$. Plugging $k=1$ into \Eref{eqn:Takahashi_condition},
one can deduce the parity
\begin{equation}
\sigma_{\rm n} = \exp{\left(\ii \pi \Big[\frac{{\rm n}-1}{{\rm p}_{0}}\Big]\right)},
\end{equation}
where $[x]$ denotes the integer part of $x$. From \Eref{eqn:Takahashi_condition}, one can then deduce the following equality
\begin{equation}
\left[\frac{k}{{\rm p}_{0}}\right] + \left[\frac{{\rm n}-k}{{\rm p}_{0}}\right]
= \left[\frac{{\rm n}-1}{{\rm p}_{0}}\right],\qquad k=1,\ldots,{\rm n}-1.
\label{eqn:TS_equality}
\end{equation}
To determine which values of ${\rm n}$, called ${\rm n}$-numbers ${\rm n}_{j}$, satisfy the above equation, Takahashi and Suzuki \cite{TS72} introduced two sequences of auxiliary numbers,
\begin{eqnarray}
{\rm m}_{0} = 0,\quad {\rm m}_{i} = \sum_{k=1}^{j}\nu_{k},\qquad {\rm m}_{i}\leq j \leq {\rm m}_{i+1},\\
{\rm y}_{-1} = 0,\quad {\rm y}_{0} = 1,\quad {\rm y}_{1} = \nu_{1},\qquad {\rm y}_{i} = {\rm y}_{i-2} + \nu_{i}{\rm y}_{i}-1,
\end{eqnarray}
where $\nu$-numbers are given by the continued-fraction expansion of ${\rm p}_{0}$,
\begin{equation}
{\rm p}_{0} = \nu_{1} + \frac{1}{\nu_{2} + 1/ \nu_{3}+\ldots}.
\end{equation} 
The solution to \Eref{eqn:TS_equality} can then be given recursively as
\begin{equation}
{\rm n}_{j}={\rm y}_{i-1}+(j-{\rm m}_{i}){\rm y}_{i}.
\end{equation}
Notice here that for given ${\rm p}_{0}$, there one generically finds representations of both positive and negative parities. The upshot of the above analysis is mostly two-fold: (i) assuming ${\rm p}_{0} \notin \mathbb{Q}$, there is an infinite sequence of finite-dimensional representations of definite parity and (ii) there is a bijective correspondence between such unitary representations of definite parity and the magnonic bound states (Bethe strings) in the Takahashi--Suzuki classification scheme.

For \emph{rational} ${\rm p}_{0}$ (i.e. for $q$ root of unity), the situation becomes exceptional. Parametrizing ${\rm p}_{0}=\ell/m$, one arrives at the conclusion that condition \eref{eqn:Takahashi_condition} can never be satisfied whenever ${\rm n}>\ell$. The maximal dimension of the representation is therefore ${\rm n}=\ell$, corresponding to spin $s_{\rm max}=(\ell-1)/2$. Those solutions to \Eref{eqn:TS_equality} with ${\rm n}>\ell$ must be discarded.

\section{Deforming quantum symmetry in interacting electrons}
\label{app:SUSY}

Here we shortly review the prospects of constructing integrable systems of interacting fermions featuring popcorn Drude weights, mainly to complement our main discussion where we concentrated on integrable spin chains and the sine-Gordon QFT.
For definiteness, we shall keep our focus on the few most prominent integrable lattice models of spinfull electrons. With this in mind, we consider quantum deformations of Yangians $Y(\mathfrak{g})$ for simple Lie algebras
$\mathfrak{g}$ of higher rank, that is $q$-deformed Hopf algebras $\mathcal{U}_{q}(\mathfrak{g})$ \cite{PS90}.
The quantum Yang--Baxter equation admits $R$-matrix solutions acting in the tensor product of two $\mathcal{U}_{q}(\mathfrak{g})$ representations
that can be used to construct integrable lattice models with nested quasiparticle spectra. We subsequently focus, as the main example, to the classical unitary series $\mathfrak{g}=\mathfrak{su}(n)$.
For rank $r=n-1>1$, we immediately encounter a problem: the integrable lattice Hamiltonians obtained from the algebraic Bethe Ansatz
construction turn out to be \emph{non-hermitian} in the regime of trigonometric deformations, i.e. for $q\in U(1)$. We are thus seemingly stuck with the simplest $r=2$ case, yielding the hierarchy of the Heisenberg XXZ chain.
Fortunately, there is a ways to get around this limitation if we allow for supersymmetry \cite{Kac77}. In the following, we shortly mention a number of possibilities for incorporating the deformed quantum symmetry
into various models of interacting fermions.

In essence, the aim is first promote a Lie algebra to a superalgebra and subsequently consider a $\mathfrak{su}(2)$ algebras as a subalgebra of a larger Lie superalgebra. The generators of superalgebra act on a vector space equipped with a
Grassmann ($\mathbb{Z}_{2}$) gradation. Since our aim is to only illustrate the basic ideas, it will suffice to consider unitary Lie superalgebras $\mathfrak{su}(m|n)$, leaving orthosymplectic or exceptional algebras aside. The signature $m|n$ refers to the number of states of fixed Grassmann parity, with even and odd parities ascribed bosonic and fermionic states, respectively. Fermionic operators accordingly correspond to operators of odd parity, connecting basis states of different parities.
The Yang--Baxter equation on such graded vector spaces, appropriately modified to account for non-trivial grading (see e.g. \cite{Hubbard_book}), provides the defining relation for graded Yangians $Y(\mathfrak{su}(m|n))$. Quantum $R$-matrices associated with the fundamental (evaluation) representations of these Yangians can then be used as local building blocks for constructing families of commuting transfer matrices using the Algebraic Bethe Ansatz procedure adapted for the grading \cite{Hubbard_book}. In this way, it is possible to construct various models of interacting spinfull fermions on a lattice with a local two-site interaction\footnote{By taking an appropriate continuous scaling limit, one arrives at integrable non-relativistic QFTs interacting fermions with contact attractive interaction known as the Yang--Gaudin models.}: in the case of $\mathfrak{su}(2|1)$, one finds the supersymmetric t-J model \cite{Schlottmann87}, while $\mathfrak{su}(2|2)$ yields the EKS model \cite{EKS92,EKS94}.

Similarly to their bosonic counterparts, graded Yangians $Y(\mathfrak{su}{(m|n)}$ also admits quantum deformations, yielding $q$-deformed graded Hopf algebras of the type $\mathcal{U}_{q}(\mathfrak{su}{(m|n)})$. The corresponding $R$-matrices permit to construct a large class of integrable lattice models known as the Perk--Schultz models \cite{Schultz81,PerkSchultz81,VegaLopes91}. The latter nonetheless still suffer from lack of hermiticity for compact $q$-deformations. This limitation can only be avoided if one considers superalgebras that involve bosonic subalgebras of rank one.
In other words, the requirement is that $\mathcal{U}_{q}(\mathfrak{sl}{(2)})$ appears as a subalgebra of the full symmetry algebras (possibly multiple times). Although this is still a quite severe restriction, many of the most prominent exactly solvable
models of one-dimensional interacting electors actually fall into this category. The most interesting instances are arguably $\mathcal{U}_{q}{(\mathfrak{su}(1|2))}$ and $\mathcal{U}_{q}{(\mathfrak{su}(2|2))}$ \cite{Beisert14}.
For root-of-unity deformations, finite-dimensional irreducible representations of these quantum groups exhibit an exceptional structure
in a direct analogy with representations theory of $\mathcal{U}_{q}{(\mathfrak{su}(2))}$.

We have not yet exhausted all the possibilities nonetheless. Lie superalgebra $\mathfrak{su}(2|2)$ indeed occupies a place among Lie
superalgebras as it allows for an exceptional \emph{central extension} to
$\mathfrak{h}\equiv \mathfrak{psu(2|2)}\ltimes \mathbb{R}^{3}\cong \mathfrak{su}(2|2)\ltimes \mathbb{R}^{2}$. Such an enlarged algebra has occupied the central stage in the study of the gauge-gravity duality,
where $\mathfrak{psu}(2|2)$ is found as a subalgebra of the full $\mathfrak{psu}(2,2|4)$ supersymmetry of an integrable quantum chain whose configurations map to single-trace operators in the large-$N$ limit of the
$\mathcal{N}=4$ super Yang--Mills theory, see e.g. Refs.~\cite{Beisert11,Beisert_review} and references therein.
In fact, Yangian $Y(\mathfrak{h})$ displays a somewhat richer structure compare to other quasitriangular Hopf algebras associated to other integrable quantum models. We refer the reader to Refs.~\cite{Beisert07,Beisert_review}, where its algebraic structure is examined in great detail.

Another remarkable fact is that in a special limit the two-body $S$-matrix associated with $\mathfrak{h}$-symmetry reduces to Shastry's $S$-matrix \cite{Shastry86} of the one-dimensional Fermi--Hubbard model\footnote{It turns out that the global femionic symmetry disappears in the process of degeneration, leaving behind only the Yangian subsymmetry $Y(\mathfrak{su}(2)\otimes \mathfrak{su}(2))$ \cite{EK_Smatrix}.}
\begin{equation}
H = -\sum_{j=1}^{L}\big(c^{\dagger}_{j,\sigma}c_{j+1} + c^{\dagger}_{j+1}c_{j}\big)
+ 4\mathfrak{u}\sum_{j=1}^{L}(n_{j,\uparrow}-\case{1}{2})(n_{j,\downarrow}-\case{1}{2}),
\end{equation}
a central paradigm of interacting spinfull electrons on a lattice with a local Coulomb repulsion \cite{Hubbard_book}.
The Coulomb strength $\mathfrak{u}$ is a free parameter of the model whose origin can be traced back to the generators spanning the center of $\mathfrak{h}$.~\footnote{In contrast, integrable fermionic models that arise from ordinary Yangians $Y(\mathfrak{g})$ do not allow for any continuous parameters.} The Hubbard model is exactly solvable by the Nested Bethe Ansatz \cite{LiebWu68}, permitting to construct the complete spectrum of eigenstates in an algebraic fashion \cite{Hubbard_book} and computing the thermodynamic free energy \cite{Takahashi_Hubbard}.
The spin and charge Drude weight in the one-dimensional Fermi--Hubbard model have been computed in Ref.~\cite{IN17}.

If we wish to accommodate for popcorn Drude weights, the Hubbard model has be deformed in a suitable way. This can be achieved by, for example, by starting with the $q$-deformed counterpart of $Y(\mathfrak{h})$. The latter has been examined in detail in Ref.~\cite{BK08}, where the authors also derive an integrable tight-binding model of interacting spinfull electrons on a one-dimensional lattice with correlated hopping exhibiting the symmetry of $\mathcal{U}_{q}(\mathfrak{su}(2)\otimes \mathfrak{su}(2))$ (reducing to Yangian symmetry of the Fermi--Hubbard model
in the limit $q\to 1$). It explicit form, Eq.(5.44) in Ref.~\cite{BK08}, is a rather formidable expression involving many terms and thus we do not 
display it here. It is worth noting that even though the fundamental $R$-matrix acting in the tensor product of $\mathcal{U}_{q}(\mathfrak{h})$ defining representations is not of the difference type, it nonetheless still yields a two-body Hamiltonian density in the usual fashion (by taking the first derivative of the $R$-matrix at a distinguished shift point). There exist another, closely related, variant of a deformed Hubbard model that has appeared previously
in the work by Alcaraz and Bariev \cite{AB99}, who obtained a family of integrable models of interacting electrons interpolating between the Fermi--Hubbard model and the `supersymmetric' t-J model. This similarity was recognized and closely examined in Ref.~\cite{BK08}, while explaining how the Alcaraz--Bariev model fits into a larger family $q$-deformed fermionic chains based on the exceptional central
extension of $\mathfrak{psu}(2|2)$. On a related note, it deserves mentioning here
two other integrable models of spinfull electrons with correlated hopping terms \cite{FQ12} that are amenable to quantum deformation.

In order to explicitly compute transport coefficient in $q$-deformed Hubbard model one has to first classify all the string solutions to the deformed Lieb--Wu equations.
Alternatively, one could follow the approach of Refs.~\cite{Arutyunov12I,Arutyunov12II} in which
the $q$-deformed Hubbard model has been retrieved from a suitable degeneration of more general $q$-deformed ${\rm AdS_{5}\times S^{5}}$ mirror $Y$-system upon decoupling the so-called `$Q$-particles'. The authors nonetheless do not treat the general case but as customarily only consider the principal series with $q=\exp{(\ii \gamma})$ for integer $\pi/\gamma=\ell$. Even though the ensuing TBA equations are structurally quite similar to those of the (undeformed) Hubbard model, it is not immediately clear in what manner does the deformed electronic dispersion relation
affect the internal structure of Bethe strings which now are either magnonic bound states of compounds that involve both charge and spin
excitations.

\section*{References}
\bibliographystyle{acm}
\bibliography{Popcorn}

\providecommand{\newblock}{}
\begin{thebibliography}{100}
\expandafter\ifx\csname url\endcsname\relax
  \def\url#1{{\tt #1}}\fi
\expandafter\ifx\csname urlprefix\endcsname\relax\def\urlprefix{URL }\fi
\providecommand{\eprint}[2][]{\url{#2}}

\bibitem{Schemmer19}
Schemmer M, Bouchoule I, Doyon B and Dubail J 2019 {\em Physical Review
  Letters\/} \href{http://dx.doi.org/10.1103/physrevlett.122.090601}{{\bf 122}}
  \urlprefix\url{https://doi.org/10.1103%2Fphysrevlett.122.090601}

\bibitem{Jepsen20}
Jepsen P~N, Amato-Grill J, Dimitrova I, Ho W~W, Demler E and Ketterle W 2020
  {\em Nature\/} \href{http://dx.doi.org/10.1038/s41586-020-3033-y}{{\bf 588}
  403--407} \urlprefix\url{https://doi.org/10.1038%2Fs41586-020-3033-y}

\bibitem{Scheie21}
Scheie A, Sherman N~E, Dupont M, Nagler S~E, Stone M~B, Granroth G~E, Moore J~E
  and Tennant D~A 2021 {\em Nature Physics\/}
  \href{http://dx.doi.org/10.1038/s41567-021-01191-6}{{\bf 17} 726--730}
  \urlprefix\url{https://doi.org/10.1038%2Fs41567-021-01191-6}

\bibitem{Malvania21}
Malvania N, Zhang Y, Le Y, Dubail J, Rigol M and Weiss D~S 2021 {\em Science\/}
  \href{http://dx.doi.org/10.1126/science.abf0147}{{\bf 373} 1129--1133}
  \urlprefix\url{https://doi.org/10.1126%2Fscience.abf0147}

\bibitem{Bloch_KPZ}
Wei D, Rubio-Abadal A, Ye B, Machado F, Kemp J, Srakaew K, Hollerith S, Rui J,
  Gopalakrishnan S, Yao N~Y {\em et~al.\/} 2021 {\em arXiv preprint
  arXiv:2107.00038\/}

\bibitem{transport_review}
Bertini B, Heidrich-Meisner F, Karrasch C, Prosen T, Steinigeweg R and
  {\v{Z}}nidari{\v{c}} M 2021 {\em Reviews of Modern Physics\/}
  \href{http://dx.doi.org/10.1103/revmodphys.93.025003}{{\bf 93}}
  \urlprefix\url{https://doi.org/10.1103%2Frevmodphys.93.025003}

\bibitem{GHD_Doyon}
Castro-Alvaredo O~A, Doyon B and Yoshimura T 2016 {\em Phys. Rev. X\/}
  \href{http://dx.doi.org/10.1103/PhysRevX.6.041065}{{\bf 6}(4) 041065}
  \urlprefix\url{https://link.aps.org/doi/10.1103/PhysRevX.6.041065}

\bibitem{GHD_Italy}
Bertini B, Collura M, {De Nardis} J and Fagotti M 2016 {\em Phys. Rev. Lett.\/}
  \href{http://dx.doi.org/10.1103/PhysRevLett.117.207201}{{\bf 117}(20) 207201}
  \urlprefix\url{https://link.aps.org/doi/10.1103/PhysRevLett.117.207201}

\bibitem{Doyon_lectures}
Doyon B 2020 {\em {SciPost} Physics Lecture Notes\/}
  \urlprefix\url{https://doi.org/10.21468%2Fscipostphyslectnotes.18}

\bibitem{GHD_review}
{Jacopo De Nardis and Benjamin Doyon and Marko Medenjak and Mi{\l}osz Panfil}
  2022 {\em Journal of Statistical Mechanics: Theory and Experiment\/}
  \href{http://dx.doi.org/10.1088/1742-5468/ac3658}{{\bf 2022} 014002}
  \urlprefix\url{https://doi.org/10.1088/1742-5468/ac3658}

\bibitem{Bastianello_review}
Bastianello A, Luca A~D and Vasseur R 2021 {\em Journal of Statistical
  Mechanics: Theory and Experiment\/}
  \href{http://dx.doi.org/10.1088/1742-5468/ac26b2}{{\bf 2021} 114003}
  \urlprefix\url{https://doi.org/10.1088%2F1742-5468%2Fac26b2}

\bibitem{CZP95}
Castella H, Zotos X and Prelov{\v{s}}ek P 1995 {\em Physical Review Letters\/}
  \href{http://dx.doi.org/10.1103/physrevlett.74.972}{{\bf 74} 972--975}
  \urlprefix\url{https://doi.org/10.1103%2Fphysrevlett.74.972}

\bibitem{Mazur69}
Mazur P 1969 {\em Physica\/}
  \href{http://dx.doi.org/10.1016/0031-8914(69)90185-2}{{\bf 43} 533--545}
  \urlprefix\url{https://doi.org/10.1016%2F0031-8914%2869%2990185-2}

\bibitem{Suzuki71}
Suzuki M 1971 {\em Physica\/}
  \href{http://dx.doi.org/10.1016/0031-8914(71)90226-6}{{\bf 51} 277--291}
  \urlprefix\url{https://doi.org/10.1016%2F0031-8914%2871%2990226-6}

\bibitem{Ilievski12}
Ilievski E and Prosen T 2012 {\em Communications in Mathematical Physics\/}
  \href{http://dx.doi.org/10.1007/s00220-012-1599-4}{{\bf 318} 809--830}
  \urlprefix\url{https://doi.org/10.1007/s00220-012-1599-4}

\bibitem{Doyon_projections}
Doyon B 2022 {\em Communications in Mathematical Physics\/}
  \urlprefix\url{https://doi.org/10.1007%2Fs00220-022-04310-3}

\bibitem{ZCP97}
Zotos X, Naef F and Prelovsek P 1997 {\em Physical Review B\/}
  \href{http://dx.doi.org/10.1103/physrevb.55.11029}{{\bf 55} 11029--11032}
  \urlprefix\url{https://doi.org/10.1103%2Fphysrevb.55.11029}

\bibitem{Zotos99}
Zotos X 1999 {\em Physical Review Letters\/}
  \href{http://dx.doi.org/10.1103/physrevlett.82.1764}{{\bf 82} 1764--1767}
  \urlprefix\url{https://doi.org/10.1103%2Fphysrevlett.82.1764}

\bibitem{Prosen11}
Prosen T 2011 {\em Physical Review Letters\/}
  \href{http://dx.doi.org/10.1103/physrevlett.106.217206}{{\bf 106}}
  \urlprefix\url{https://doi.org/10.1103%2Fphysrevlett.106.217206}

\bibitem{Marko11}
{\v{Z}}nidari{\v{c}} M 2011 {\em Physical Review Letters\/}
  \href{http://dx.doi.org/10.1103/physrevlett.106.220601}{{\bf 106}}
  \urlprefix\url{https://doi.org/10.1103%2Fphysrevlett.106.220601}

\bibitem{PI13}
Prosen T and Ilievski E 2013 {\em Phys. Rev. Lett.\/}
  \href{http://dx.doi.org/10.1103/PhysRevLett.111.057203}{{\bf 111}(5) 057203}
  \urlprefix\url{https://link.aps.org/doi/10.1103/PhysRevLett.111.057203}

\bibitem{DeLuca17}
{De Luca} A, Collura M and {De Nardis} J 2017 {\em Phys. Rev. B\/}
  \href{http://dx.doi.org/10.1103/PhysRevB.96.020403}{{\bf 96}(2) 020403}
  \urlprefix\url{https://link.aps.org/doi/10.1103/PhysRevB.96.020403}

\bibitem{IN_Drude}
Ilievski E and {De Nardis} J 2017 {\em Physical Review Letters\/}
  \href{http://dx.doi.org/10.1103/physrevlett.119.020602}{{\bf 119}}
  \urlprefix\url{https://doi.org/10.1103%2Fphysrevlett.119.020602}

\bibitem{IN17}
Ilievski E and {De Nardis} J 2017 {\em Phys. Rev. B\/}
  \href{http://dx.doi.org/10.1103/PhysRevB.96.081118}{{\bf 96}(8) 081118}
  \urlprefix\url{https://link.aps.org/doi/10.1103/PhysRevB.96.081118}

\bibitem{Bulchandani18}
Bulchandani V~B, Vasseur R, Karrasch C and Moore J~E 2018 {\em Physical Review
  B\/} \href{http://dx.doi.org/10.1103/physrevb.97.045407}{{\bf 97}}
  \urlprefix\url{https://doi.org/10.1103/physrevb.97.045407}

\bibitem{Ilievski18}
Ilievski E, {De Nardis} J, Medenjak M and Prosen T 2018 {\em Physical Review
  Letters\/} \href{http://dx.doi.org/10.1103/physrevlett.121.230602}{{\bf 121}}
  \urlprefix\url{https://doi.org/10.1103%2Fphysrevlett.121.230602}

\bibitem{Gopalakrishnan18}
Gopalakrishnan S, Huse D~A, Khemani V and Vasseur R 2018 {\em Physical Review
  B\/} \href{http://dx.doi.org/10.1103/physrevb.98.220303}{{\bf 98}}
  \urlprefix\url{https://doi.org/10.1103/physrevb.98.220303}

\bibitem{Ljubotina19}
Ljubotina M, {\v Z}nidari{\v c} M and Prosen T 2019 {\em Physical Review
  Letters\/} \href{http://dx.doi.org/10.1103/physrevlett.122.210602}{{\bf 122}}
  \urlprefix\url{https://doi.org/10.1103%2Fphysrevlett.122.210602}

\bibitem{GV19}
Gopalakrishnan S and Vasseur R 2019 {\em Physical Review Letters\/}
  \href{http://dx.doi.org/10.1103/physrevlett.122.127202}{{\bf 122}}
  \urlprefix\url{https://doi.org/10.1103%2Fphysrevlett.122.127202}

\bibitem{GVW19}
Gopalakrishnan S, Vasseur R and Ware B 2019 {\em Proceedings of the National
  Academy of Sciences\/} \href{http://dx.doi.org/10.1073/pnas.1906914116}{{\bf
  116} 16250--16255} \urlprefix\url{https://doi.org/10.1073%2Fpnas.1906914116}

\bibitem{Urichuk19}
Urichuk A, Oez Y, Klümper A and Sirker J 2019 {\em {SciPost} Physics\/}
  \href{http://dx.doi.org/10.21468/scipostphys.6.1.005}{{\bf 6}}
  \urlprefix\url{https://doi.org/10.21468%2Fscipostphys.6.1.005}

\bibitem{LZP19}
Ljubotina M, Zadnik L and Prosen T 2019 {\em Physical Review Letters\/}
  \href{http://dx.doi.org/10.1103/physrevlett.122.150605}{{\bf 122}}
  \urlprefix\url{https://doi.org/10.1103%2Fphysrevlett.122.150605}

\bibitem{Urichuk21}
Urichuk A, Sirker J and Kl\"{u}mper A 2021 {\em Physical Review B\/}
  \href{http://dx.doi.org/10.1103/physrevb.103.245108}{{\bf 103}}
  \urlprefix\url{https://doi.org/10.1103%2Fphysrevb.103.245108}

\bibitem{DeNardis18}
{De Nardis} J, Bernard D and Doyon B 2018 {\em Physical Review Letters\/}
  \href{http://dx.doi.org/10.1103/physrevlett.121.160603}{{\bf 121}}
  \urlprefix\url{https://doi.org/10.1103/physrevlett.121.160603}

\bibitem{DeNardis_SciPost}
{De Nardis} J, Bernard D and Doyon B 2019 {\em {SciPost} Physics\/}
  \href{http://dx.doi.org/10.21468/scipostphys.6.4.049}{{\bf 6}}
  \urlprefix\url{https://doi.org/10.21468%2Fscipostphys.6.4.049}

\bibitem{superuniversality}
Ilievski E, Nardis J~D, Gopalakrishnan S, Vasseur R and Ware B 2021 {\em
  Physical Review X\/} \href{http://dx.doi.org/10.1103/physrevx.11.031023}{{\bf
  11}} \urlprefix\url{https://doi.org/10.1103%2Fphysrevx.11.031023}

\bibitem{KPZ}
Kardar M, Parisi G and Zhang Y~C 1986 {\em Phys. Rev. Lett.\/}
  \href{http://dx.doi.org/10.1103/PhysRevLett.56.889}{{\bf 56}(9) 889--892}
  \urlprefix\url{https://link.aps.org/doi/10.1103/PhysRevLett.56.889}

\bibitem{NGIV20}
{De Nardis} J, Gopalakrishnan S, Ilievski E and Vasseur R 2020 {\em Physical
  Review Letters\/}
  \href{http://dx.doi.org/10.1103/physrevlett.125.070601}{{\bf 125}}
  \urlprefix\url{https://doi.org/10.1103%2Fphysrevlett.125.070601}

\bibitem{Vir20}
Bulchandani V~B 2020 {\em Physical Review B\/}
  \href{http://dx.doi.org/10.1103/physrevb.101.041411}{{\bf 101}}
  \urlprefix\url{https://doi.org/10.1103%2Fphysrevb.101.041411}

\bibitem{superdiffusion_review}
Bulchandani V~B, Gopalakrishnan S and Ilievski E 2021 {\em Journal of
  Statistical Mechanics: Theory and Experiment\/}
  \href{http://dx.doi.org/10.1088/1742-5468/ac12c7}{{\bf 2021} 084001}
  \urlprefix\url{https://doi.org/10.1088%2F1742-5468%2Fac12c7}

\bibitem{QLreview}
Ilievski E, Medenjak M, Prosen T and Zadnik L 2016 {\em Journal of Statistical
  Mechanics: Theory and Experiment\/}
  \href{http://dx.doi.org/10.1088/1742-5468/2016/06/064008}{{\bf 2016} 064008}
  \urlprefix\url{https://doi.org/10.1088%2F1742-5468%2F2016%2F06%2F064008}

\bibitem{Faddeev95}
Faddeev L 1995 {\em International Journal of Modern Physics A\/}
  \href{http://dx.doi.org/10.1142/s0217751x95000905}{{\bf 10} 1845--1878}
  \urlprefix\url{https://doi.org/10.1142%2Fs0217751x95000905}

\bibitem{Faddeev16}
Faddeev L 2016 {How the Algebraic Bethe Ansatz Works for Integrable Models}
  {\em Fifty Years of Mathematical Physics\/} ({WORLD} {SCIENTIFIC}) pp
  370--439 \urlprefix\url{https://doi.org/10.1142%2F9789814340960_0031}

\bibitem{Benz05}
Benz J, Fukui T, Kl\"{u}mper A and Scheeren C 2005 {\em Journal of the Physical
  Society of Japan\/} \href{http://dx.doi.org/10.1143/jpsjs.74s.181}{{\bf 74}
  181--190} \urlprefix\url{https://doi.org/10.1143%2Fjpsjs.74s.181}

\bibitem{DS17}
Doyon B and Spohn H 2017 {\em SciPost Phys.\/}
  \href{http://dx.doi.org/10.21468/SciPostPhys.3.6.039}{{\bf 3}(6) 039}
  \urlprefix\url{https://scipost.org/10.21468/SciPostPhys.3.6.039}

\bibitem{StringCharge}
Ilievski E, Quinn E, Nardis J~D and Brockmann M 2016 {\em Journal of
  Statistical Mechanics: Theory and Experiment\/}
  \href{http://dx.doi.org/10.1088/1742-5468/2016/06/063101}{{\bf 2016} 063101}
  \urlprefix\url{https://doi.org/10.1088%2F1742-5468%2F2016%2F06%2F063101}

\bibitem{Collura_wall}
Collura M, Luca A~D and Viti J 2018 {\em Physical Review B\/}
  \href{http://dx.doi.org/10.1103/physrevb.97.081111}{{\bf 97}}
  \urlprefix\url{https://doi.org/10.1103%2Fphysrevb.97.081111}

\bibitem{Urichuk22}
{Urichuk, Andrew and Kl{\"u}mper, Andreas and Sirker, Jesko} 2022 {\em arXiv
  preprint arXiv:2202.08410\/}

\bibitem{Prosen14}
Prosen T 2014 {\em Nuclear Physics B\/}
  \href{http://dx.doi.org/10.1016/j.nuclphysb.2014.07.024}{{\bf 886}
  1177--1198} \urlprefix\url{https://doi.org/10.1016%2Fj.nuclphysb.2014.07.024}

\bibitem{Pereira14}
Pereira R~G, Pasquier V, Sirker J and Affleck I 2014 {\em Journal of
  Statistical Mechanics: Theory and Experiment\/} {\bf 2014} P09037
  \urlprefix\url{http://stacks.iop.org/1742-5468/2014/i=9/a=P09037}

\bibitem{Zadnik16}
Zadnik L, Medenjak M and Prosen T 2016 {\em Nuclear Physics B\/}
  \href{http://dx.doi.org/10.1016/j.nuclphysb.2015.11.023}{{\bf 902} 339--353}
  \urlprefix\url{https://doi.org/10.1016%2Fj.nuclphysb.2015.11.023}

\bibitem{Zadnik17}
Zadnik L and Prosen T 2017 {\em Journal of Physics A: Mathematical and
  Theoretical\/} \href{http://dx.doi.org/10.1088/1751-8121/aa6e09}{{\bf 50}
  265203} \urlprefix\url{https://doi.org/10.1088%2F1751-8121%2Faa6e09}

\bibitem{Ilievski_GGE}
Ilievski E, {De Nardis} J, Wouters B, Caux J~S, Essler F~H~L and Prosen T 2015
  {\em Phys. Rev. Lett.\/}
  \href{http://dx.doi.org/10.1103/PhysRevLett.115.157201}{{\bf 115}(15) 157201}
  \urlprefix\url{http://link.aps.org/doi/10.1103/PhysRevLett.115.157201}

\bibitem{Landscape}
Ilievski E and Quinn E 2019 {\em {SciPost} Physics\/}
  \href{http://dx.doi.org/10.21468/scipostphys.7.3.033}{{\bf 7}}
  \urlprefix\url{https://doi.org/10.21468%2Fscipostphys.7.3.033}

\bibitem{PVC16}
Piroli L, Vernier E and Calabrese P 2016 {\em Physical Review B\/}
  \href{http://dx.doi.org/10.1103/physrevb.94.054313}{{\bf 94}}
  \urlprefix\url{https://doi.org/10.1103%2Fphysrevb.94.054313}

\bibitem{IQC17}
Ilievski E, Quinn E and Caux J~S 2017 {\em Physical Review B\/}
  \href{http://dx.doi.org/10.1103/physrevb.95.115128}{{\bf 95}}
  \urlprefix\url{https://doi.org/10.1103%2Fphysrevb.95.115128}

\bibitem{VR16}
Vidmar L and Rigol M 2016 {\em Journal of Statistical Mechanics: Theory and
  Experiment\/} \href{http://dx.doi.org/10.1088/1742-5468/2016/06/064007}{{\bf
  2016} 064007}
  \urlprefix\url{https://doi.org/10.1088%2F1742-5468%2F2016%2F06%2F064007}

\bibitem{Alba_review}
Alba V, Bertini B, Fagotti M, Piroli L and Ruggiero P 2021 {\em Journal of
  Statistical Mechanics: Theory and Experiment\/}
  \href{http://dx.doi.org/10.1088/1742-5468/ac257d}{{\bf 2021} 114004}
  \urlprefix\url{https://doi.org/10.1088%2F1742-5468%2Fac257d}

\bibitem{Abbott_book}
Abbott S 2015 {\em {Understanding Analysis}\/} (Springer New York)
  \urlprefix\url{https://doi.org/10.1007%2F978-1-4939-2712-8}

\bibitem{Herbrych11}
Herbrych J, Prelov{\v{s}}ek P and Zotos X 2011 {\em Physical Review B\/}
  \href{http://dx.doi.org/10.1103/physrevb.84.155125}{{\bf 84}}
  \urlprefix\url{https://doi.org/10.1103%2Fphysrevb.84.155125}

\bibitem{Mierzejewski21}
Mierzejewski M, Herbrych J and Prelov{\v{s}}ek P 2021 {\em Physical Review B\/}
  \href{http://dx.doi.org/10.1103/physrevb.103.235115}{{\bf 103}}
  \urlprefix\url{https://doi.org/10.1103%2Fphysrevb.103.235115}

\bibitem{Karrasch12}
Karrasch C, Bardarson J~H and Moore J~E 2012 {\em Physical Review Letters\/}
  \href{http://dx.doi.org/10.1103/physrevlett.108.227206}{{\bf 108}}
  \urlprefix\url{https://doi.org/10.1103%2Fphysrevlett.108.227206}

\bibitem{Karrasch13}
Karrasch C, Hauschild J, Langer S and Heidrich-Meisner F 2013 {\em Physical
  Review B\/} \href{http://dx.doi.org/10.1103/physrevb.87.245128}{{\bf 87}}
  \urlprefix\url{https://doi.org/10.1103%2Fphysrevb.87.245128}

\bibitem{Karrasch17}
Karrasch C 2017 {\em New Journal of Physics\/}
  \href{http://dx.doi.org/10.1088/1367-2630/aa631a}{{\bf 19} 033027}
  \urlprefix\url{https://doi.org/10.1088%2F1367-2630%2Faa631a}

\bibitem{Azbel64}
Azbel M~Y 1964 {\em Sov. Phys. JETP\/} {\bf 19} 634--645

\bibitem{Hofstadter76}
Hofstadter D~R 1976 {\em Physical Review B\/}
  \href{http://dx.doi.org/10.1103/physrevb.14.2239}{{\bf 14} 2239--2249}
  \urlprefix\url{https://doi.org/10.1103%2Fphysrevb.14.2239}

\bibitem{WZ94}
Wiegmann P and Zabrodin A 1994 {\em Nuclear Physics B\/}
  \href{http://dx.doi.org/10.1016/0550-3213(94)90443-x}{{\bf 422} 495--514}
  \urlprefix\url{https://doi.org/10.1016%2F0550-3213%2894%2990443-x}

\bibitem{FK95}
Faddeev L~D and Kashaev R~M 1995 {\em Communications in Mathematical Physics\/}
  \href{http://dx.doi.org/10.1007/bf02101600}{{\bf 169} 181--191}
  \urlprefix\url{https://doi.org/10.1007%2Fbf02101600}

\bibitem{ATW98}
Abanov A, Talstra J and Wiegmann P 1998 {\em Nuclear Physics B\/}
  \href{http://dx.doi.org/10.1016/s0550-3213(98)00346-0}{{\bf 525} 571--596}
  \urlprefix\url{https://doi.org/10.1016%2Fs0550-3213%2898%2900346-0}

\bibitem{TS72}
Takahashi M and Suzuki M 1972 {\em Progress of Theoretical Physics\/}
  \href{http://dx.doi.org/10.1143/ptp.48.2187}{{\bf 48} 2187--2209}
  \urlprefix\url{https://doi.org/10.1143%2Fptp.48.2187}

\bibitem{KR87I}
Kirillov A~N and Reshetikhin N~Y 1987 {\em Journal of Physics A: Mathematical
  and General\/} \href{http://dx.doi.org/10.1088/0305-4470/20/6/038}{{\bf 20}
  1565--1585}
  \urlprefix\url{https://doi.org/10.1088%2F0305-4470%2F20%2F6%2F038}

\bibitem{KR87II}
Kirillov A~N and Reshetikhin N~Y 1987 {\em Journal of Physics A: Mathematical
  and General\/} \href{http://dx.doi.org/10.1088/0305-4470/20/6/039}{{\bf 20}
  1587--1597}
  \urlprefix\url{https://doi.org/10.1088%2F0305-4470%2F20%2F6%2F039}

\bibitem{YY69}
Yang C~N and Yang C~P 1969 {\em Journal of Mathematical Physics\/}
  \href{http://dx.doi.org/10.1063/1.1664947}{{\bf 10} 1115--1122}
  \urlprefix\url{https://doi.org/10.1063%2F1.1664947}

\bibitem{Takahashi_Heisenberg}
Takahashi M 1971 {\em Physics Letters A\/}
  \href{http://dx.doi.org/10.1016/0375-9601(71)90531-7}{{\bf 36} 325--326}
  \urlprefix\url{https://doi.org/10.1016%2F0375-9601%2871%2990531-7}

\bibitem{Gaudin71}
Gaudin M 1971 {\em Physical Review Letters\/}
  \href{http://dx.doi.org/10.1103/physrevlett.26.1301}{{\bf 26} 1301--1304}
  \urlprefix\url{https://doi.org/10.1103%2Fphysrevlett.26.1301}

\bibitem{Araki69}
Araki H 1969 {\em Communications in Mathematical Physics\/}
  \href{http://dx.doi.org/10.1007/bf01645134}{{\bf 14} 120--157}
  \urlprefix\url{https://doi.org/10.1007%2Fbf01645134}

\bibitem{Takahashi_Hubbard}
Takahashi M 1972 {\em Progress of Theoretical Physics\/}
  \href{http://dx.doi.org/10.1143/ptp.47.69}{{\bf 47} 69--82}
  \urlprefix\url{https://doi.org/10.1143%2Fptp.47.69}

\bibitem{KP92}
Kl\"{u}mper A and Pearce P~A 1992 {\em Physica A: Statistical Mechanics and its
  Applications\/} \href{http://dx.doi.org/10.1016/0378-4371(92)90149-k}{{\bf
  183} 304--350}
  \urlprefix\url{https://doi.org/10.1016%2F0378-4371%2892%2990149-k}

\bibitem{equivalence}
Takahashi M, Shiroishi M and Klümper A 2001 {\em Journal of Physics A:
  Mathematical and General\/}
  \href{http://dx.doi.org/10.1088/0305-4470/34/13/105}{{\bf 34} L187--L194}
  \urlprefix\url{https://doi.org/10.1088%2F0305-4470%2F34%2F13%2F105}

\bibitem{QTM_review}
Klümper A 2004 {Integrability of quantum chains: Theory and applications to
  the spin-1/2 XXZ chain} {\em Quantum Magnetism\/} (Springer Berlin
  Heidelberg) pp 349--379 \urlprefix\url{https://doi.org/10.1007%2Fbfb0119598}

\bibitem{Mestyan14}
Mesty{\'{a}}n M and Bal{\'{a}}zsgay 2014 {\em Journal of Statistical Mechanics:
  Theory and Experiment\/}
  \href{http://dx.doi.org/10.1088/1742-5468/2014/09/p09020}{{\bf 2014} P09020}
  \urlprefix\url{https://doi.org/10.1088%2F1742-5468%2F2014%2F09%2Fp09020}

\bibitem{Pozsgay_PRX}
Pozsgay B 2020 {\em {SciPost} Physics\/}
  \href{http://dx.doi.org/10.21468/scipostphys.8.2.016}{{\bf 8}}
  \urlprefix\url{https://doi.org/10.21468%2Fscipostphys.8.2.016}

\bibitem{Pozsgay_currents}
Pozsgay B 2020 {\em Physical Review Letters\/}
  \href{http://dx.doi.org/10.1103/physrevlett.125.070602}{{\bf 125}}
  \urlprefix\url{https://doi.org/10.1103%2Fphysrevlett.125.070602}

\bibitem{FZ82}
Fateev V and Zamolodchikov A 1982 {\em Physics Letters A\/}
  \href{http://dx.doi.org/10.1016/0375-9601(82)90736-8}{{\bf 92} 37--39}
  \urlprefix\url{https://doi.org/10.1016%2F0375-9601%2882%2990736-8}

\bibitem{KRS90}
Kulish P, Reshetikhin N and Sklyanin E 1990
  \href{http://dx.doi.org/10.1142/9789812798336_0027}{ 498--508}
  \urlprefix\url{https://doi.org/10.1142%2F9789812798336_0027}

\bibitem{Frahm90}
Frahm H, Yu N~C and Fowler M 1990 {\em Nuclear Physics B\/}
  \href{http://dx.doi.org/10.1016/0550-3213(90)90435-g}{{\bf 336} 396--434}
  \urlprefix\url{https://doi.org/10.1016%2F0550-3213%2890%2990435-g}

\bibitem{FY90}
Frahm H and Yu N~C 1990 {\em Journal of Physics A: Mathematical and General\/}
  \href{http://dx.doi.org/10.1088/0305-4470/23/11/032}{{\bf 23} 2115--2132}
  \urlprefix\url{https://doi.org/10.1088%2F0305-4470%2F23%2F11%2F032}

\bibitem{Korepin79}
Korepin V~E 1979 {\em Theoretical and Mathematical Physics\/}
  \href{http://dx.doi.org/10.1007/bf01028501}{{\bf 41} 953--967}
  \urlprefix\url{https://doi.org/10.1007%2Fbf01028501}

\bibitem{Miao21}
Miao Y, Lamers J and Pasquier V 2021 {\em {SciPost} Physics\/}
  \href{http://dx.doi.org/10.21468/scipostphys.11.3.067}{{\bf 11}}
  \urlprefix\url{https://doi.org/10.21468%2Fscipostphys.11.3.067}

\bibitem{Korff01}
Korff C and McCoy B~M 2001 {\em Nuclear Physics B\/}
  \href{http://dx.doi.org/10.1016/s0550-3213(01)00417-5}{{\bf 618} 551--569}
  \urlprefix\url{https://doi.org/10.1016%2Fs0550-3213%2801%2900417-5}

\bibitem{Korff03}
Korff C 2003 {\em Journal of Physics A: Mathematical and General\/}
  \href{http://dx.doi.org/10.1088/0305-4470/36/19/305}{{\bf 36} 5229--5266}
  \urlprefix\url{https://doi.org/10.1088%2F0305-4470%2F36%2F19%2F305}

\bibitem{Deguchi06}
Deguchi T 2006 {\em Symmetry, Integrability and Geometry: Methods and
  Applications\/} \urlprefix\url{https://doi.org/10.3842%2Fsigma.2006.021}

\bibitem{FM1}
Fabricius K and McCoy B~M 2001 {\em Journal of Statistical Physics\/}
  \href{http://dx.doi.org/10.1023/a:1010380116927}{{\bf 103} 647--678}
  \urlprefix\url{https://doi.org/10.1023%2Fa%3A1010380116927}

\bibitem{FM2}
Fabricius K and McCoy B~M 2001 {\em Journal of Statistical Physics\/}
  \href{http://dx.doi.org/10.1023/a:1010372504158}{{\bf 104} 573--587}
  \urlprefix\url{https://doi.org/10.1023%2Fa%3A1010372504158}

\bibitem{BA99}
Braak D and Andrei N 1999 {\em Nuclear Physics B\/}
  \href{http://dx.doi.org/10.1016/s0550-3213(98)00811-6}{{\bf 542} 551--580}
  \urlprefix\url{https://doi.org/10.1016%2Fs0550-3213%2898%2900811-6}

\bibitem{Deguchi07}
Deguchi T 2007 {\em Journal of Physics A: Mathematical and Theoretical\/}
  \href{http://dx.doi.org/10.1088/1751-8113/40/27/005}{{\bf 40} 7473--7508}
  \urlprefix\url{https://doi.org/10.1088%2F1751-8113%2F40%2F27%2F005}

\bibitem{Miao_Onsager}
Miao Y 2021 {\em {SciPost} Physics\/}
  \href{http://dx.doi.org/10.21468/scipostphys.11.3.066}{{\bf 11}}
  \urlprefix\url{https://doi.org/10.21468%2Fscipostphys.11.3.066}

\bibitem{PV_XXZ}
Piroli L and Vernier E 2016 {\em Journal of Statistical Mechanics: Theory and
  Experiment\/} \href{http://dx.doi.org/10.1088/1742-5468/2016/05/053106}{{\bf
  2016} 053106}
  \urlprefix\url{https://doi.org/10.1088%2F1742-5468%2F2016%2F05%2F053106}

\bibitem{Sklyanin79}
Sklyanin E 1979

\bibitem{Faddeev_book}
Faddeev L~D and Takhtajan L~A 1987 {\em Hamiltonian Methods in the Theory of
  Solitons\/} (Springer Berlin Heidelberg)
  \urlprefix\url{https://doi.org/10.1007/978-3-540-69969-9}

\bibitem{MIG21}
Miao Y, Ilievski E and Gamayun O 2021 {\em {SciPost} Physics\/}
  \href{http://dx.doi.org/10.21468/scipostphys.10.4.086}{{\bf 10}}
  \urlprefix\url{https://doi.org/10.21468%2Fscipostphys.10.4.086}

\bibitem{Sklyanin83}
Sklyanin E~K 1983 {\em Functional Analysis and Its Applications\/}
  \href{http://dx.doi.org/10.1007/bf01077848}{{\bf 16} 263--270}
  \urlprefix\url{https://doi.org/10.1007%2Fbf01077848}

\bibitem{Sklyanin88}
Sklyanin E~K 1988 {\em Journal of Soviet Mathematics\/}
  \href{http://dx.doi.org/10.1007/bf01084941}{{\bf 40} 93--107}
  \urlprefix\url{https://doi.org/10.1007%2Fbf01084941}

\bibitem{Sklyanin89XXZ}
Sklyanin E~K 1989 {\em Journal of Soviet Mathematics\/}
  \href{http://dx.doi.org/10.1007/bf01096094}{{\bf 46} 2104--2111}
  \urlprefix\url{https://doi.org/10.1007%2Fbf01096094}

\bibitem{LLMM}
Krajnik {\v{Z}}, Ilievski E, Prosen T and Pasquier V 2021 {\em {SciPost}
  Physics\/} \href{http://dx.doi.org/10.21468/scipostphys.11.3.051}{{\bf 11}}
  \urlprefix\url{https://doi.org/10.21468%2Fscipostphys.11.3.051}

\bibitem{Bargheer08}
Bargheer T, Beisert N and Gromov N 2008 {\em New Journal of Physics\/}
  \href{http://dx.doi.org/10.1088/1367-2630/10/10/103023}{{\bf 10} 103023}
  \urlprefix\url{https://doi.org/10.1088%2F1367-2630%2F10%2F10%2F103023}

\bibitem{Bojan}
Prosen T and {\v Z}unkovi{\v c} B 2013 {\em Physical Review Letters\/}
  \href{http://dx.doi.org/10.1103/physrevlett.111.040602}{{\bf 111}}
  \urlprefix\url{https://doi.org/10.1103%2Fphysrevlett.111.040602}

\bibitem{Koch21}
Koch R, Caux J~S and Bastianello A 2021 {\em arXiv preprint arXiv:2110.14574\/}

\bibitem{DamleSachdev05}
Damle K and Sachdev S 2005 {\em Physical Review Letters\/}
  \href{http://dx.doi.org/10.1103/physrevlett.95.187201}{{\bf 95}}
  \urlprefix\url{https://doi.org/10.1103%2Fphysrevlett.95.187201}

\bibitem{KormosZaradi16}
Kormos M and Zar{\'{a}}nd G 2016 {\em Physical Review E\/}
  \href{http://dx.doi.org/10.1103/physreve.93.062101}{{\bf 93}}
  \urlprefix\url{https://doi.org/10.1103%2Fphysreve.93.062101}

\bibitem{Kukuljan18}
Kukuljan I, Sotiriadis S and Takacs G 2018 {\em Physical Review Letters\/}
  \href{http://dx.doi.org/10.1103/physrevlett.121.110402}{{\bf 121}}
  \urlprefix\url{https://doi.org/10.1103%2Fphysrevlett.121.110402}

\bibitem{Horvath19}
Horv{\'{a}}th D~X, Lovas I, Kormos M, Tak{\'{a}}cs G and Zar{\'{a}}nd G 2019
  {\em Physical Review A\/}
  \href{http://dx.doi.org/10.1103/physreva.100.013613}{{\bf 100}}
  \urlprefix\url{https://doi.org/10.1103%2Fphysreva.100.013613}

\bibitem{Bertini_sG}
Bertini B, Piroli L and Kormos M 2019 {\em Physical Review B\/}
  \href{http://dx.doi.org/10.1103/physrevb.100.035108}{{\bf 100}}
  \urlprefix\url{https://doi.org/10.1103%2Fphysrevb.100.035108}

\bibitem{Coleman75}
Coleman S 1975 {\em Physical Review D\/}
  \href{http://dx.doi.org/10.1103/physrevd.11.2088}{{\bf 11} 2088--2097}
  \urlprefix\url{https://doi.org/10.1103%2Fphysrevd.11.2088}

\bibitem{Pohlmeyer77}
Pohlmeyer K 1977 {The Classical Sine Gordon Theory} {\em New Developments in
  Quantum Field Theory and Statistical Mechanics Carg{\`{e}}se 1976\/}
  (Springer {US}) pp 307--338
  \urlprefix\url{https://doi.org/10.1007%2F978-1-4615-8918-1_13}

\bibitem{Z77}
Zamolodchikov A~B 1977 {\em Communications in Mathematical Physics\/}
  \href{http://dx.doi.org/10.1007/bf01626520}{{\bf 55} 183--186}
  \urlprefix\url{https://doi.org/10.1007%2Fbf01626520}

\bibitem{Pohlmeyer78}
Pohlmeyer K 1978 {Solitons and Breathers} {\em Many Degrees of Freedom in Field
  Theory\/} (Springer {US}) pp 63--116
  \urlprefix\url{https://doi.org/10.1007%2F978-1-4615-8924-2_3}

\bibitem{ZZ79}
Zamolodchikov A~B and Zamolodchikov A~B 1979 {\em Annals of Physics\/}
  \href{http://dx.doi.org/10.1016/0003-4916(79)90391-9}{{\bf 120} 253--291}
  \urlprefix\url{https://doi.org/10.1016%2F0003-4916%2879%2990391-9}

\bibitem{Klassen93}
KLASSEN T~R and MELZER E 1993 {\em International Journal of Modern Physics A\/}
  \href{http://dx.doi.org/10.1142/s0217751x93001703}{{\bf 08} 4131--4174}
  \urlprefix\url{https://doi.org/10.1142%2Fs0217751x93001703}

\bibitem{Feher11}
Feh{\'{e}}r G and Tak{\'{a}}cs G 2011 {\em Nuclear Physics B\/}
  \href{http://dx.doi.org/10.1016/j.nuclphysb.2011.06.020}{{\bf 852} 441--467}
  \urlprefix\url{https://doi.org/10.1016%2Fj.nuclphysb.2011.06.020}

\bibitem{Haldane80}
Haldane F~D~M 1980 {\em Physical Review Letters\/}
  \href{http://dx.doi.org/10.1103/physrevlett.45.1358}{{\bf 45} 1358--1362}
  \urlprefix\url{https://doi.org/10.1103%2Fphysrevlett.45.1358}

\bibitem{Z95}
Zamolodchikov A~B 1995 {\em International Journal of Modern Physics A\/}
  \href{http://dx.doi.org/10.1142/s0217751x9500053x}{{\bf 10} 1125--1150}
  \urlprefix\url{https://doi.org/10.1142%2Fs0217751x9500053x}

\bibitem{IMP15}
Ilievski E, Medenjak M and Prosen T 2015 {\em Physical Review Letters\/}
  \href{http://dx.doi.org/10.1103/physrevlett.115.120601}{{\bf 115}}
  \urlprefix\url{https://doi.org/10.1103%2Fphysrevlett.115.120601}

\bibitem{DDV87}
Destri C and Vega H~D 1987 {\em Nuclear Physics B\/}
  \href{http://dx.doi.org/10.1016/0550-3213(87)90193-3}{{\bf 290} 363--391}
  \urlprefix\url{https://doi.org/10.1016%2F0550-3213%2887%2990193-3}

\bibitem{DeVega90}
Vega H~J~D 1990 {Yang-Baxter Algebras, Integrable Theories and Quantum Groups}
  {\em {NATO} {ASI} Series\/} (Springer {US}) pp 387--434
  \urlprefix\url{https://doi.org/10.1007%2F978-1-4615-3802-8_10}

\bibitem{Vernier17}
Vernier E and Cubero A~C 2017 {\em Journal of Statistical Mechanics: Theory and
  Experiment\/} \href{http://dx.doi.org/10.1088/1742-5468/aa5288}{{\bf 2017}
  023101} \urlprefix\url{https://doi.org/10.1088%2F1742-5468%2Faa5288}

\bibitem{Tateo95}
Tateo R 1995 {\em Physics Letters B\/}
  \href{http://dx.doi.org/10.1016/0370-2693(95)00751-6}{{\bf 355} 157--164}
  \urlprefix\url{https://doi.org/10.1016%2F0370-2693%2895%2900751-6}

\bibitem{NakanishiStella}
Nakanishi T and Stella S 2016 {\em Transactions of the American Mathematical
  Society\/} \href{http://dx.doi.org/10.1090/tran/6505}{{\bf 368} 6835--6886}
  \urlprefix\url{https://doi.org/10.1090%2Ftran%2F6505}

\bibitem{LeClair89}
Leclair A 1989 {\em Physics Letters B\/}
  \href{http://dx.doi.org/10.1016/0370-2693(89)91661-4}{{\bf 230} 103--107}
  \urlprefix\url{https://doi.org/10.1016%2F0370-2693%2889%2991661-4}

\bibitem{BL90}
Bernard D and Leclair A 1990 {\em Nuclear Physics B\/}
  \href{http://dx.doi.org/10.1016/0550-3213(90)90466-q}{{\bf 340} 721--751}
  \urlprefix\url{https://doi.org/10.1016%2F0550-3213%2890%2990466-q}

\bibitem{Babelon96}
Babelon O, Bernard D and Smirnov F~A 1996 {\em Communications in Mathematical
  Physics\/} \href{http://dx.doi.org/10.1007/bf02517893}{{\bf 182} 319--354}
  \urlprefix\url{https://doi.org/10.1007%2Fbf02517893}

\bibitem{Balog04}
Balog J and s {\'{A}}~H 2004 {\em Journal of Physics A: Mathematical and
  General\/} \href{http://dx.doi.org/10.1088/0305-4470/37/5/028}{{\bf 37}
  1903--1925}
  \urlprefix\url{https://doi.org/10.1088%2F0305-4470%2F37%2F5%2F028}

\bibitem{Mezincescu90}
Mezincescu L and Nepomechie R~I 1990 {\em Physics Letters B\/}
  \href{http://dx.doi.org/10.1016/0370-2693(90)90622-d}{{\bf 246} 412--416}
  \urlprefix\url{https://doi.org/10.1016%2F0370-2693%2890%2990622-d}

\bibitem{PS90}
Pasquier V and Saleur H 1990 {\em Nuclear Physics B\/}
  \href{http://dx.doi.org/10.1016/0550-3213(90)90122-t}{{\bf 330} 523--556}
  \urlprefix\url{https://doi.org/10.1016%2F0550-3213%2890%2990122-t}

\bibitem{Kac77}
Kac V 1977 {\em Advances in Mathematics\/}
  \href{http://dx.doi.org/10.1016/0001-8708(77)90017-2}{{\bf 26} 8--96}
  \urlprefix\url{https://doi.org/10.1016%2F0001-8708%2877%2990017-2}

\bibitem{Hubbard_book}
{Fabian H L Essler and Holger Frahm and Frank G{\"o}hmann and Andreas
  Kl{\"u}mper and Vladimir E Korepin} 2005 {\em {The One-Dimensional Hubbard
  Model}\/} (Cambridge University Press ({CUP}))
  \urlprefix\url{https://doi.org/10.1017%2Fcbo9780511534843}

\bibitem{Schlottmann87}
Schlottmann P 1987 {\em Physical Review B\/}
  \href{http://dx.doi.org/10.1103/physrevb.36.5177}{{\bf 36} 5177--5185}
  \urlprefix\url{https://doi.org/10.1103%2Fphysrevb.36.5177}

\bibitem{EKS92}
Essler F~H~L, Korepin V~E and Schoutens K 1992 {\em Physical Review Letters\/}
  \href{http://dx.doi.org/10.1103/physrevlett.68.2960}{{\bf 68} 2960--2963}
  \urlprefix\url{https://doi.org/10.1103%2Fphysrevlett.68.2960}

\bibitem{EKS94}
E{\ss}ler F~H, Korepin V~E and Schoutens K 1994
  \href{http://dx.doi.org/10.1142/s0217979294001354}{{\bf 08} 3205--3242}
  \urlprefix\url{https://doi.org/10.1142%2Fs0217979294001354}

\bibitem{Schultz81}
Schultz C~L 1981 {\em Physical Review Letters\/}
  \href{http://dx.doi.org/10.1103/physrevlett.46.629}{{\bf 46} 629--632}
  \urlprefix\url{https://doi.org/10.1103%2Fphysrevlett.46.629}

\bibitem{PerkSchultz81}
Perk J~H and Schultz C~L 1981 {\em Physics Letters A\/}
  \href{http://dx.doi.org/10.1016/0375-9601(81)90994-4}{{\bf 84} 407--410}
  \urlprefix\url{https://doi.org/10.1016%2F0375-9601%2881%2990994-4}

\bibitem{VegaLopes91}
de~Vega H~J and Lopes E 1991 {\em Physical Review Letters\/}
  \href{http://dx.doi.org/10.1103/physrevlett.67.489}{{\bf 67} 489--492}
  \urlprefix\url{https://doi.org/10.1103%2Fphysrevlett.67.489}

\bibitem{Beisert14}
Beisert N and de~Leeuw M 2014 {\em Journal of Physics A: Mathematical and
  Theoretical\/} \href{http://dx.doi.org/10.1088/1751-8113/47/30/305201}{{\bf
  47} 305201}
  \urlprefix\url{https://doi.org/10.1088%2F1751-8113%2F47%2F30%2F305201}

\bibitem{Beisert11}
Beisert N 2011 {\em Letters in Mathematical Physics\/}
  \href{http://dx.doi.org/10.1007/s11005-011-0479-8}{{\bf 99} 529--545}
  \urlprefix\url{https://doi.org/10.1007%2Fs11005-011-0479-8}

\bibitem{Beisert_review}
Beisert N 2011 {\em Letters in Mathematical Physics\/}
  \href{http://dx.doi.org/10.1007/s11005-011-0479-8}{{\bf 99} 529--545}
  \urlprefix\url{https://doi.org/10.1007%2Fs11005-011-0479-8}

\bibitem{Beisert07}
Beisert N 2007 {The S-Matrix of AdS/CFT and Yangian Symmetry} {\em Proceedings
  of {BETHE} {ANSATZ}: 75 {YEARS} {LATER} {\textemdash} {PoS}(Solvay)\/} (Sissa
  Medialab) \urlprefix\url{https://doi.org/10.22323%2F1.038.0002}

\bibitem{Shastry86}
Shastry B~S 1986 {\em Physical Review Letters\/}
  \href{http://dx.doi.org/10.1103/physrevlett.56.2334.3}{{\bf 56} 2334--2335}
  \urlprefix\url{https://doi.org/10.1103%2Fphysrevlett.56.2334.3}

\bibitem{EK_Smatrix}
Essler F~H and Korepin V~E 1994 {\em Nuclear Physics B\/}
  \href{http://dx.doi.org/10.1016/0550-3213(94)90019-1}{{\bf 426} 505--533}
  \urlprefix\url{https://doi.org/10.1016%2F0550-3213%2894%2990019-1}

\bibitem{LiebWu68}
Lieb E~H and Wu F~Y 1968 {\em Physical Review Letters\/}
  \href{http://dx.doi.org/10.1103/physrevlett.21.192.2}{{\bf 21} 192--192}
  \urlprefix\url{https://doi.org/10.1103%2Fphysrevlett.21.192.2}

\bibitem{BK08}
Beisert N and Koroteev P 2008 {\em Journal of Physics A: Mathematical and
  Theoretical\/} \href{http://dx.doi.org/10.1088/1751-8113/41/25/255204}{{\bf
  41} 255204}
  \urlprefix\url{https://doi.org/10.1088%2F1751-8113%2F41%2F25%2F255204}

\bibitem{AB99}
Alcaraz F~C and Bariev R~Z 1999 {\em Journal of Physics A: Mathematical and
  General\/} \href{http://dx.doi.org/10.1088/0305-4470/32/35/101}{{\bf 32}
  L387--L392}
  \urlprefix\url{https://doi.org/10.1088%2F0305-4470%2F32%2F35%2F101}

\bibitem{FQ12}
Frolov S and Quinn E 2012 {\em Journal of Physics A: Mathematical and
  Theoretical\/} \href{http://dx.doi.org/10.1088/1751-8113/45/9/095004}{{\bf
  45} 095004}
  \urlprefix\url{https://doi.org/10.1088%2F1751-8113%2F45%2F9%2F095004}

\bibitem{Arutyunov12I}
Arutyunov G, de~Leeuw M and van Tongeren S~J 2012 {\em Journal of High Energy
  Physics\/} \href{http://dx.doi.org/10.1007/jhep10(2012)090}{{\bf 2012}}
  \urlprefix\url{https://doi.org/10.1007%2Fjhep10%282012%29090}

\bibitem{Arutyunov12II}
Arutyunov G, de~Leeuw M and van Tongeren S~J 2013 {\em Journal of High Energy
  Physics\/} \href{http://dx.doi.org/10.1007/jhep02(2013)012}{{\bf 2013}}
  \urlprefix\url{https://doi.org/10.1007%2Fjhep02%282013%29012}

\end{thebibliography}

\end{document}